\documentclass[12pt,twoside]{article}

\usepackage[default]{jasa_harvard}    
\usepackage{JASA_manu}

\RequirePackage[OT1]{fontenc}
\RequirePackage{amsthm,amsmath}
\RequirePackage{amsthm,amsmath,fancyhdr}
\RequirePackage[colorlinks,citecolor=blue,urlcolor=blue]{hyperref}
\RequirePackage{hypernat}
\RequirePackage{graphicx,epstopdf,subfigure,latexsym,amssymb}
\RequirePackage{float,epsfig,multirow,rotating,times}
\RequirePackage{upgreek,wrapfig}
\RequirePackage{comment}
\newcommand {\ctn}{\citeasnoun} 

\newcommand{\btheta}{\boldsymbol{\theta}}
\newcommand{\bTheta}{\boldsymbol{\Theta}}
\newcommand{\bLambda}{\boldsymbol{\Lambda}}

\newcommand{\bbeta}{\boldsymbol{\beta}}

\newcommand{\bSigma}{\boldsymbol{\Sigma}}

\newcommand{\bmu}{\boldsymbol{\mu}}
\newcommand{\bzeta}{\boldsymbol{\zeta}}

\newcommand{\bG}{\boldsymbol{G}}

\newcommand{\bS}{\boldsymbol{S}}

\newcommand{\bx}{\boldsymbol{x}}
\newcommand{\bX}{\boldsymbol{X}}
\newcommand{\by}{\boldsymbol{y}}
\newcommand{\bY}{\boldsymbol{Y}}

\begin{document}

\normalsize

\title{\vspace{-0.8in}
An Improved Bayesian Semiparametric Model for Palaeoclimate Reconstruction:
Cross-validation Based Model Assessment}
\author{Sabyasachi Mukhopadhyay and Sourabh Bhattacharya\thanks{
Sabyasachi Mukhopadhyay is a postdoctoral researcher in Southampton Statistical Sciences Research Institute,
University of Southampton, U. K. and Sourabh Bhattacharya
is an Assistant Professor in
Bayesian and Interdisciplinary Research Unit, Indian Statistical
Institute, 203, B. T. Road, Kolkata 700108.
Corresponding e-mail: sourabh@isical.ac.in.}}
\date{\vspace{-0.5in}}
\maketitle

\begin{abstract}

Fossil-based palaeoclimate reconstruction is an important area of ecological science that has gained momentum 
in the backdrop of the global climate change debate. The hierarchical Bayesian paradigm provides an interesting 
platform for studying such important scientific issue. However, our cross-validation based assessment of the existing 
Bayesian hierarchical models with respect to two modern proxy data sets based on chironomid and pollen, respectively, 
revealed that the models are inadequate for the data sets.

In this paper, we model the species assemblages (compositional data) by the zero-inflated multinomial distribution, 
while modelling the species response functions using Dirichlet process based Gaussian mixtures. This modelling strategy 
yielded significantly improved performances, and a formal Bayesian test of model adequacy, developed recently, showed 
that our new model is adequate for both the modern data sets. Furthermore, combining together the zero-inflated assumption, 
Importance Resampling Markov Chain Monte Carlo (IRMCMC) and the recently developed Transformation-based Markov Chain 
Monte Carlo (TMCMC), we develop a powerful and efficient computational methodology.
\\[2mm]
{\bf Keywords:} {\it Cross-validation; Dirichlet Process; Palaeoclimate Reconstruction; Response Function;
Transformation Based Markov Chain Monte Carlo; Zero-inflated Multinomial.} 

\end{abstract}

\maketitle

\section{Introduction}
\label{sec:intro}

The science of palaeoclimate reconstruction involves predicting prehistoric climate changes by studying fossil records of
species abundances (assemblages) preserved in lake sediments and a `modern, training data set' consisting of known records of 
species abundances and climate values
at different sites in the `modern time', where modern time is conventionally defined as the time period from the year 1950
till present. Broadly, methods of palaeoclimate reconstruction consist of two steps. 
The first step is to calibrate a relationship between the observed species abundances
and the observed climates using the modern, training data. 
It is generally assumed that the species abundances depend upon climate, not the other way. In this sense, 
the calibration step is a `forward' problem.
Then, assuming that the calibrated relationship holds good even in the past ages where fossil records of the species are
available but not the prehistoric climates, the calibrated relationship is `inverted' to obtain reconstructions of the past climates.
Thus, the problem of climate reconstruction is an inverse problem. 

In the current scenario of the climate change discussion, 
the problem of palaeoclimate reconstruction has gained much importance. 
In this context, the Bayesian model-based attempt of the Irish climate reconstruction
using pollen assemblages by \ctn{Haslett06} (henceforth, HWB), is a particularly welcome contribution. 
The model
builds upon the palaeoclimate model of \ctn{Vasko00} (henceforth, VTK) who considered the multinomial Dirichlet 
model for the compositional data of chironomid assemblages (non-biting midges, well-known for providing
accurate information regarding past climates; see \ctn{Battarbee00}), and used the 
unimodal Gaussian function to describe the responses of the different species to climate. 
By unimodal Gaussian response function we mean that the expectation of the number of any particular species 
is a bell-shaped function of climate; there is an optimum climate value at which the species is expected to 
thrive the most, and deviation from the optimum climate leads to an exponential decrease in the expected number of 
the species. 

The main
modeling contribution of HWB is to propose a nonparametric approach to modelling the species response 
function. The reason for considering a new approach to modeling the response surfaces
is that the unimodal Gaussian response function is too simplistic and may not be adequate for
most of the species since the species are expected to respond differently to environmental changes,
indicating that the response functions may vary from species to species, apart from being complex
in nature. For a detailed discussion regarding these issues, see \ctn{Ohl12}.

But in spite of the commendable attempt and the sensible results related to Irish climate reconstruction, 
some issues related to the model of HWB should not be overlooked. 
Firstly, their nonparametric model for the response surface, which is based on lattice Gaussian Markov Random 
Field (GMRF)
(see, for example, \ctn{Rue05}), introduces a lot of parameters (around 10,000) which 
makes computation burdensome. Secondly, for higher dimensional climate variables the climate grid may
not be feasible to construct; moreover, this would involve too many parameters, rendering computation infeasible
as well.
Thirdly, the unknown past climate variables are assumed to take 
values in the region formed by
the modern climate values, which need not be an appropriate assumption for general palaeoclimate problems. 

In an effort to rectify these problems, \ctn{Bhatta06} (henceforth, SB) modeled the response functions as 
a mixture of unknown number of Gaussian functions, while using the multinomial Dirichlet
distribution to model the compositional data. He applied this model to the modern training data set
consisting of (modern) chironomid counts obtained from 62 lakes of Finland along with the corresponding modern
temperatures, also analysed by VTK. The results of leave-one-out cross-validation showed that
in 83\% cases the true temperature values are included in the 95\% credible intervals associated
with the posteriors of SB. This was a significant improvement over the model of VTK,
which had just 43\% coverage of the true temperature values.

However, before applying any potential palaeoclimate model to climate reconstruction, it is
desirable to validate it as rigorously as possible.
Indeed, with respect to the chironomid data neither the model of VTK nor that of SB satisfy the model adequacy test
developed in \ctn{Bhattacharya12} (see also \ctn{Bhatta04}). It is shown in \ctn{Bhatta04} that the model of
HWB, involving the pollen data, also fails the model adequacy test, even though coverage of the observed climate values 
GDD5 (growing degree days
above $5^\circ$C) and MTCO (mean temperature of the coldest month)
have been quite satisfactory. As demonstrated in \ctn{Bhatta04} (Chapter 7), the model of HWB
overfits the pollen data. 
In fact, although the predicted climates (modes of the posterior distributions) and the observed climates agree well with each other, 
the posterior distributions have large credible regions, indicating high uncertainty. Such large credible regions are responsible
for the poor fit (overfit). Presumably, many of the parameters related to the response surfaces were not adequately informed by the data. 
Indeed, as can be seen from Figure 5 of HWB, many of the small lattice squares of the climate grid hardly contain any data point.
Due to the Markov property of the GMRF assumption the parameters associated with such lattice squares do not depend upon distant 
lattice squares containing enough data; hence, these parameters do not have information from the data to reduce their posterior 
variabilities. Hence, the credible regions turned out to be too large, resulting in overfit.


In this paper, we shall conern ourselves with assessment of model adequacy via cross-validation of
the training data. We shall not attempt actual climate reconstruction in this paper.
In particular, we present a hierarchical zero-inflated multinomial model for the compositional fossil data
and, following SB, propose a mixture of unknown number of Gaussian functions
to model the response function of each
species. The only difference between this model and that of SB is the zero-inflated
multinomial model in place of the ordinary mutinomial model.
But importantly, this apparently simple modification resulted in quite significant improvement
of the results previously obtained by SB.
Indeed, with our zero-inflated multinomial model and mixtures of unknown number
of Gaussian functions, in the case of the chironomid data of VTK
we have been able to include approximately 97\% of the observed temperature values in our respective 95\%
highest posterior density (HPD) credible regions, 3 cases only marginally missing the HPD regions.
More encouragingly, our model satisfies
the model adequacy test proposed in \ctn{Bhattacharya12}.
Generalising our ideas to the pollen data case of HWB 
we show that our model satisfies the test of adequacy even for the pollen data -- the cross-validation exercise
associated with the pollen data showed inclusion of approximately 95\% observed climate values in the respective
95\% HPD regions.
Indeed, in the aforementioned previous works on palaeoclimate reconstruction, the count data, characterized
by a large number of zeroes (about 59\% zeroes in the chironomid case and about 37\% zeroes in the 
case of pollen), rendered the ordinary multinomial distribution inappropriate. 

Apart from the very much improved results, our model and methods facilitate very fast and efficient
computation, which is crucial for palaeoclimate reconstruction where
the data sets tend to be (at least moderately) large.
For the cross-validation purpose we combine the Importance Resampling Markov Chain Monte Carlo (IRMCMC) methodology of \ctn{Bhatta07}
with the recently developed Transformation based Markov Chain Monte Carlo (TMCMC) (\ctn{Dutta11}) to further
improve computational efficiency. 
A brief overview of TMCMC is provided in Section \ref{subsec:tmcmc_overview};
here we just note that TMCMC allows updating high-dimensional 
parameter vectors using simple deterministic transformations of one-dimensional random variables having arbitrary 
distributions on some relevant support. 

It is worth mentioning that recently \ctn{Salter12} have developed a nested Dirichlet-Multinomial model for multivariate
pollen counts data. Their work is motivated by \ctn{Ohl12}; however, their need to use the integrated nested Laplace approximation (INLA)
(\ctn{Rue08}) for the purpose of fast computation, also played a very significant role in their model-building procedure.
In particular, \ctn{Salter12} specify a model which exploits the nested structure within the pollen species based on botanic
similarities; within each level of the nested structure the species proportions are assumed to be Beta/Dirichlet, and conditionally independent
of the other levels consisting of the other species, given their GMRF prior on the 
two-dimensional climate grid (same as that of HWB, and so this model also precludes extrapolation and is difficult to
generalize for high-dimensional climate variables) and other hyperparameters.
At each level, the count data is then assumed to be zero-inflated Binomial/Multinomial, given the proportions at that level of the nested
structure. The conditional independencies, although undesirable, are necessary for INLA implementation. Thus, although INLA has
greatly sped up their computation, the method did demand sacrifice of model flexibility. Also, although INLA has been appropriate
for the cross-validation summary statistics that \ctn{Salter12} consider, it is perhaps the case that INLA, being a deterministic
approach, can not approximate the posterior distrbutions of arbitrary discrepancy measures, for example, those that we consider
in this paper; see also \ctn{Banerjee08} for a brief discussion.

The rest of our paper is structured as follows. In Section \ref{sec:chironomid_model} we propose our new model 
for the chironomid data. Fitting our model using MCMC is discussed in detail in 
Section \ref{subsec:fullcond}, and our method of leave-one-out cross-validation using IRMCMC
is provided in Section \ref{sec:cross_validation}. Cross-validation of the chironomid data
and detailed analysis of the results of the cross-validation are presented in Section \ref{sec:application}.
The formal model adequacy test, along with its application to the chironomid data using posterior 
samples from the cross-validation exercise, are discussed in Section \ref{sec:model_assess_results}.
In Section \ref{sec:pollen_data} we generalize our model and methods to the pollen data of HWB, while
cross-validation of the pollen data and subsequently the model adequacy test are discussed in 
Sections \ref{sec:haslett_sim} and \ref{sec:model_glendalough}, respectively. We finally conclude
with some discussion on future work in Section \ref{sec:conclusions}.
Additional details are provided in the supplement \ctn{Sabya13supp}, whose sections and figures have the prefix ``S-" when referred
to in this paper.

\section{An improved model for the chironomid data}
\label{sec:chironomid_model}
Before proceeding we briefly review the data set, the full description of which can be found in
\ctn{Olander99}; see also VTK.
\subsection{Brief description of the data set}
\label{subsec:data_description}
As already mentioned in the introduction, chironomids are non-biting midges, and considered very suitable 
for past climate reconstruction. 
The modern, training data set analysed by VTK
consists of counts of chironomid head capsules present in the top 1 cm surface-sediment 
from 62 lakes located mainly in northwestern
Finnish Lapland. Recorded also are site-specific mean July air temperatures, estimated for each lake using 1961--1990 Climate Normals data
from 11 nearby climate stations (2 in Norway, 5 in Finland, and 4 in Sweden) and applying consistent regional lapse
rates and linear interpolation (see \ctn{Olander99} for details). After excluding rare species, 52 taxa of chironomid were finally selected.

Thus, the chironomid data of VTK consists of modern time assemblages for $m = 52$ species of chironomid, along with the mean July
temperature values at each of $n = 62$ lakes (sites) in Finland. This modern, training
data set has been used by \ctn{Korhola02} for reconstructing past climates of Finland using VTK's model. 

In the following subsections of this present section we provide details of semiparametrically modelling this data. 
The same model will be generalised
to the case of the pollen data of HWB in Section \ref{sec:pollen_data}.
In what follows, we begin with the zero-inflated Poisson model for the count data, 
finally deriving from it the zero-inflated
multinomial model.

\subsection{Hierarchical model specification starting with zero-inflated Poisson model}
\label{subsec:hierarchical}
For $i=1,\ldots,n$ and $k=1,\ldots,m$, let $y_{ik}$ denote the count of the $k$-th chironomid species available at the $i$-th site;
let $\bY$ denote the complete count data set. Also, let $x_i$ denote the temperature at site $i$.
Let $\bX=\{x_1,\ldots,x_n\}$ denote
the complete set of temperature values. With these we consider the following
mixture model for $y_{ik}$:
\begin{equation}
[y_{ik}\mid\lambda_{ik}]\sim\pi_{ik}\delta_{\{0\}}+(1-\pi_{ik})\mathbb P(\lambda_{ik}),
\label{eq:mixture}
\end{equation}
where $\lambda_{ik}>0$, $0\leq\pi_{ik}\leq 1$, $\delta_{\{0\}}$ denotes point mass at zero, and 
$\mathbb P(\lambda_{ik})$ denotes the Poisson distribution
with parameter $\lambda_{ik}$. Further,
\begin{align}
\lambda_{ik} &\sim Gamma(\xi_{ik},1/\psi),\ \ \mbox{where}\label{eq:lambda_prior}\\
\xi_{ik}&=\sum_{j=1}^{M_k}\frac{1}{\sqrt{2\pi}\gamma_{kj}}
\exp\left\{-\frac{1}{2}\left(\frac{x_i-\beta_{kj}}{\gamma_{kj}}\right)^2\right\},
\label{eq:response}
\end{align}
In (\ref{eq:lambda_prior}) $Gamma(\xi_{ik},1/\psi)$ denotes the Gamma distribution with mean $\psi\xi_{ik}$
and variance $\psi^2\xi_{ik}$, where $\psi>0$ is a fixed constant. Here $\xi_{ik}$ and $\psi$ are
shape and scale parameters, respectively.
In (\ref{eq:response}) $\beta_{kj}$ and $\gamma_{kj}$ stand for the $j$-th optimum
temperature ($j$-th optimum of the $k$-th species) and the $j$-th tolerance level
(a measure of temperature within the vicinity of the optimum temperature that the species can withstand); 
$M_k$ is the {\it maximum} number of optima and the tolerance levels of the $k$-th species.
These will be further elucidated in Section \ref{subsec:response}.

\subsection{Viewing species optima and tolerance levels as samples from Dirichlet processes}
\label{subsec:dp}
Writing 
$\btheta_{kj}=(\beta_{kj},\gamma_{kj})$, we assume that
for each $k$, $\bTheta_k=\{\btheta_{k1},\ldots,\btheta_{kM_k}\}$ is a sample
from the Dirichlet process (see, for example, \ctn{Ferguson73}):
\begin{align}
\btheta_{kj}&\stackrel{iid}{\sim}G; \ \ \ \ j=1,\ldots,M_k; \ \ k=1,\ldots,m,\ \ \mbox{where}\label{eq:dp1}\\
G & \sim DP(\alpha G_0),\label{eq:dp2}
\end{align}
In (\ref{eq:dp2}), $DP(\alpha G_0)$ denotes the Dirichlet process  
with $\alpha>0$ representing the strength
of the belief in the central distribution $G_0$. Here we assume that
under $G_0$, the joint distribution of $\btheta_{kj}$ is normal-inverse-gamma, given by
\begin{equation}
[\btheta_{kj}\mid G_0]\propto \exp\{-b/\gamma_{kj}\}\gamma^{-a-1}_{kj}\times
\frac{\exp\{-(\beta_{kj}-\mu_{\beta})^2/2\gamma_{kj}^2\}}{\gamma_{kj}}.
\label{eq:G_0}
\end{equation}
The values of the parameters $a,b$, and $\mu_{\beta}$ will be specified in the context of the application.

\subsection{Response function}
\label{subsec:response}
Introducing the allocation variables $z_{ik}$ (these can also be thought of as auxiliary or latent variables) 
helps ascertain whether the corresponding count $y_{ik}$
is zero or arose randomly from $\mathbb P(\lambda_{ik})$. Formally, $z_{ik}=1$ with probability $\pi_{ik}$
and 0 with probability $1-\pi_{ik}$. 
Observe that
\begin{align}
E[y_{ik}\mid z_{ik}=0]&=E\left\{E[y_{ik}\mid z_{ik}=0,\lambda_{ik}]\right\}\notag\\
&=E(\lambda_{ik})=\psi\xi_{ik}\notag\\
&=\psi\sum_{j=1}^{M_k}\frac{1}{\sqrt{2\pi}\gamma_{kj}}
\exp\left\{-\frac{1}{2}\left(\frac{x_i-\beta_{kj}}{\gamma_{kj}}\right)^2\right\},
\label{eq:response1}
\end{align}
showing that the response function of the $k$-th species at the $i$-th site is given by (\ref{eq:response1}).
Now, since the Dirichlet process is discrete with probability one, it follows that with positive
probability, the parameters
$\{\btheta_{kj};j=1,\ldots,M_k\}$ are equal. A consequence of this
is the reduction of (\ref{eq:response1}) to the following:
\begin{equation}
E[y_{ik}\mid z_{ik}=0]=\psi\sum_{j=1}^{M^*_k}\frac{N_{kj}}{\sqrt{2\pi}\gamma^*_{kj}}
\exp\left\{-\frac{1}{2}\left(\frac{x_i-\beta^*_{kj}}{\gamma^*_{kj}}\right)^2\right\},
\label{eq:response2}
\end{equation}
where, with $\btheta^*_{kj}=(\beta^*_{kj},\gamma^*_{kj})$, the set 
$\{\btheta^*_{kj};j=1,\ldots,M^*_k\}$ is the set of distinct values among
$\{\btheta_{kj};j=1,\ldots,M_k\}$, and $N_{kj}$ is the frequency of the occurrence
of $\btheta^*_{kj}$. Of course, $\sum_{j=1}^{M^*_k}N_{kj}=M_k$.
Since the number of, and the frequencies of coincidences among the parameters is random, it is clear that
(\ref{eq:response2}) is a mixture of Gaussian functions with unknown number of components.
Moreover, it is also clear that all the $m$ species have different response functions, 
with different number of mixture components. 
This is important, since different taxa may require different numbers of components to adequately model the
response surface. 

An alternative to our mixture representation of the response surfaces are spline based models for the same.
For this modeling style, for different species, the orders of the splines (orders of the polynomial parts), 
the numbers and locations of the knots, 
must be treated as unknown and different. 
Although the part of the spline associated with the knots can be
modeled using Dirichlet process, the same is not appropriate for modeling the polynomial part of the spline. 
The reason is that Dirichlet process can only force the polynomial coefficients to be equal
with positive probability, but coincidences among the polynomial coefficients can not decrease the order of the polynomial.
As such, the polynomial part must be handled using complicated variable-dimensional MCMC methods,
for example, reversible jump MCMC (RJMCMC). Since complicated RJMCMC has to be carried out for all the species, this would
very significantly increase the computational burden. 
But such computational difficulties can be overcome by a new, general, MCMC methodology for variable dimensional models, which
is being developed by \ctn{Das13}. The methdogology, which we refer to as Transdimensional TMCMC (TTMCMC) is an extension
of TMCMC for variable dimensional cases, and can update all the (random number of) parameters in a single block, using deterministic
transformations of some arbitrary one-dimensional random variable. 
This would greatly assist in computation associated with spline-based response
functions that we hope to pursue in the future.

\subsection{From zero-inflated Poisson to zero-inflated multinomial}
\label{subsec:multinomial}
Letting $y_{i\cdot}=\sum_{k=1}^my_{ik}$, it follows that the joint distribution of $\by_i=(y_{i1},\ldots,y_{im})$
is zero-inflated multinomial, given by:
\\[2mm]
$[\by_i\mid y_{i\cdot},z_{i1},\ldots,z_{im},\lambda_{i1},\ldots,\lambda_{im}]$
\begin{equation}
=\left(\frac{y_{i\cdot}}{\prod_{k:z_{ik}=0}y_{ik}!}\right)\prod_{k:z_{ik}=0}
\left(\frac{\lambda_{ik}}{\sum_{\ell:z_{i\ell}=0}\lambda_{i\ell}}\right)^{y_{ik}}.
\label{eq:multinomial}
\end{equation}

Now note that $p_{ik}=\frac{\lambda_{ik}}{\sum_{\ell:z_{i\ell=0}}\lambda_{i\ell}}$ denotes the unknown proportion
of the $k$-th species at the $i$-th site, whenever $z_{ik}=0$, that is, whenever $y_{ik}\neq 0$. These proportions are clearly dependent
since all of them are scaled by the same sum $\sum_{\ell:z_{i\ell=0}}\lambda_{i\ell}$. In fact,
since {\it a priori} $\lambda_{ik}\sim Gamma(\xi_{ik},1/\psi)$, it follows that 
$[\{p_{ik}:z_{ik}=0\}]\sim Dirichlet(\{\xi_{ik}:z_{ik}=0\})$. 
In other words, even though the species parameters $\bTheta_k$ are considered independent at the Poisson level, the species
proportions $\{p_{ik};k=1,\ldots,m\}$ are dependent at the multinomial level for each $i=1,\ldots,n$.
Thus, we have the following Multinomial-Dirichlet structure: for $i=1,\ldots,n$,
\begin{align}
[\{y_{ik}:z_{ik}=0\}\mid y_{i\cdot},z_{i1},\ldots,z_{im},\lambda_{i1},\ldots,\lambda_{im}] &\sim Multinomial\left(y_{i\cdot},\{p_{ik}:z_{ik}=0\}\right);\notag\\
[\{p_{ik}:z_{ik}=0\}]&\sim Dirichlet(\{\xi_{ik}:z_{ik}=0\}).\notag
\end{align}


Although it is possible to express our Bayesian model in terms of the Dirichlet parameters $p_{ik}$ and then 
analytically integrate out the latter, so that $\lambda_{ik}$ no longer needs to be simulated by MCMC methods,
there are two reasons to retain $\lambda_{ik}$. Firstly, $\lambda_{ik}$ are the Poisson parameters
associated with the first stage of our modeling, which does not condition on $y_{i\cdot}$; hence it may be
of interest to learn $\lambda_{ik}$. Here note that the model in terms of $p_{ik}$ (even if $p_{ik}$ are retained),
is not identifiable with respect to $\lambda_{ik}$, since multiplying $\{\lambda_{ik};k=1,\ldots,m\}$ with some constant 
yields the same $p_{ik}$. Hence, if $\lambda_{ik}$ are of interest, the model must be expressed in terms of $\lambda_{ik}$,
not $p_{ik}$.

Secondly, and more importantly, retaining these parameters 
expand the parameter space, which may allow free movement of the MCMC sampler, thereby facilitating improved mixing. 
One such instance is reported in \ctn{Bhatta07}, where the MCMC sampler associated with the marginalized model 
failed to discover a minor mode of a bimodal cross-validation posterior associated with VTK's model, 
but the expanded model of VTK with the Dirichlet parameters 
allowed the MCMC sampler to explore the mode adequately. Since multimodality plays very important
roles in both of our examples, we resort to modeling in terms of $\lambda_{ik}$.  
Since we update the $\lambda_{ik}$ parameters in a single step using TMCMC, retaining these parameters
does not cause computational burden.

We have pointed out that although the species parameters are independent at the Poisson level, dependence is induced
at the multinomial stage, via conditioning on $y_{i\cdot}$.
However, it is possible to induce dependence between the species parameters $\bTheta_k$ even at the Poisson level, by considering the 
hierarchical Dirichlet process (\ctn{Teh06}). In other words, we could assume that, for $k=1,\ldots,m$, 
$\btheta_{k1},\ldots,\btheta_{kM_k}\stackrel{iid}{\sim} G_k$;
$G_1,\ldots,G_m\stackrel{iid}{\sim}G_0$; $G_0\sim DP(\gamma H)$, where $\gamma>0$ and $H$ is a specified distribution. 
The implication of such a hierarchical structure is that the parameters $\btheta_{kj}$
associated with the species response functions will be shared with positive probability by the various species, inducing
dependence. However, in our set-up this would create severe computational difficulties. 
Again, such computation difficulties can perhaps be overcome by TTMCMC of \ctn{Das13}.
We intend to explore the issues related to the new modelling ideas and computational methods in the future.


\subsection{Joint posterior}
\label{subsec:joint}
Now, letting $\bTheta=\{\bTheta_k;k=1,\ldots,m\}$, $\Pi=\{\pi_{ik};i=1,\ldots,n;
k=1,\ldots,m\}$, $\Lambda=\{\lambda_{ik};i=1,\ldots,n;k=1,\ldots,m\}$, 
$Z=\{z_{ik};i=1,\ldots,n;k=1,\ldots,m\}$, 
the posterior of
$(\btheta,\Pi,\Lambda,Z)$ is given by
\begin{align}
[\bTheta,\Pi,\Lambda,Z\mid \bX,\bY]
&\propto
\prod_{i=1}^n\left(\frac{y_{i\cdot}}{\prod_{k:z_{ik}=0}y_{ik}!}\right)\prod_{k:z_{ik}=0}
\left(\frac{\lambda_{ik}}{\sum_{\ell:z_{i\ell}=0}\lambda_{i\ell}}\right)^{y_{ik}}\notag\\
& \ \ \ \ \times\prod_{i=1}^n\prod_{k=1}^m\pi^{z_{ik}}_{ik}(1-\pi_{ik})^{1-z_{ik}}
\times \prod_{k=1}^m\exp\left\{-\lambda_{ik}/\psi\right\}\lambda^{\xi_{ik}-1}_{ik}\notag\\
& \ \ \ \ \times \prod_{k=1}^m[\bTheta_k],
\label{eq:joint_posterior}
\end{align}
where $[\bTheta_k]$ is given by the following Polya urn scheme (\ctn{Blackwell73}):
\begin{align}
[\btheta_{k1}]&\sim G_0;\label{eq:polya1}\\
[\btheta_{kj}\mid\btheta_{k1},\ldots,\btheta_{k,j-1}]
&\sim \frac{\alpha G_0(\btheta_{kj})}{\alpha+j-1}
+\sum_{\ell=1}^{j-1}\frac{\delta_{\btheta_{k\ell}}(\btheta_{kj})}{\alpha+j-1}; \ \ j=2,\ldots,M_k.\label{eq:polya2}
\end{align}
In the expression for the joint posterior (\ref{eq:joint_posterior}), we assumed that $\pi_{ik}\stackrel{iid}{\sim}Uniform (0,1)$, for each $i,k$. 
A few remarks regarding this prior choice is in order. 

It is natural to choose a subjective prior on the zero-inflation probabilities $\Pi$ which depends upon climate.
However, the zero-inflation probabilities directly affect the number of zeroes in the data, and so any subjective prior,
which may depend upon the climate must be chosen with great care because mis-specification in this case can easily give rise to a conflict
between the data and the prior.  
An instance of mis-specification may be that at several locations several taxa may be
completely outside its range boundary which gives rise to excess zeroes, even though
the climate on which the prior of $\pi_{ik}$ for such locations and species depend, may be optimal for those taxa. In this case the
prior would not indicate excess zeroes, even though the observed data may contain excess zeroes, suggesting a conflict between
the prior and the data.
The objective prior $Uniform (0,1)$ cuts down such risk, as is evident from 
Figure \ref{fig:cred_chiro} and \ref{fig:check_haslett},
which indicate that the observed values
are fitted well by our model and the associated priors. Moreover, the $Uniform(0,1)$ prior also serves to simplify the computations
to a large extent, since the associated Gibbs step involves a simple simulation exercise from the relevant Beta distributions.
It is worth mentioning that $\Pi$ could be easily integrated out analytically from the joint posterior (\ref{eq:joint_posterior})
to simplify the model, but since we are interested in the posterior of $\Pi$ and since retaining these parameters
may induce better mixing of our MCMC sampler, we did not marginalize the joint posterior with respect to $\Pi$.

\section{Model fitting using Markov chain Monte Carlo (MCMC)}
\label{subsec:fullcond}
For MCMC purposes the full conditionals of the unknowns $z_{ik}$ and $\pi_{ik}$ are available in standard
forms for sampling using simple Gibbs steps. 
It will also be observed that the full conditionals do not involve the complete likelihood thanks to the zero-inflated multinomial
distribution, involving only those terms which are associated with strictly positive count data points. Since a large number of counts
are zero, this provides the very important advantage of very fast and efficient computation.
Updating $\btheta_{kj}$ using the Polya urn distribution as the proposal for Metropolis-Hastings steps 
as in SB turned out to to be quite effective here.
Finally, we update $\Lambda$ in a single block using TMCMC to further enhance computational
efficiency. Before proceeding further we first provide a brief overview of TMCMC. 

\subsection{Overview of TMCMC}
\label{subsec:tmcmc_overview}
TMCMC enables updating an entire block of parameters using
deterministic bijective transformations of some arbitrary low-dimensional random variable.
Thus very high-dimensional parameter spaces can be explored using simple transformations
of very low-dimensional random variables.
In fact, transformations
of some one-dimensional random variable always suffices, which we shall adopt in our examples.
Quite clearly, the underlying idea also greatly improves computational speed and acceptance rate compared to block Metropolis-Hastings
methods.  
Interestingly, the TMCMC acceptance ratio is indepenent of the proposal distribution chosen for
the arbitrary low-dimensional random variable. For implementation in our cases, we shall consider the additive
transformation, since it is shown in \ctn{Dutta11} that many fewer number of ``move types" are required by this
transformation compared to non-additive transformations.

To elaborate the additive TMCMC mechanism, assume that a block of parameters $\bzeta=(\zeta_1,\ldots,\zeta_r)$
is to be updated simultaneously using additive TMCMC, where $r~(\geq 2)$ is some positive integer. 
At the $t$-th iteration we shall then simulate 
$\eta\sim g(\eta)I_{\{\eta>0\}}$, where $g(\cdot)$ is some arbitrary distribution
and $I_{\{\eta>0\}}$ is the indicator function of the set $\{\eta>0\}$. In our examples we shall
choose $g(\cdot)$ to be $N(0,1)$ density, so that $\eta$ is simulated from a truncated normal distribution. 
We then 
propose, for $j=1,\ldots,r$, $\zeta^{(t)}_{j}=\zeta^{(t-1)}_{j}\pm a_{j}\eta$ with equal probability 
(although equal probability is a convenience, not a necessity), where $(a_1,\ldots,a_r)$
are appropriate scaling constants. Thus, using additive transformations of a single, one-dimensional $\eta$,
we update the entire block $\bzeta$ at once. In our examples, we select the tuning parameters
$(a_1,\ldots,a_r)$ using information from several pilot runs of our TMCMC algorithm.
In other words, we run our TMCMC algorithm several times for $20,000$ iterations, each time
with a set of possible trial values of $(a_1,\ldots,a_r)$; in fact, we begin with all the trial values
set equal to 0.5, and then observing the mixing properties of the associated pilot run, we modify the
trial values accordingly. We continue this for several pilot runs until the mixing is reasonable.
We ascertain mixing informally using trace and autocorrelation plots of the sample path of the TMCMC.

The aforementioned procedure of selecting the tuning parameters, although yielded reasonable mixing, 
is evidently somewhat ad-hoc. 
A more rigorous method for choosing the tuning parameters in additive TMCMC can be based on 
the recenly developed optimal scaling theory for additive TMCMC by \ctn{Dey13}. Since \ctn{Dey13} show that
the optimal acceptance rate for additive TMCMC under various set-ups is 0.439, one can tune the scaling
constants to achieve about 44\% acceptance rate. Note that for random walk Metropolis, the corresponding
optimal acceptance rate is 0.234, much lower than that of additive TMCMC. Comparisons between additive TMCMC and
random walk Metropolis in terms of optimal scaling are thoroughly explored in \ctn{Dey13}.

In Section S-1 of the supplement we descibe an MCMC algorithm, which is a combination of Gibbs steps, Metropolis-Hastings
and TMCMC steps, for updating the unknowns. 
The updating procedure will be used to cross-validate our model, which we discuss below.

\section{Leave-one-out cross-validation}
\label{sec:cross_validation}

In order to assess the validity of our model we successively leave out data point $i$ (that is, we leave out
both $x_i$ and the assemblage $\by_i$) from the training data set, and using
the remaining data set along with $\by_i$, the latter regarded as the test data, attempt to predict $x_i$.
So, we must now include a new parameter, which we denote by $x$, corresponding to the left out climate value $x_i$.
Now, this new parameter $x$ requires a prior. 
We set a prior $N(\mu_x,\sigma^2_x)$ for this new parameter.  

As a referee suggests, one could also look upon $x$ as the true measurement
of the climate value at the $i$-th site, where $x_i$ is the observed value of the climate subject to a measurement error at site $i$. 
From this perspective, the prior on $x$ can be interpreted as the prior on the true measurement of the climate variable at 
site $i$. We write $x_i=x+\zeta_i$, where $\zeta_i\sim N(0,\sigma^2_{\zeta})$ denotes the measurement error. The modified
likelihood associated with this perspective is the original likelihood conditional on the observed climate values, multiplied
with this normal likelihood contributed by the measurement error at the $i$-th site. The prior for $x$ must then be duly multiplied
with the joint likelihood and the priors for the other parameters to arrive at the form of the joint posterior. The observed
climate $x_i$ coincides with the true climate $x$ if and only if $\sigma^2_{\zeta}=0$, that is, when there is no measurement
error. In that case, the posterior of $x$ coincides with our cross-validation posterior when $x_i$ is held out. Indeed, we are
not aware of any evidence to suggest that there is significant climate measurement error in either the chironomid data or the pollen data.
Hence, for both the applications we shall assume that the observed climate values are the true climate values, and the prior
on the new parameter corresponding to the held out climate value makes sense from this perspective.

\subsection{Full conditional of $x$}
\label{subsec:fullcond_x}
The full conditional of $x$ given the rest is given by
\begin{equation}
[x\mid\cdots]\propto\prod_{i=1}^n\frac{\lambda^{\xi_{ik}-1}_{ik}}{\psi^{\xi_{ik}}\Gamma(\xi_{ik})}
\times\exp\left\{-\left(\frac{x-\mu_x}{\sigma_x}\right)^2\right\},
\label{eq:fullcond_x}
\end{equation}
where in $\xi_{ik}$, $x_i$ must be replaced with $x$.
For updating the one-dimensional variable $x$, random walk Metropolis with appoximately optimized scaling constant will be used. 
In fact, \ctn{Dutta11}
show that a TMCMC step for updating one-dimensional parameter coincides with a Metropolis-Hasting step;
in this case, the additive TMCMC step is equivalent to a random walk Metropolis step.
All the other variables will be updated in the way described in Section S-1. 

Now observe that since we need to perform an MCMC run for each left out data point, $n$ many computationally
burdensome MCMC implementations are necessary, thus calling for innovative computational shortcuts. 
The usual importance sampling based ideas (see, for example, \ctn{Gelfand92}, \ctn{Gelfand96}) do not work
in inverse problem set-ups such as in our case. In an inverse problem the response variable (say, $\by$) is modeled conditional
on some covariates (say, $x$), but prediction of some future $x_{n+1}$ given $\by_{n+1}$ and the training data set 
$\{(x_i,\by_i);i=1,\ldots,n\}$, is of interest. This is a much more complicated problem compared to the usual forward situation,
where prediction of $\by_{n+1}$ is of interest, given the training data set and $x_{n+1}$.
Details are provided in \ctn{Bhatta07}.
To meet the challenges of cross-validation in inverse problems, \ctn{Bhatta07} (see also \ctn{Bhatta04}) proposed a 
very fast and efficient methodology
by judiciously combining importance re-sampling (IR) and MCMC. Here we adopt their methodology, which 
has been termed IRMCMC
by the above authors. 
Details, for our current problem, are provided in Section S-2 of the supplement.

\section{Cross-validation of chironomid data}
\label{sec:application}




For our application we fixed $\alpha=10$, $\psi=1$, $\mu_{\beta}=11.19$, $a=11$, $b=30$, $\mu_x=11.19$, $M_{k}=10$ for all $k=1,\ldots,52$. 
These choices are motivated by VTK and SB who attempted to incorporate ecological knowledge into their
priors; in particular, the choice $\alpha=10,M_k=10$ implies that {\it a priori}
the probability of a multimodal response function for the $k$-th species is $0.53$, which is slightly 
higher than the probability
of a unimodal response function. It is also worth mentioning that using fixed value of $\alpha$ in the
context of Dirichlet process is commonplace; see, for example, \ctn{Escobar95}, \ctn{Neal00}, \ctn{Green01}, \ctn{Dahl09},
\ctn{Jensen08}, \ctn{Ishwaran01a}, \ctn{Ishwaran01c}, 
\ctn{Fearnhead04}, \ctn{Daume07},
\ctn{Kurihara06}, \ctn{Kurihara07}. 

Some remarks regarding the choice of $M_k$ and $\alpha$ in general palaeoclimate 
problems is in order. In the paradigm of regular mixtures, that is, when the data arise
from some mixture model with unknown number of components, \ctn{Bhatta08}, \ctn{Sabya11}, \ctn{Sabya12} 
consider Dirichlet process based mixture models of the form (\ref{eq:response}) 
(see also \ctn{Bhattacharya08} for Dirichlet process based mixtures in the context of circular data, 
and \ctn{Majumdar13} in the case of genetics), where
upper bounds on the number of mixture components were required. For normal mixtures based
on Dirichlet process, a detailed asymptotic 
investigation regarding asymptotic choice of the upper bound has been carried
out by \ctn{Sabya13}; it turned out, under suitable regularity conditions, 
that the form of $M_n$ (the upper bound allowed to increase with the sample size $n$) satisfying 
$M_n/\sqrt{n}\rightarrow 0$
as $n\rightarrow\infty$, is adequate. Thus, for fixed sample size $n$, one may choose $M_n$ to be
less than $\sqrt{n}$. Although our current set-up is very different from regular mixture problems, 
as a rule of thumb, we can select $M_k$ to be less than $\sqrt{n}$, the number of sites. The asymptotic
choices of $\alpha_n$ ($\alpha$ allowed to depend upon $n$), again increasing with $n$ but a rate slower than that of $M_n$, are shown 
by \ctn{Sabya13} to be adequate.

For details regarding the other prior choices, see VTK and SB.
We choose $\sigma_x^2=10$ to allow a reasonably wide range of possible values of $x$ to be considered. 
As we report in Section \ref{subsec:cv_results}, our cross-validation results are remarkably robust with respect
to other choices of $\alpha$ and $\sigma^2_x$.

For the purpose of IRMCMC we first selected $i^*$ as $i^*=\{i:x_i=median(\bX)\}$. Since $n=62$ is even, there are two choices of the median.
Following \ctn{Bhatta07} we chose $i^*=38$. For this importance sampling density, we simulated a sample of size $L=10,000$ 
after discarding a burn-in period of length $20,000$. From these stored $10,000$ MCMC realizations we re-sampled, without replacement, 
$K_1=200$ realizations
for each of the 62 cases. For each case, given each of the 200 re-sampled realizations, we then simulated, using MCMC, $K_2=50$ samples from
$[x,\pi_{i1},\ldots,\pi_{im},\lambda_{i1},\ldots,\lambda_{im},z_{i1},\ldots,z_{im}\mid\cdots]$, thus obtaining 
$10,000$ IRMCMC realizations from each
of the 62 posteriors associated with leave-one-out cross-validation. The entire cross-validation exercise using IRMCMC took just about an hour.
For computing the 95\% HPD regions of the cross-validation posteriors $[x\vert \bX_{-i},\bY];i=1,\ldots,62$, we implemented the well-known
line-pushing method; see, for example, \ctn{Carlin96}.

\subsection{Results of cross-validation}
\label{subsec:cv_results}

In 96.67\% of 62 cases, the observed temperature values $x_i$ fell within their respective 95\% HPD regions, 
suggesting very substantial improvement of our model over
those of VTK and SB.  
The reason for such high percentage of inclusion of the observed temperature values in the respective HPD's 
is due to taking into account large number of zero counts of the data by using zero-inflated multinomial model
and also due to using an appropriate species-temperature response function. 
The percentage of coverage remained almost unchanged for different choices of $\sigma^2_x$ and $\alpha$, suggesting
remarkable robustness of our cross-validation results with respect to these prior choices.

\begin{figure}
\centering
\subfigure[Site 6.]{ \label{fig:site1}
\includegraphics[width=4.5cm,height=4.5cm]{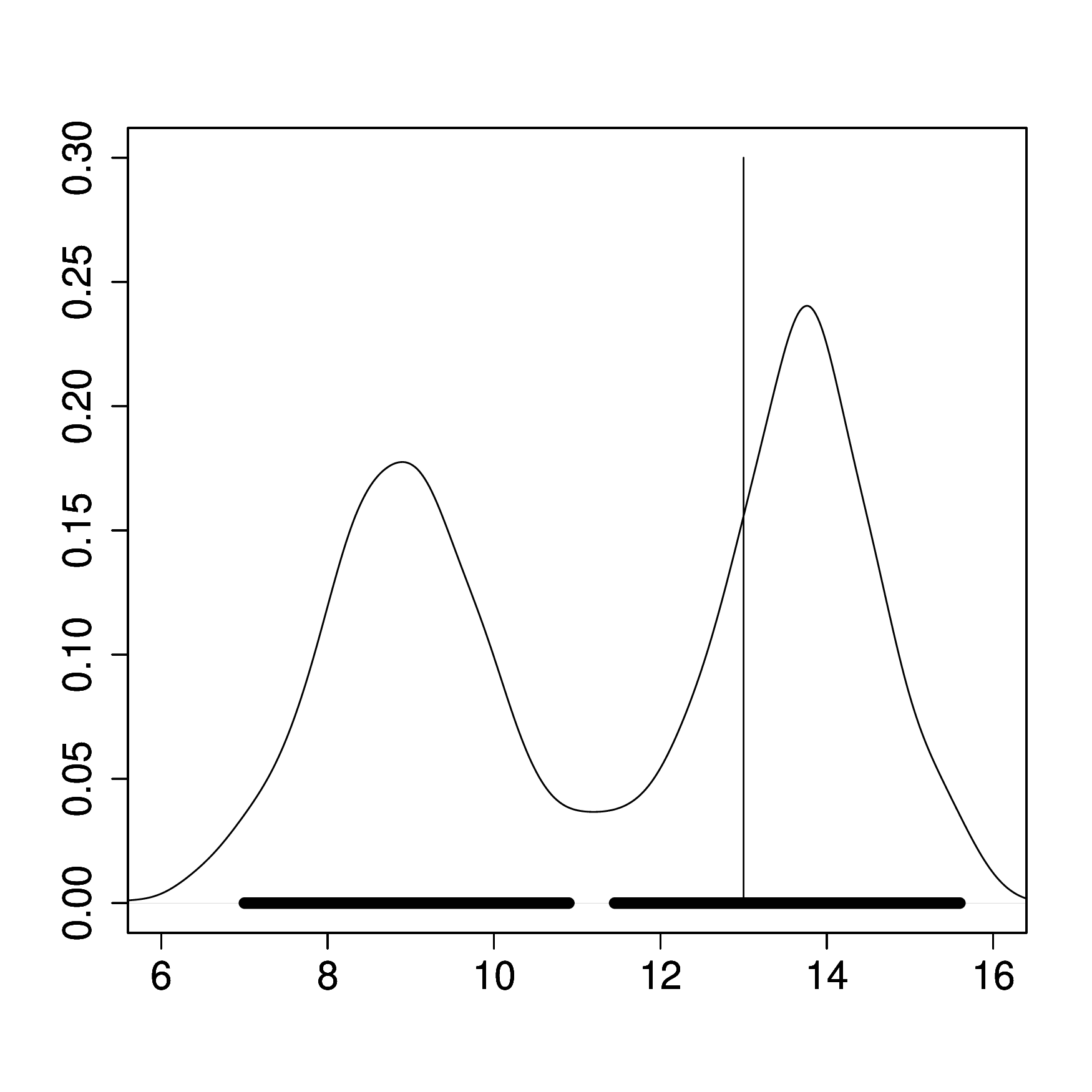}}
\hspace{2mm}
\subfigure[Site 15.]{ \label{fig:site7} 
\includegraphics[width=4.5cm,height=4.5cm]{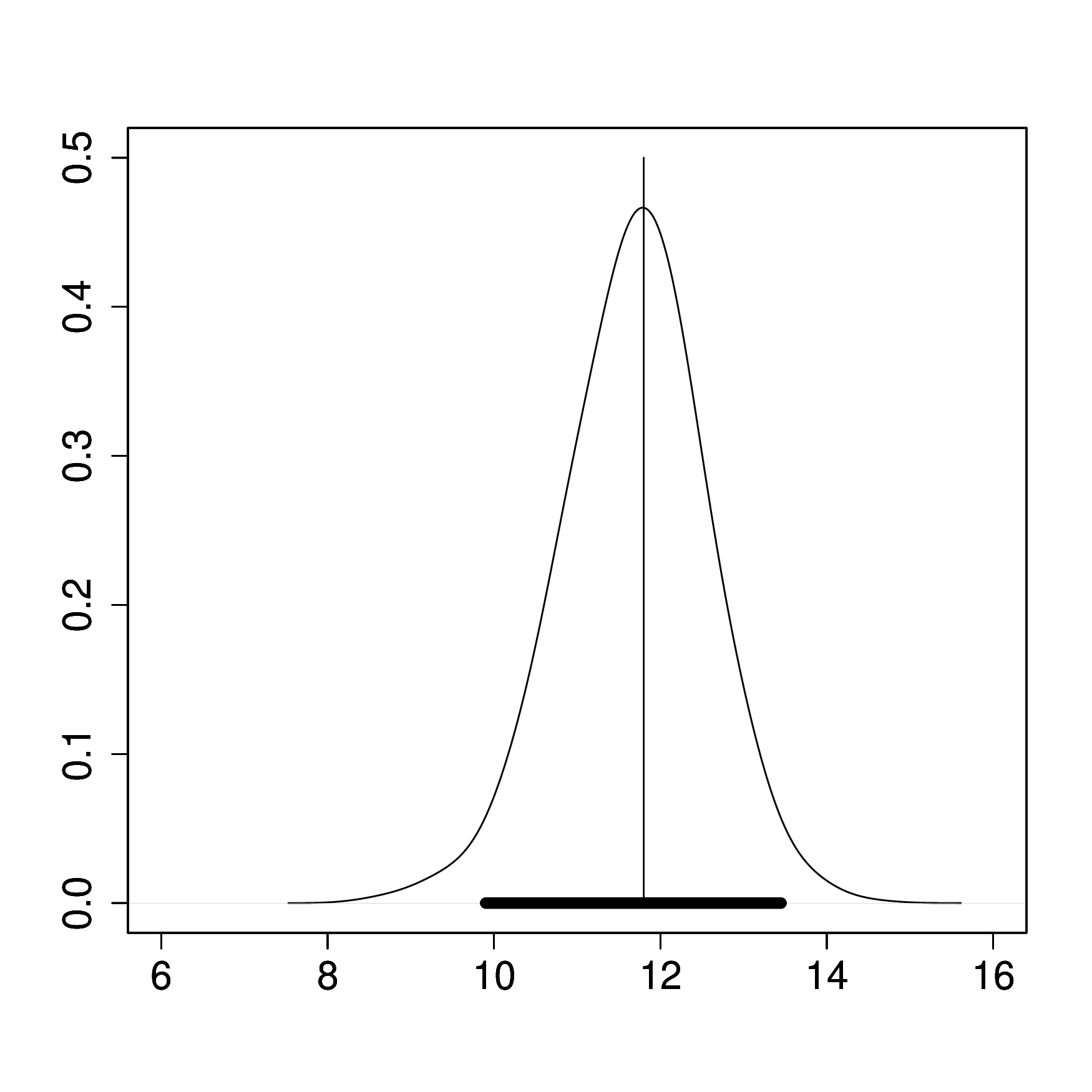}}
\hspace{2mm}
\subfigure[Site 23.]{ \label{fig:site14}
\includegraphics[width=4.5cm,height=4.5cm]{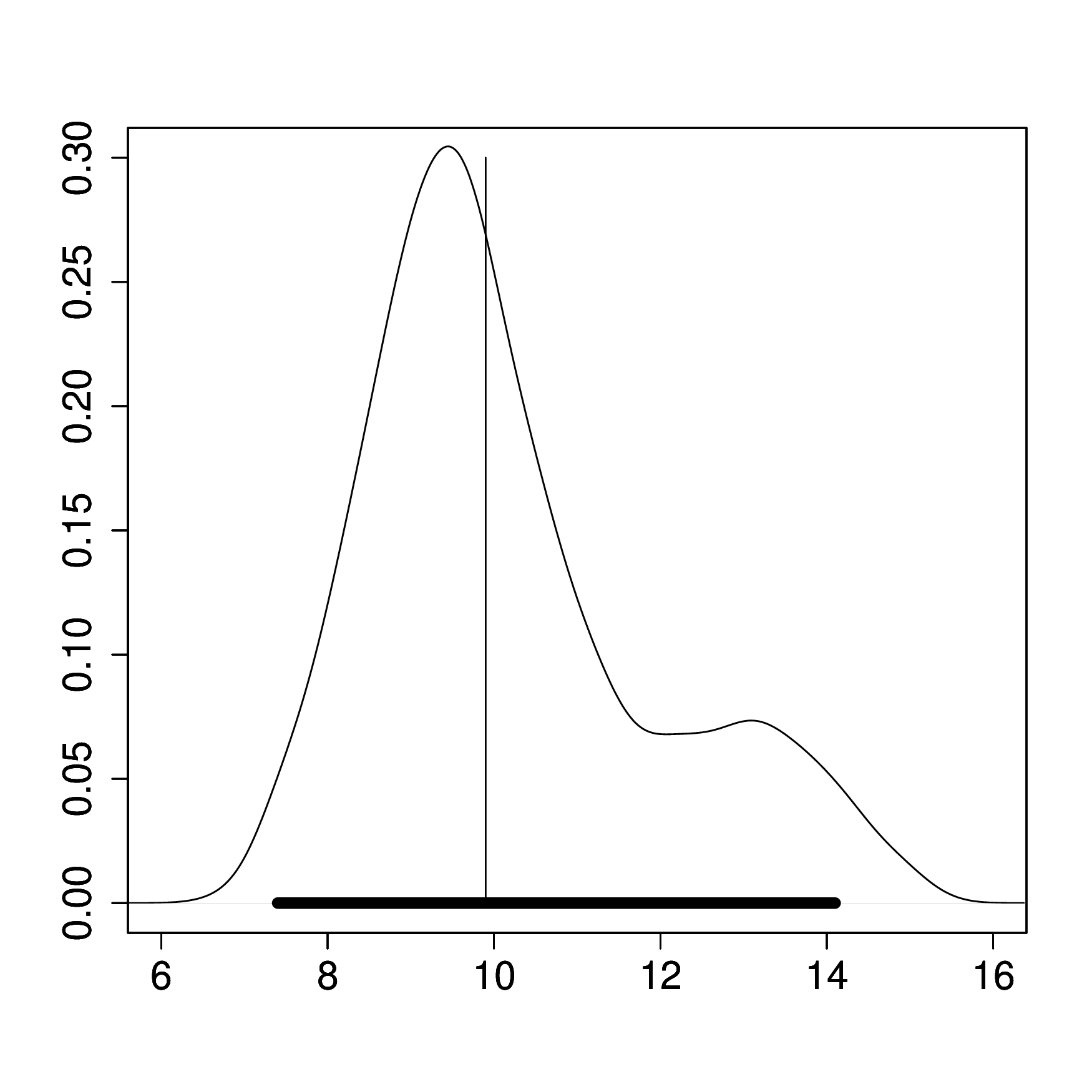}}\\
\vspace{2mm}
\subfigure[Site 24.]{ \label{fig:site21} 
\includegraphics[width=4.5cm,height=4.5cm]{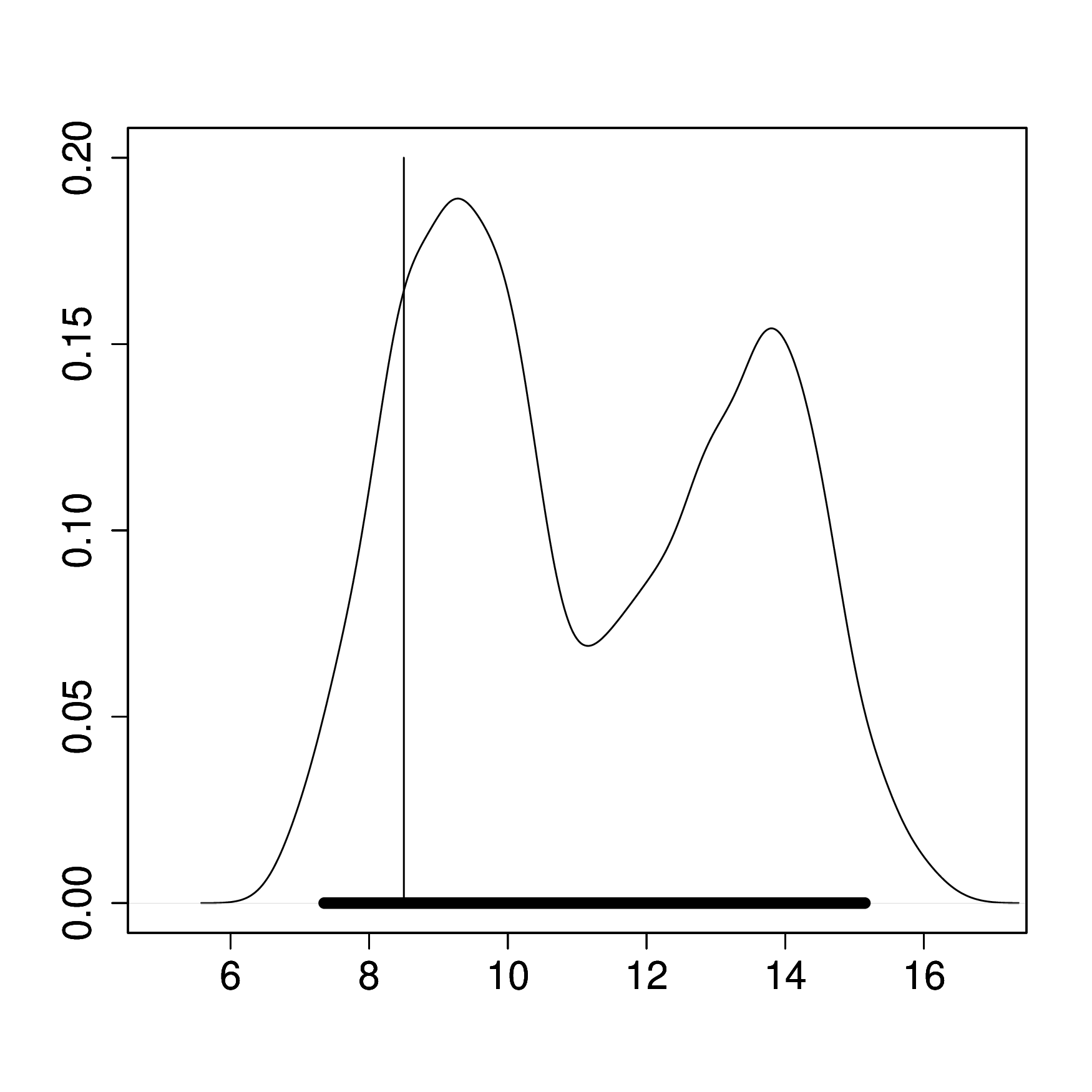}}
\hspace{2mm}
\subfigure[Site 29.]{ \label{fig:site28} 
\includegraphics[width=4.5cm,height=4.5cm]{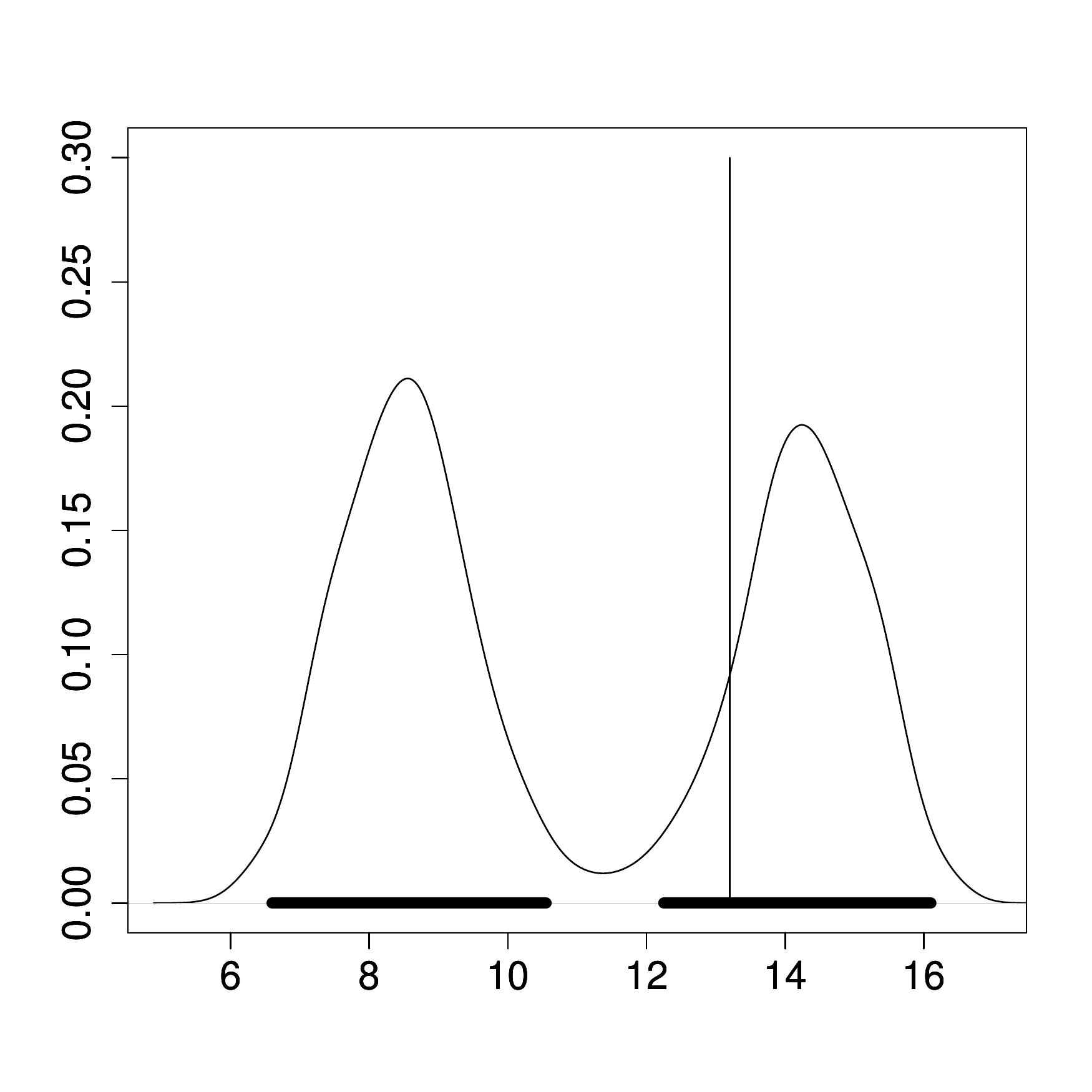}}
\hspace{2mm}
\subfigure[Site 35.]{ \label{fig:site35} 
\includegraphics[width=4.5cm,height=4.5cm]{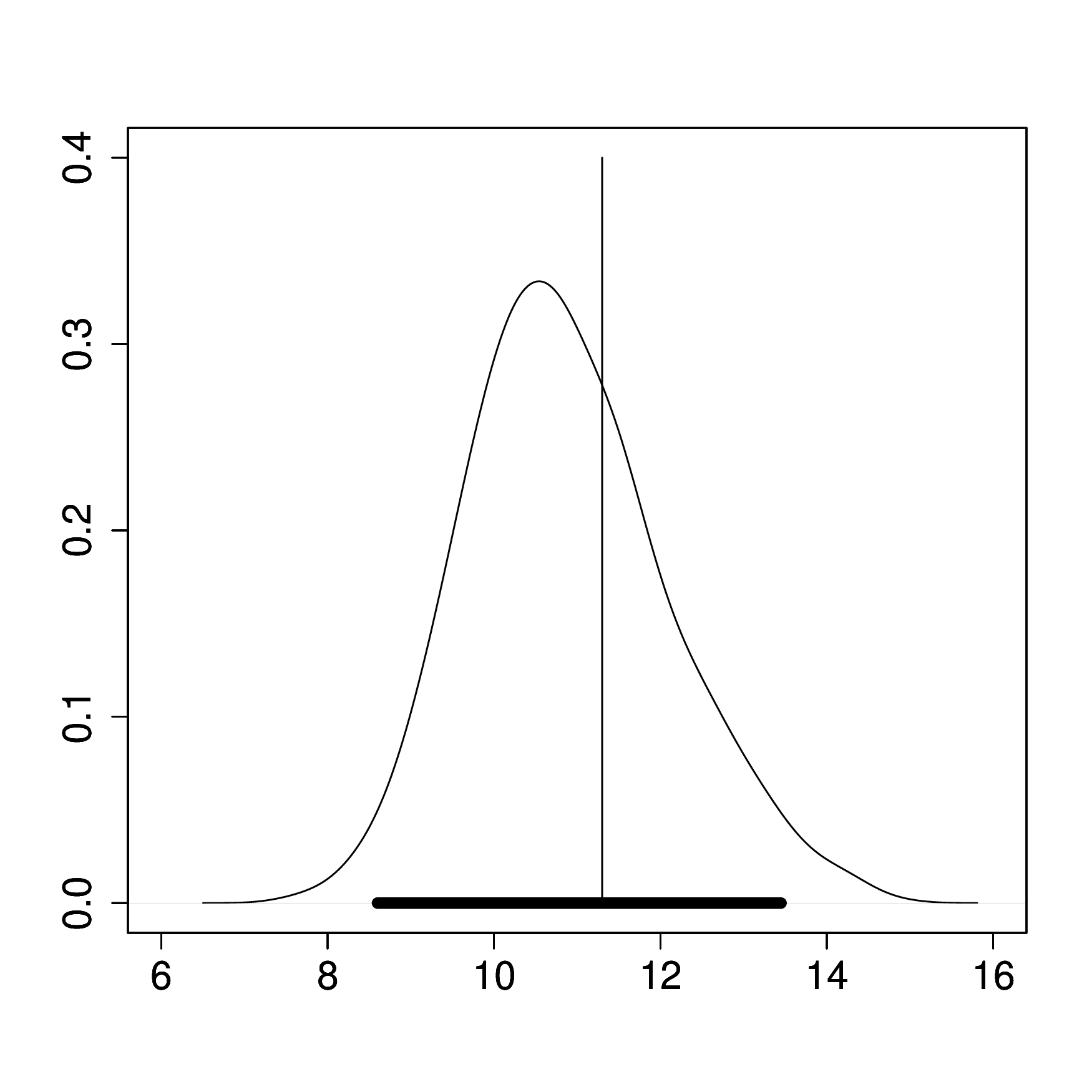}}\\
\vspace{2mm}
\subfigure[Site 45.]{ \label{fig:site45} 
\includegraphics[width=4.5cm,height=4.5cm]{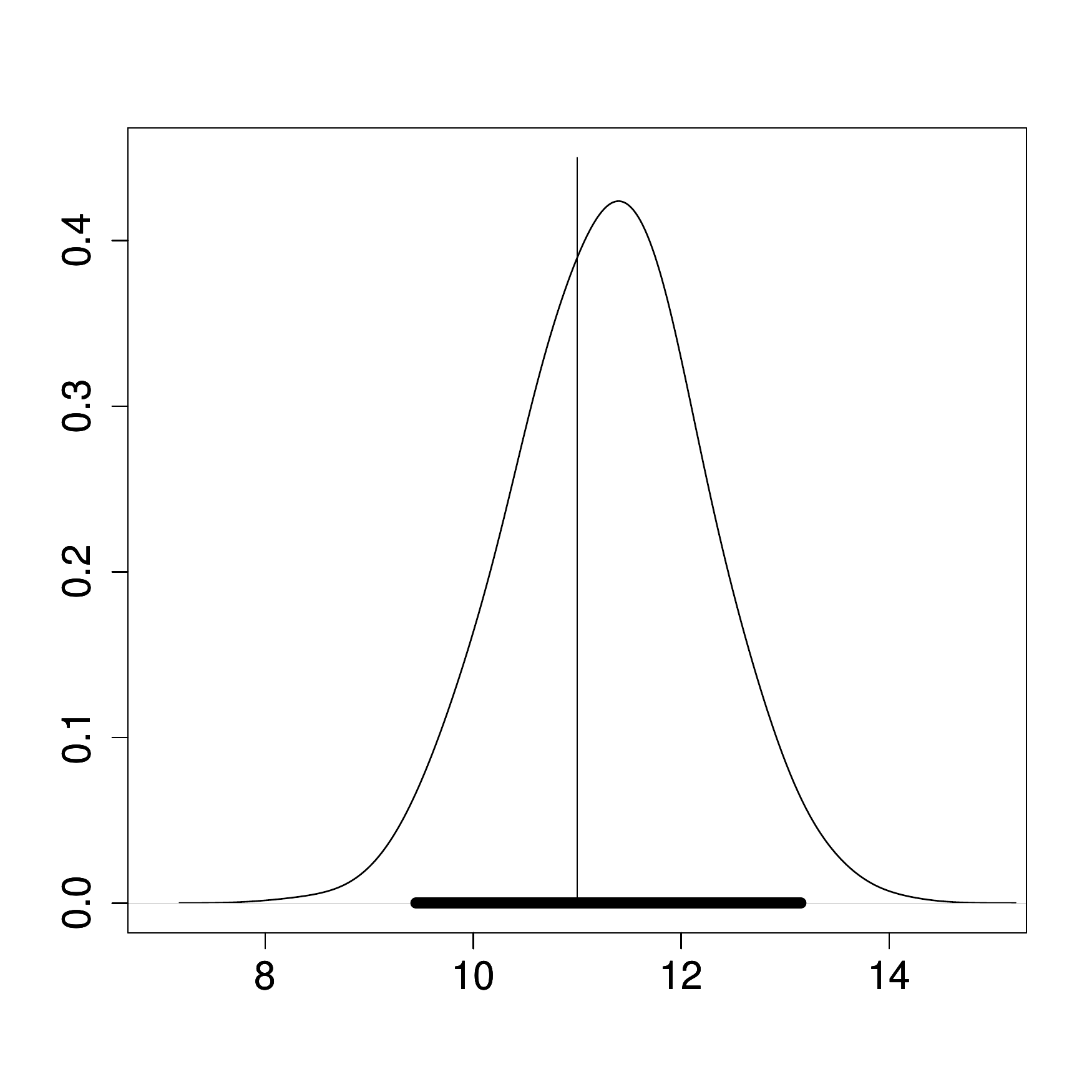}}
\hspace{2mm}
\subfigure[Site 58.]{ \label{fig:site58} 
\includegraphics[width=4.5cm,height=4.5cm]{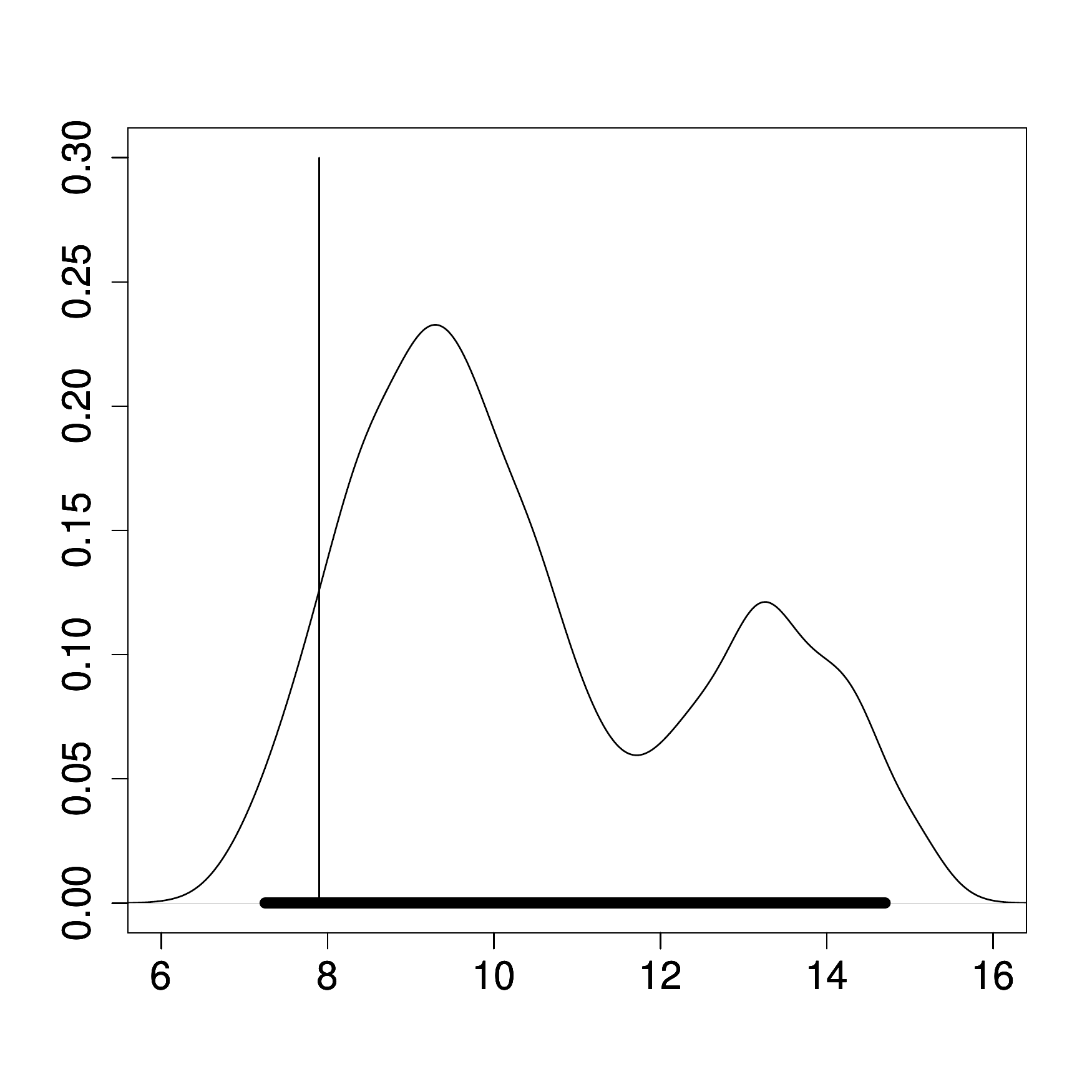}}
\hspace{2mm}
\subfigure[Site 60.]{ \label{fig:site60} 
\includegraphics[width=4.5cm,height=4.5cm]{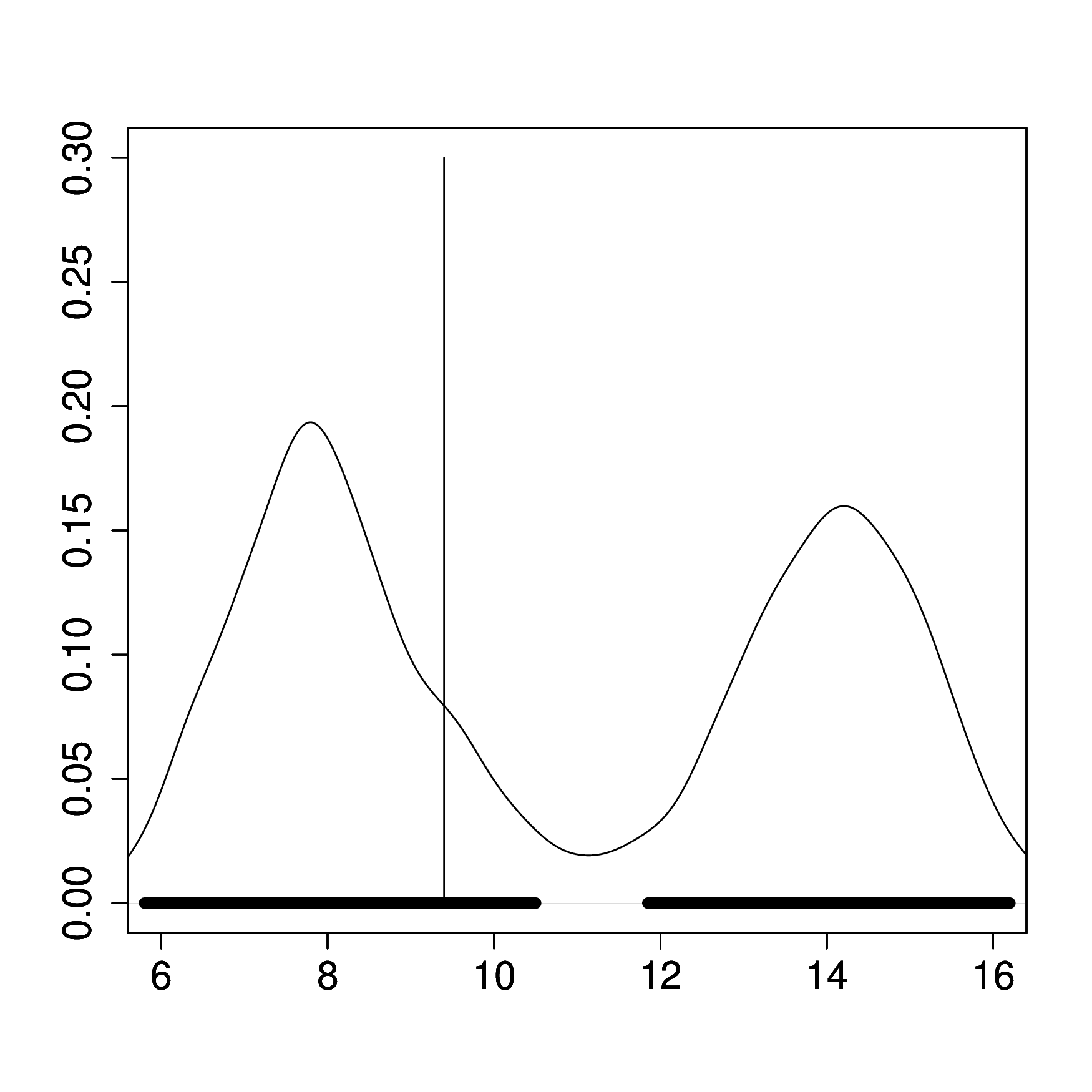}}
\caption{{\bf Chironomid data:} Leave-one-out cross-validation posteriors of temperature; 
the vertical line indicates the true (observed) value $\{x_i\}$. The thick, horizontal line within the support of the cross-validation posterior
indicates the 95\% HPD.} 
\label{fig:cv1}
\end{figure}

Some of the cross-validation posteriors, along with the corresponding observed $x_i$, and the 95\% HPD regions, are shown in Figure \ref{fig:cv1}. 
Many of the cross-validation posteriors are multimodal, which are consequences of multiple climate preferences of the different species.

\begin{figure}
\centering
\subfigure[Site 1, Species 1.]{ \label{fig:prop1}
\includegraphics[width=4.5cm,height=4.5cm]{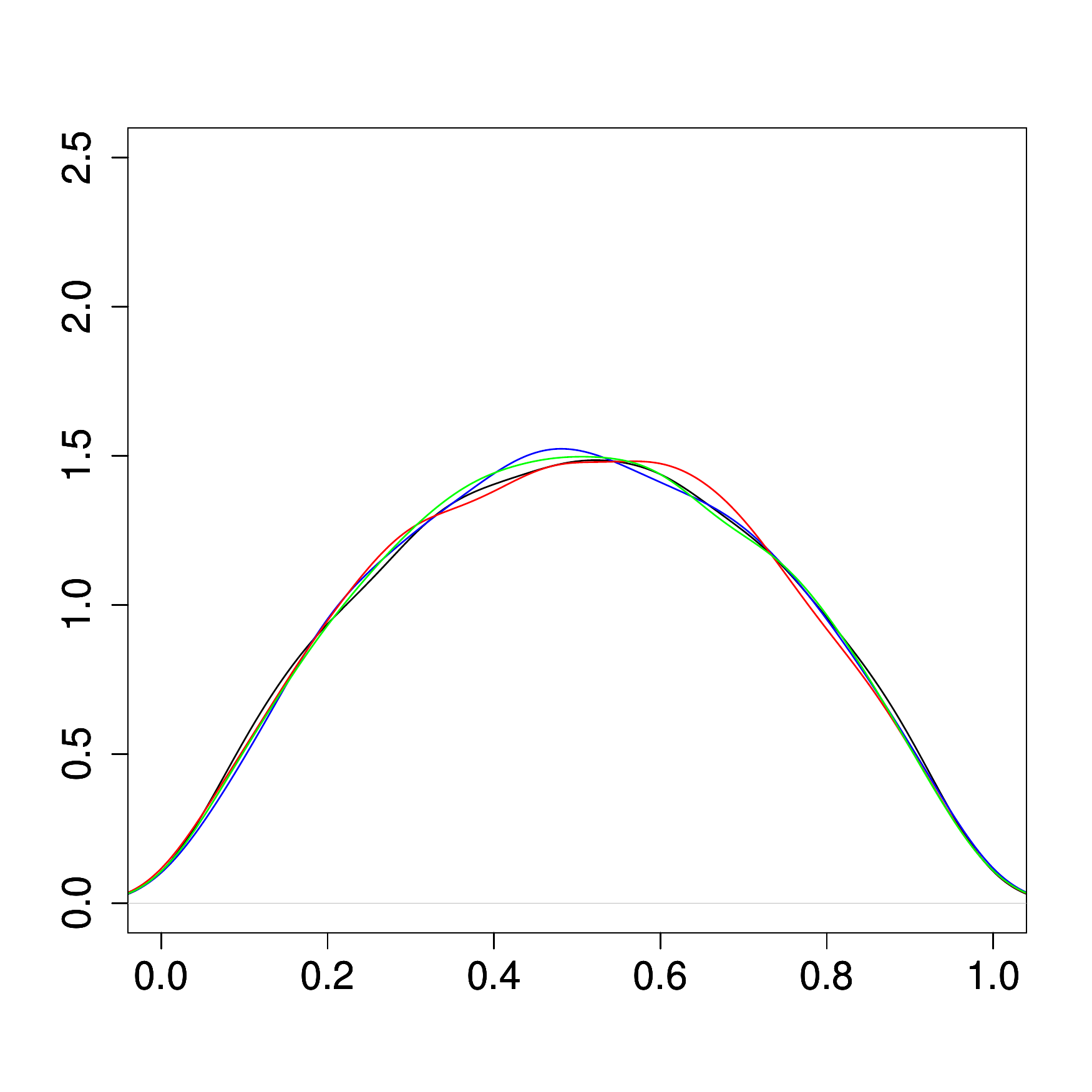}}
\hspace{2mm}
\subfigure[Site 1, Species 35.]{ \label{fig:prop2} 
\includegraphics[width=4.5cm,height=4.5cm]{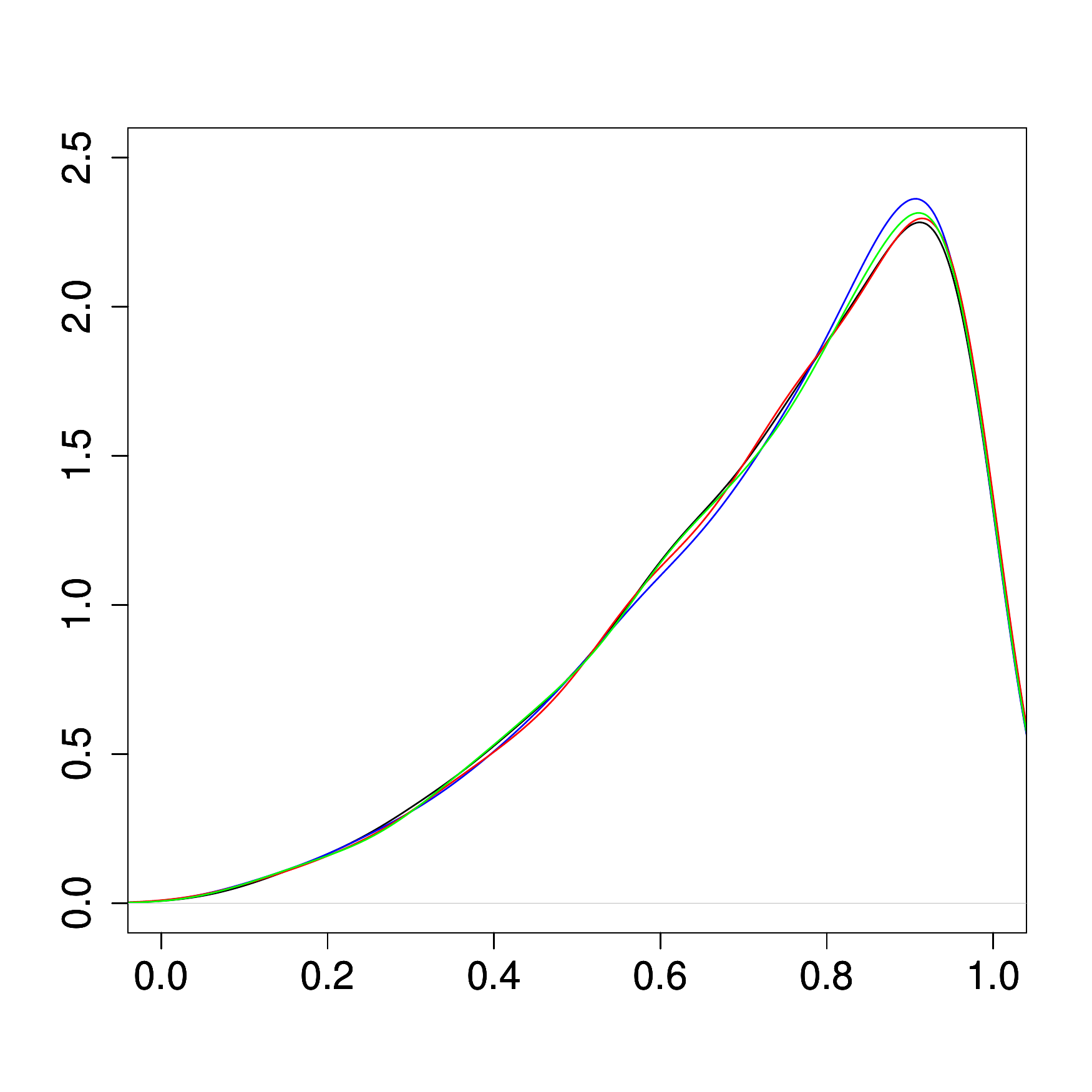}}
\hspace{2mm}
\subfigure[Site 1, Species 45.]{ \label{fig:prop3}
\includegraphics[width=4.5cm,height=4.5cm]{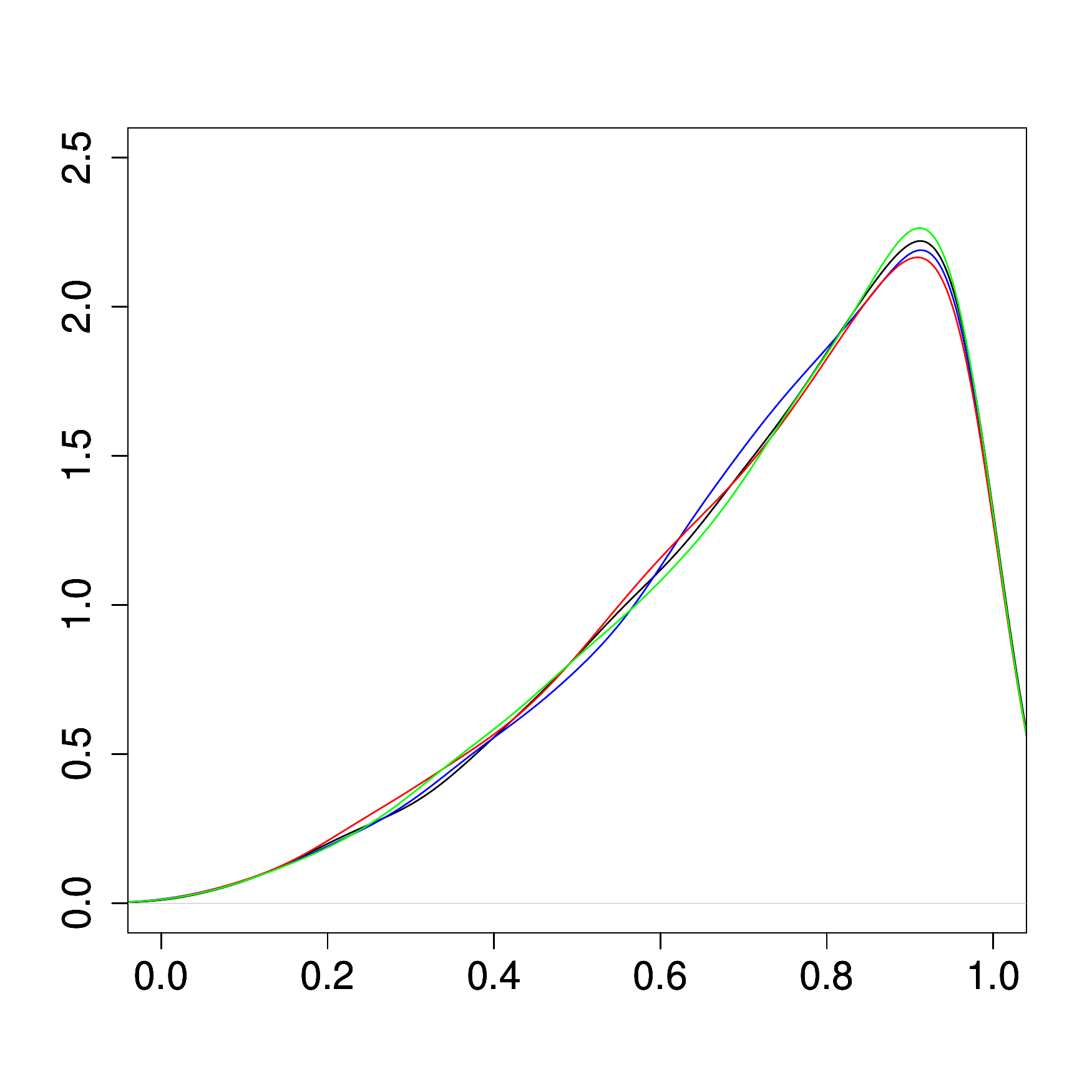}}\\
\vspace{2mm}
\subfigure[Site 30, Species 5.]{ \label{fig:prop4} 
\includegraphics[width=4.5cm,height=4.5cm]{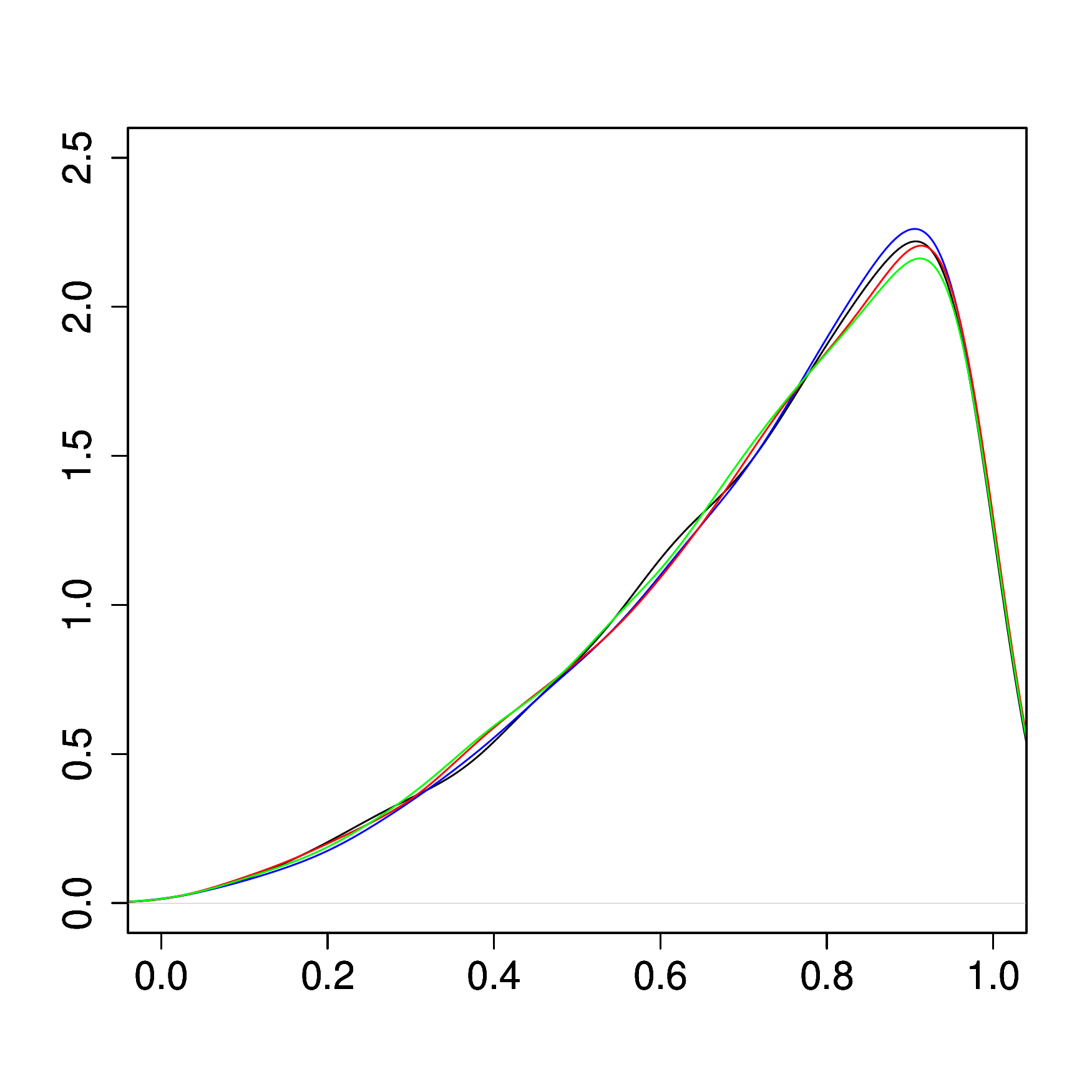}}
\hspace{2mm}
\subfigure[Site 30, Species 30.]{ \label{fig:prop5} 
\includegraphics[width=4.5cm,height=4.5cm]{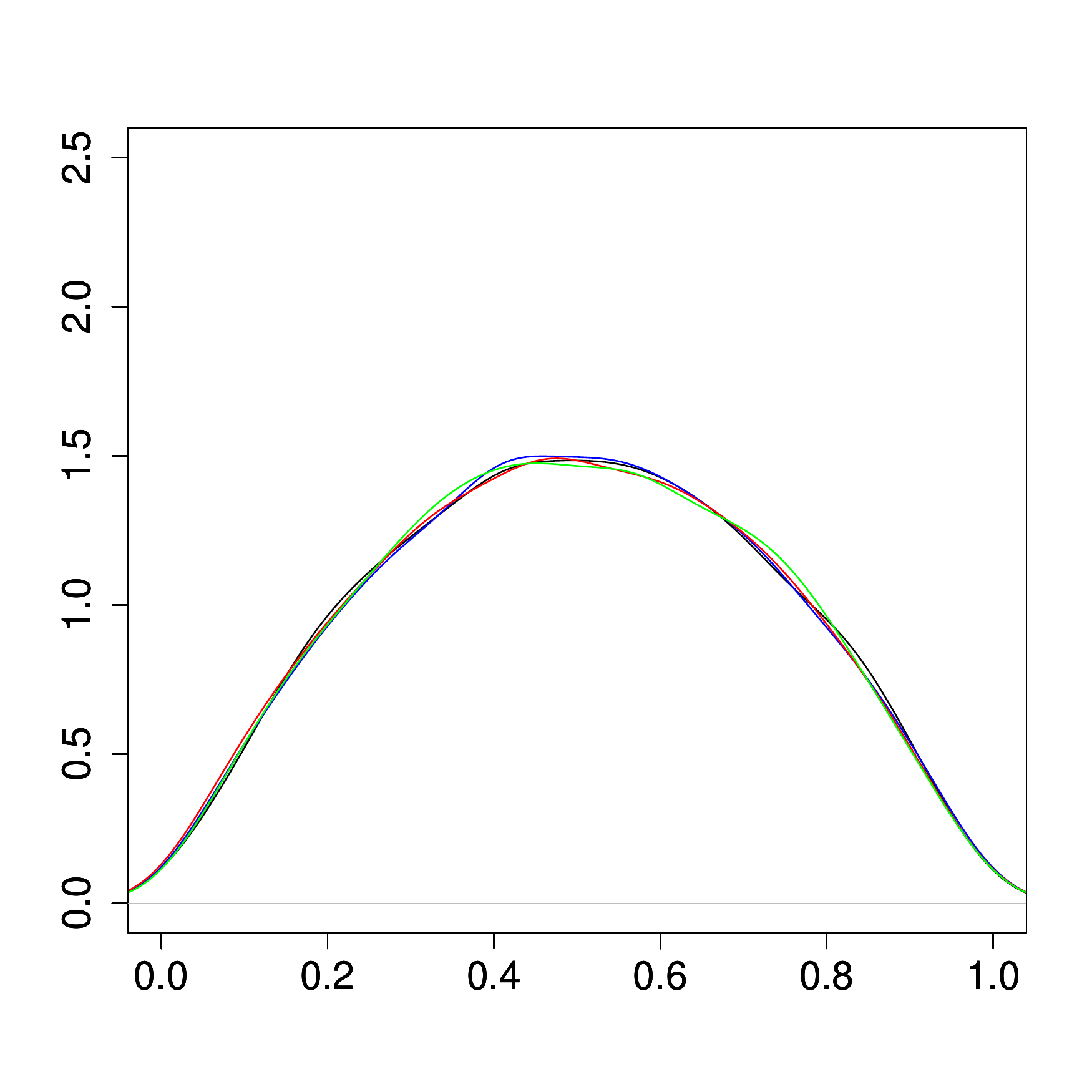}}
\hspace{2mm}
\subfigure[Site 30, Species 45.]{ \label{fig:prop6} 
\includegraphics[width=4.5cm,height=4.5cm]{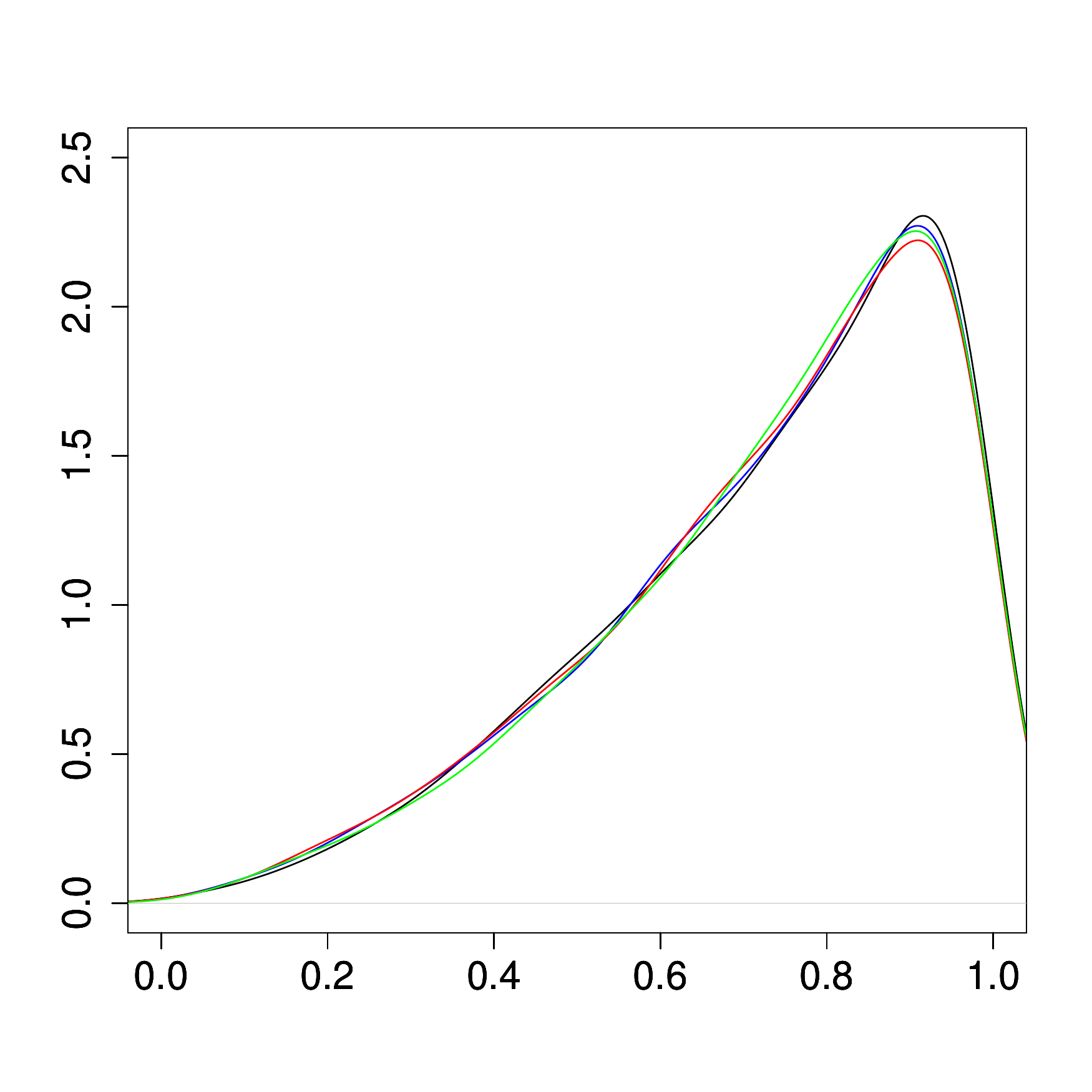}}\\
\vspace{2mm}
\subfigure[Site 60, Species 5.]{ \label{fig:prop7} 
\includegraphics[width=4.5cm,height=4.5cm]{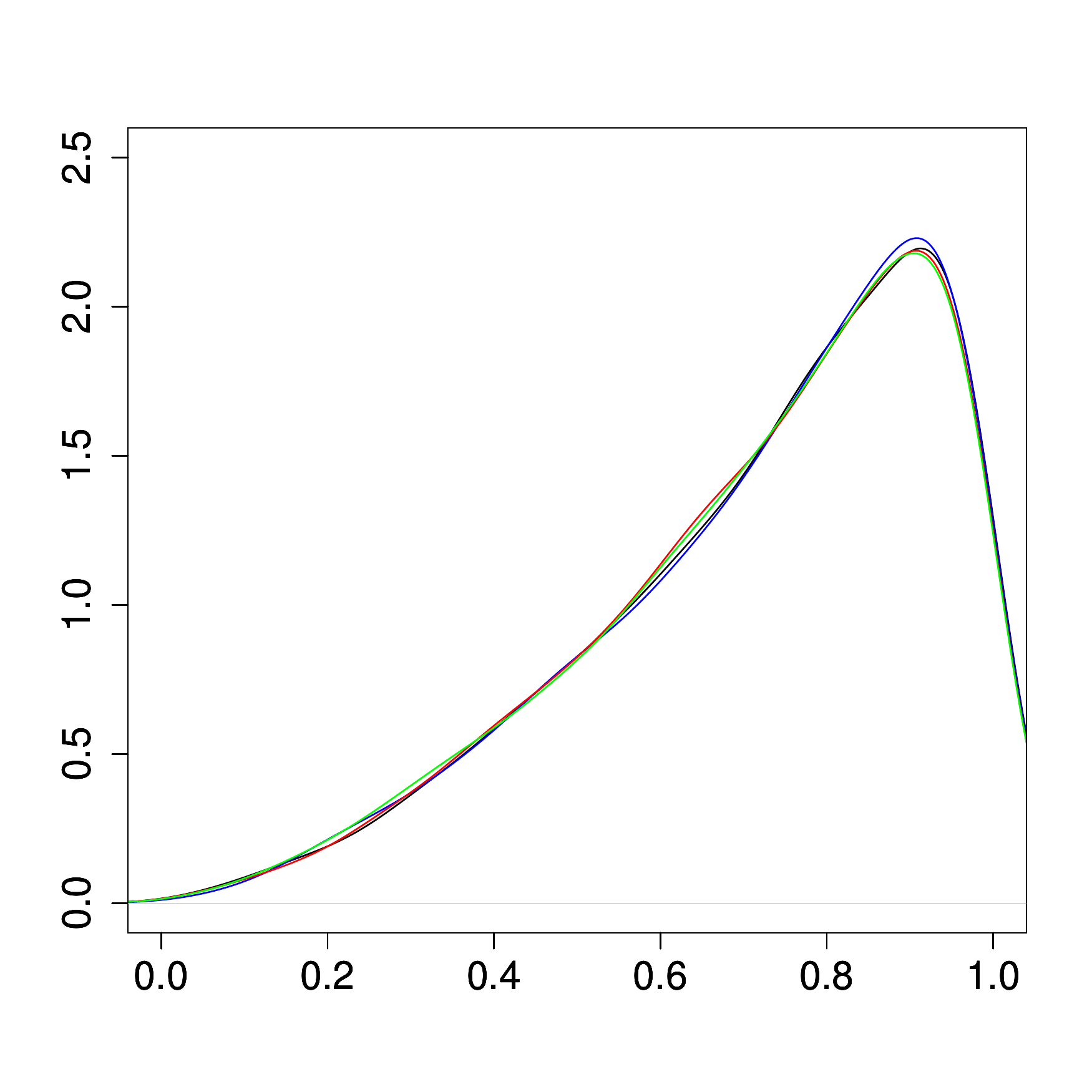}}
\hspace{2mm}
\subfigure[Site 60, Species 35.]{ \label{fig:prop8} 
\includegraphics[width=4.5cm,height=4.5cm]{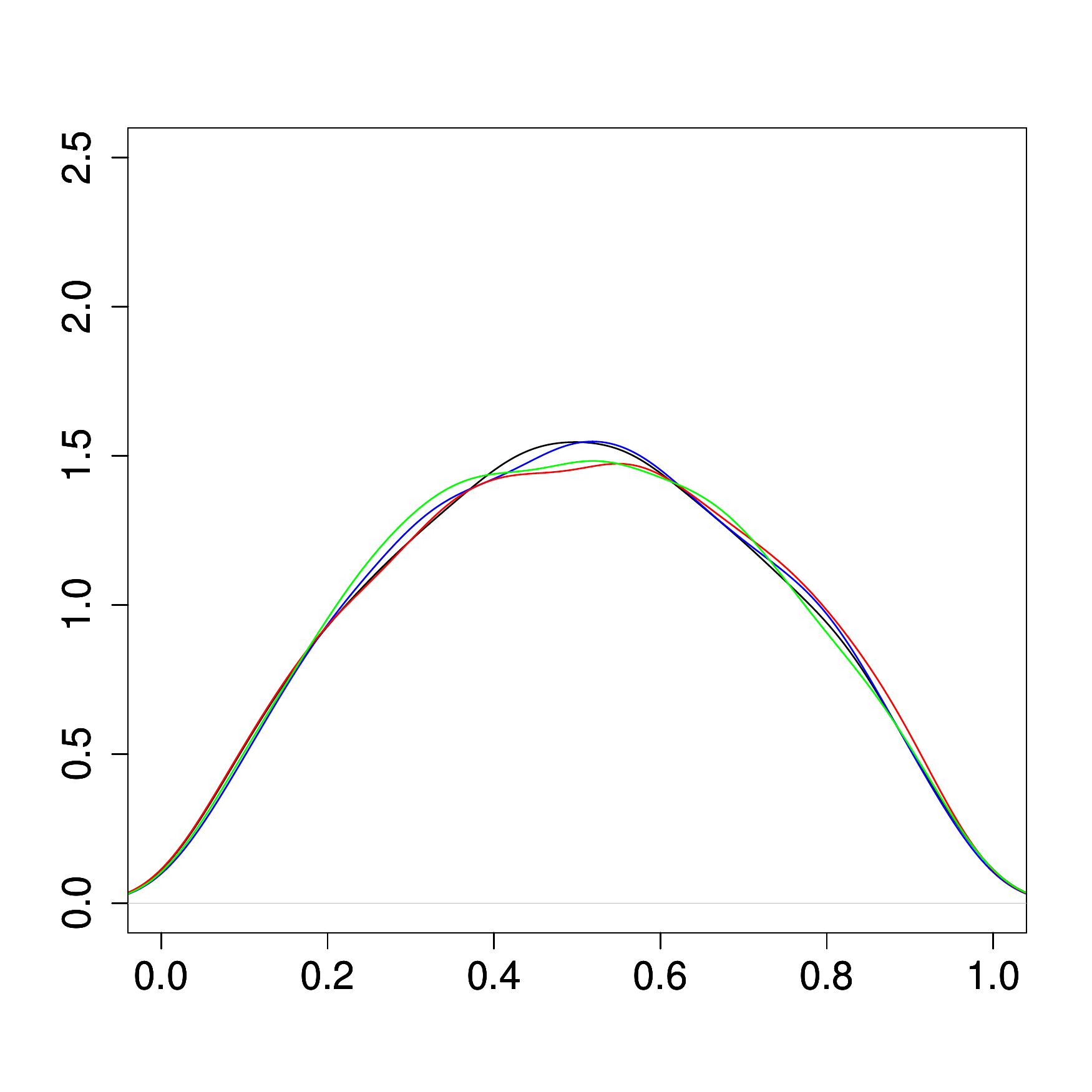}}
\hspace{2mm}
\subfigure[Site 60, Species 45.]{ \label{fig:prop9} 
\includegraphics[width=4.5cm,height=4.5cm]{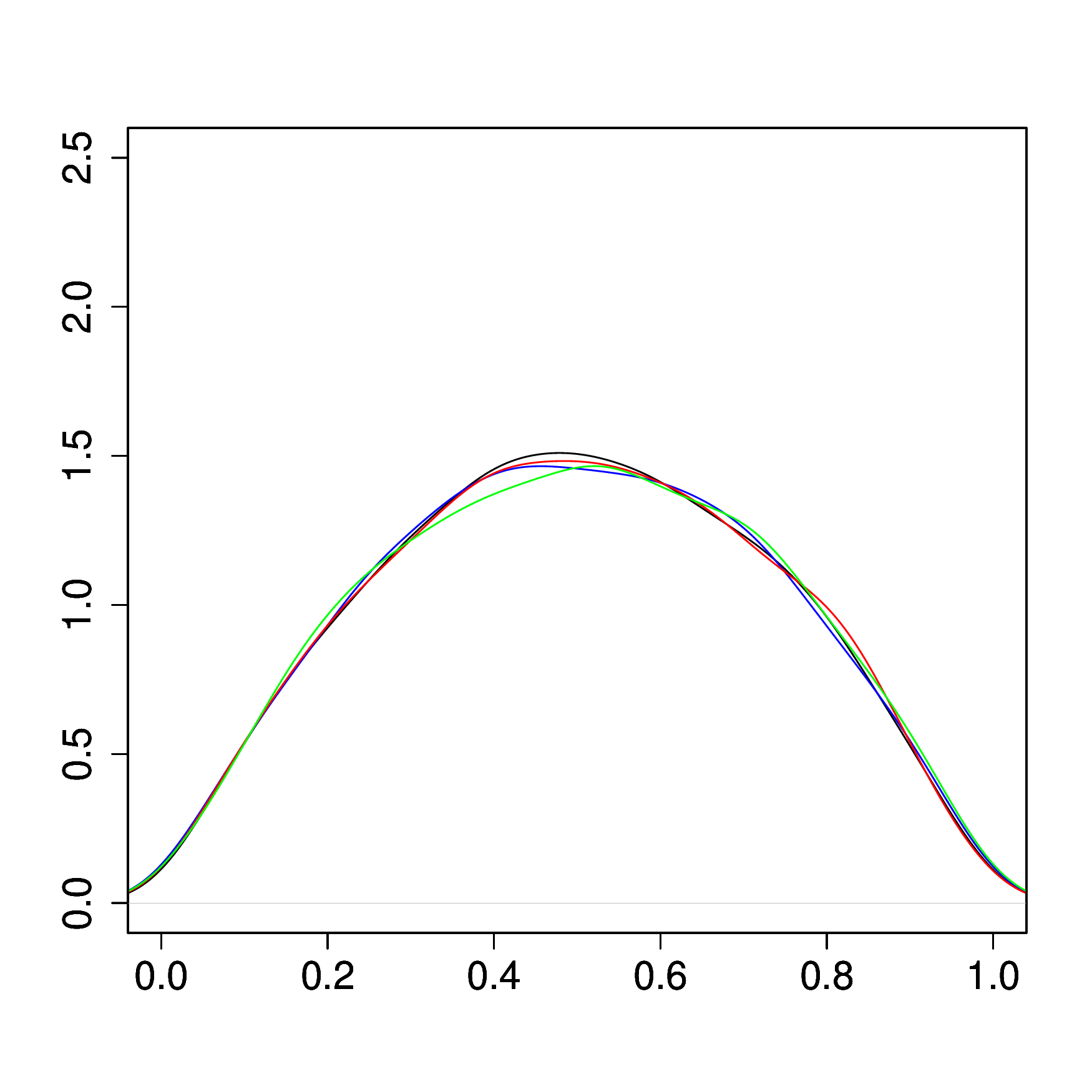}}
\caption{{\bf Chironomid data:} Posterior distributions of $\pi_{ik}$ corresponding to the full MCMC run for the 
joint posterior associated with $i^*=38$ with respect to different choices of $\alpha$ and $\sigma^2_x$.
Different colours represent posteriors with respect to different prior choices; 
black corresponds to ($\alpha=25$, $\sigma^2_x=10$), blue to ($\alpha=25$, $\sigma^2_x=5$),
red to ($\alpha=10$, $\sigma^2_x=10$), and green to ($\alpha=10$, $\sigma^2_x=5$).} 
\label{fig:cv2}
\end{figure}

Figure \ref{fig:cv2} shows the posteriors of some of the $\pi_{ik}$, the probabilities of zero counts, associated with our model,
under different choices of $\alpha$ and $\sigma^2_x$. The displayed figures correspond
to the full MCMC run for the joint posterior associated with $i^*=38$. 
Considerable robustness of the posteriors of $\pi_{ik}$ with respect to different choices of $\alpha$ and $\sigma^2_x$
is exhibited by the plots. Importantly, it is clearly seen that the posteriors of $\pi_{ik}$
have modes closer to 1 than to 0 indicating that it is indeed really important to model the count data with zero-inflated multinomial distribution
to account for such large proportion of zeros. 

\subsection{Goodness of fit of the response functions}
\label{subsec:response_functions}

Apart from the cross-validation results, it is also of interest to ascertain how well 
our Dirichlet process based response functions perform. Since this is directly related to the question
of predicting the species abundances, here we consider predicting the observed species abundances using
the posterior expectations of $\tilde y_{ik}$ conditional on $y_{i\cdot}$, where $\tilde y_{ik}$
is the random variable associated with (or, a replicate of) the observed data point $y_{ik}$. 


It follows by conditional independence, that
\begin{eqnarray}
&&\left[\tilde y_{ik}\mid  y_{i\cdot},\bX,\bY\right]\nonumber\\
&&=\sum_{z_{i1},\ldots,z_{im}}\int \left[\tilde y_{ik}\mid y_{i\cdot},z_{i1},\ldots,z_{im},
\lambda_{i1},\ldots,\lambda_{im}\right]\nonumber\\
&&\quad\quad\times\left[z_{i1},\ldots,z_{im},\lambda_{i1},\ldots,\lambda_{im}\mid \bX,\bY\right]
d\lambda_{i1}\ldots d\lambda_{im},
\label{eq:response_function2}
\end{eqnarray}
where $$\left[\tilde y_{ik}\mid y_{i\cdot},z_{i1},\ldots,z_{im},\lambda_{i1},\ldots,\lambda_{im}\right]
\sim Binomial\left(y_{i\cdot},\frac{\lambda_{ik}}{\sum_{\ell:z_{i\ell}=0}\lambda_{i\ell}}\right)$$
if $z_{ik}=0$. On the other hand, if $z_{ik}=1$, then
$$\left[\tilde y_{ik}\mid y_{i\cdot},z_{i1},\ldots,z_{im},\lambda_{i1},\ldots,\lambda_{im}\right]
\sim\delta_{\{0\}}.$$
Thus, the posterior distribution $\left[\tilde y_{ik}\mid  y_{i\cdot},\bX,\bY\right]$ 
can be studied by drawing samples from 
$\left[\tilde y_{ik}\mid y_{i\cdot},z_{i1},\ldots,z_{im},\lambda_{i1},\ldots,\lambda_{im}\right]$,
given available MCMC samples drawn from 
$\left[z_{i1},\ldots,z_{im},\lambda_{i1},\ldots,\lambda_{im}\mid \bX,\bY\right]$.

\begin{figure}
\centering
\subfigure[Species 1.]{ \label{fig:cred_ade_our2}
\includegraphics[width=5.5cm,height=5.5cm]{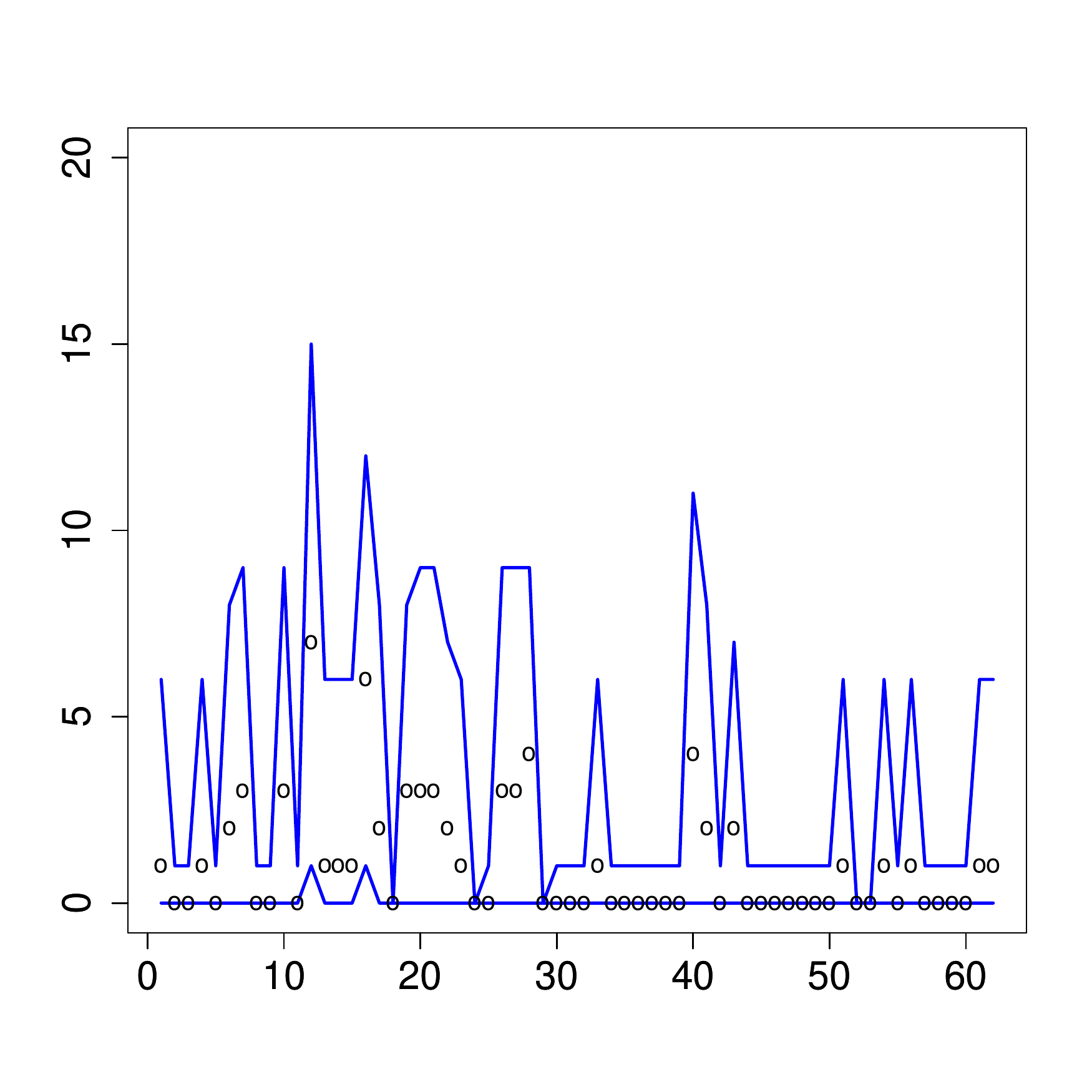}}
\hspace{2mm}
\subfigure[Species 26.]{ \label{fig:cred_ade_our1} 
\includegraphics[width=5.5cm,height=5.5cm]{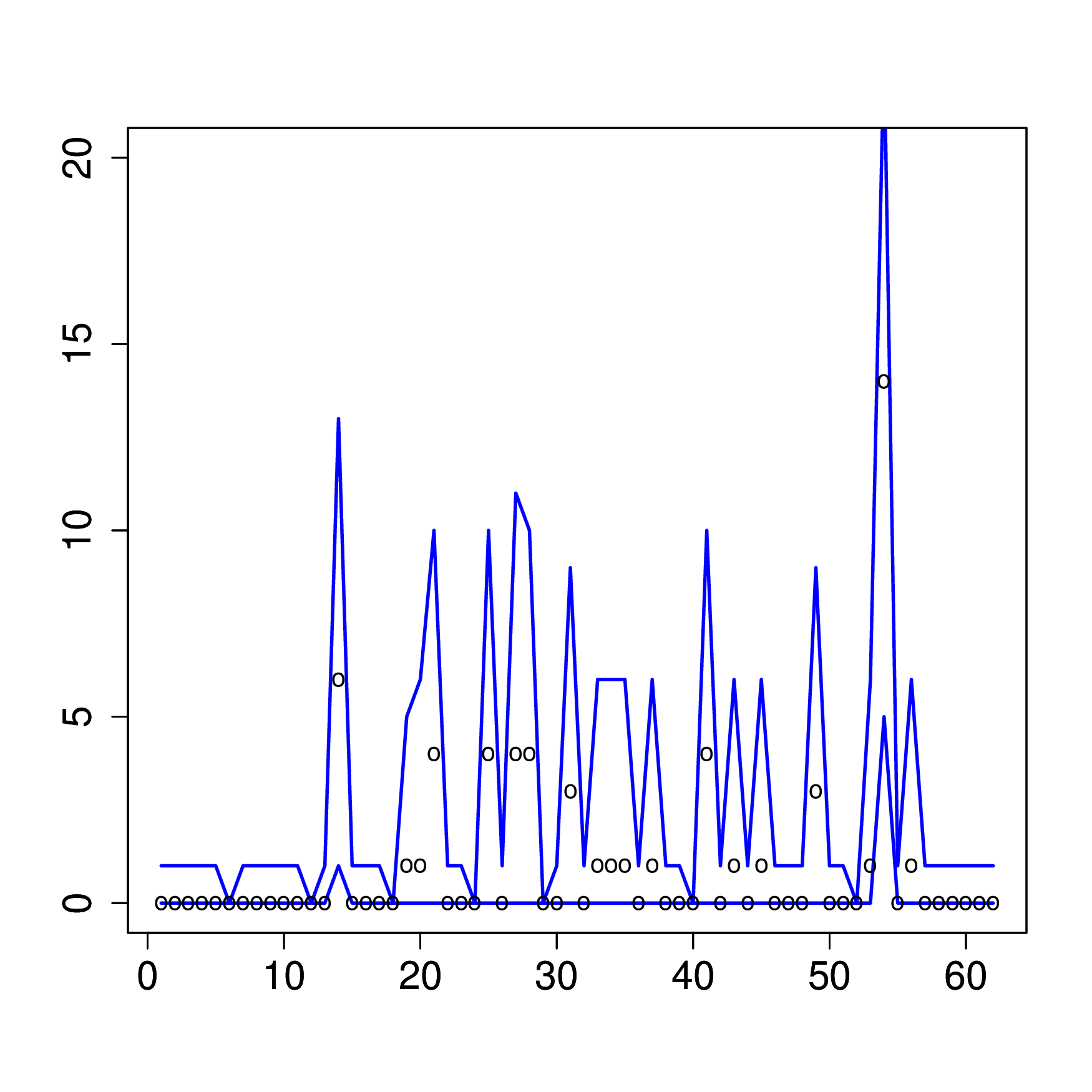}}\\
\vspace{2mm}
\subfigure[Species 51.]{ \label{fig:cred_ade_our3}
\includegraphics[width=5.5cm,height=5.5cm]{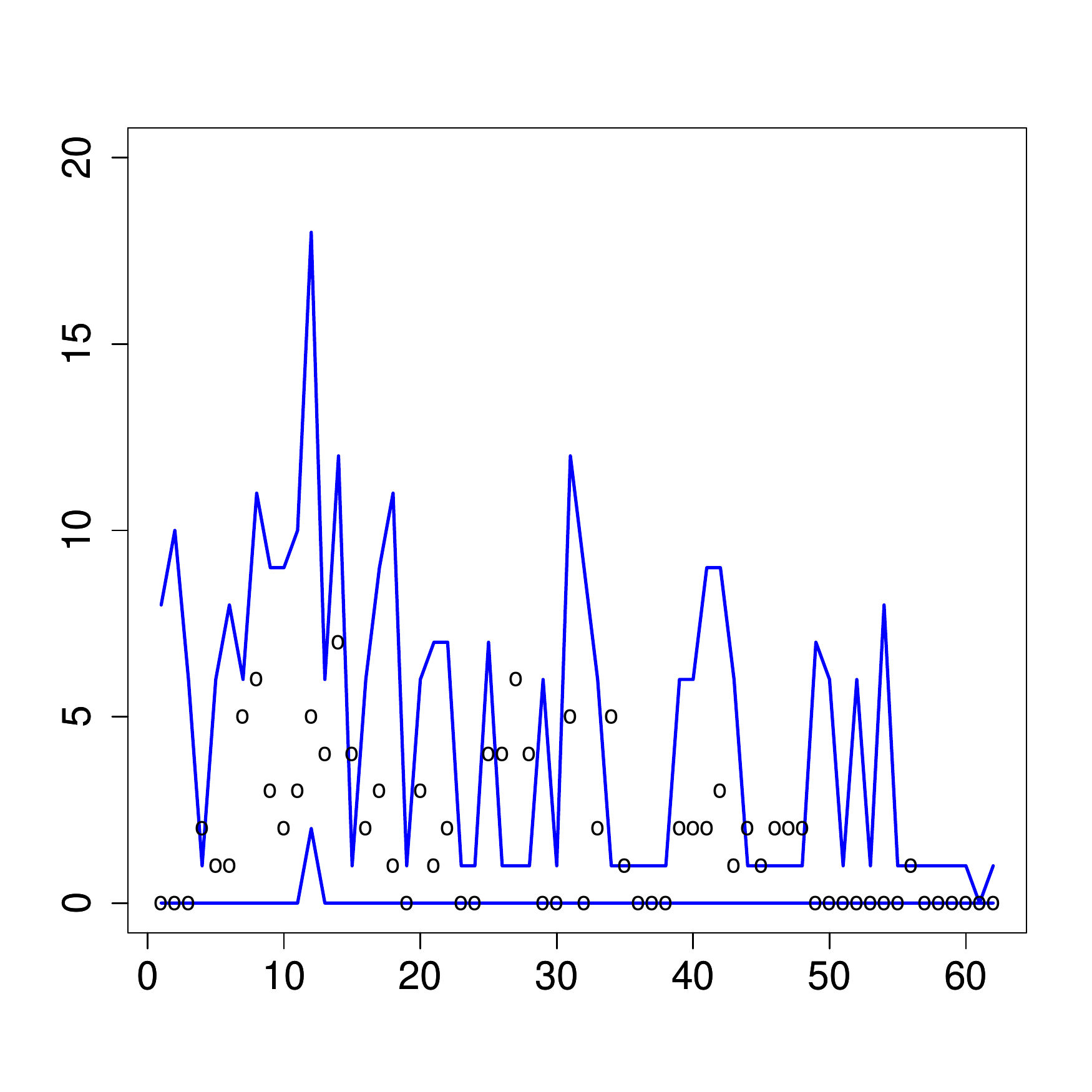}}
\caption{{\bf Credible intervals of species abundances for the chironomid data:} 
The circles represent the observed abundances and the lower and the upper curves 
represent lower and upper 95\% credible intervals of $\tilde y_{ik}$, joined by lines.}
\label{fig:cred_chiro}
\end{figure}

We construct the predicted version of the count data for the $k$-th species using the posterior distributions
of $\{\tilde y_{ik};i=1,\ldots,n\}$. 
Figure \ref{fig:cred_chiro} 
shows the respective 95\% credible intervals of $\{\tilde y_{ik};i=1,\ldots,62\}$, joined by lines; the
circles denote the count data.
It is clear
from the figures that a reasonably good fit 
is provided by our response function model.

The results of cross-validation and the fit of the response functions to the observed data may seem to be satisfactory, 
but a test of overall model adequacy is necessary
to formally certify our new model. In the next section we address the issue of model adequacy test. 

\section{A test for overall model adequacy}
\label{sec:model_assess_results}

To quote \ctn{Gelman96}, assessing the plausibility of a posited model (or of assumptions in general)
is always fundamental, especially in Bayesian data analysis. \ctn{Gelman96} seem to be the first to attempt an extension of 
the essence of the classical
approach of model assesment to the Bayesian framework. Their approach is based on computing the posterior distribution of
the parameters given the data and then to compute a $P$-value, involving a discrepancy measure, which is a
function of the data as well as the parameters. Their approach differs from the available classical approaches
mainly in introducing a discrepancy measure that depends on the parameters as well. \ctn{Bayarri99} introduced two alternative 
$P$-values and demonstrated that they are advantageous compared to
the $P$-value of \ctn{Gelman96}.

Motivated by the palaeoclimate reconstruction problem in ``modern data'' on fossil pollen assemblages, 
\ctn{Bhattacharya12} proposed a novel approach to model assesment based on ``inverse reference distributions'' (IRD).
He has shown that his approach is suitable for assessing Bayesian model fit in inverse problems but may be extended to quite
general Bayesian framework and has some distinct advantages compared to the other approaches. Here we will use the idea of 
\ctn{Bhattacharya12} for assessing the plausibility of our model. 

The idea of \ctn{Bhattacharya12} is based on the philosophy that the model fits the data if the posterior distribution of 
the random variables corresponding to the non-random covariates capture the observed values of the covariates. Otherwise, the model does not 
fit the data. It is worth noting that although the values of the covariates are known, the model is to be fitted
{\it assuming that the values are unknown and the random variables that stand for the unknown covariates are to be predicted}.
The covariates predicted in this manner can then be compared with the originally observed values to assess model fit in
a fully Bayesian manner.  

The key idea can be mathematically formulated in the following way.
Suppose $\bY = \{{\by_i}, i=1, \ldots, n\}$ represent the data and $\bX = \{{x_i}, i=1, \ldots, n\}$ represent the non-random covariates.
Let $\tilde \bX$ stand for the random vector associated with $\bX$; the former may also be thought of as a replicate of 
$\bX$
but must be predicted conditionally on $\bY$ in an inverse sense.
If the posterior distribution of $\tilde \bX$ is consistent with observed $\bX$ then the model is said to have fit the data adequately. 
Otherwise the model is considered inadequate for the data. The fully Bayesian approach to this prediction requires 
computation of an inverse reference
distribution based on the posterior
\begin{equation}
\pi\left(\tilde{\bX}\mid \bY\right) \propto \int \pi\left(\tilde{\bX}, \btheta\right) \mathcal L\left(\bY, \tilde{\bX}, \btheta\right)d\btheta,
\nonumber
\end{equation}
where $\mathcal L$ denotes the likelihood of the unknowns $\left(\tilde{\bX}, \btheta\right)$, $\btheta$ being the set of model
parameters. \ctn{Bhattacharya12} discuss in details the advantages of using this reference distrbution. 
He also shows how the reference distribution may turn out to be improper and demonstrated how the 
leave-one-out cross-validation idea
may overcome the problem of impropriety. 
To assess consistency of the simulated covariates with the observed values \ctn{Bhattacharya12} suggests appropriate
discrepancy measures $T(\cdot)$ -- a reference distribution of the random discrepancy measure $T(\tilde{\bX})$ is to be
constructed using the simulated covariates $\tilde{\bX}$; then if $T(\bX)$, the observed discrepancy measure corresponding to the observed
covariates $\bX$, falls within the appropriate credible region of $T(\tilde{\bX})$, the model is to be accepted, otherwise it should be rejected.  
The decision theoretic justification of the procedure is provided in \ctn{Bhattacharya12}.

\begin{figure}
\centering
\subfigure[$\alpha=25$ and $\sigma^2_x=10$.]{ \label{fig:adeq_our2}
\includegraphics[width=5.5cm,height=5.5cm]{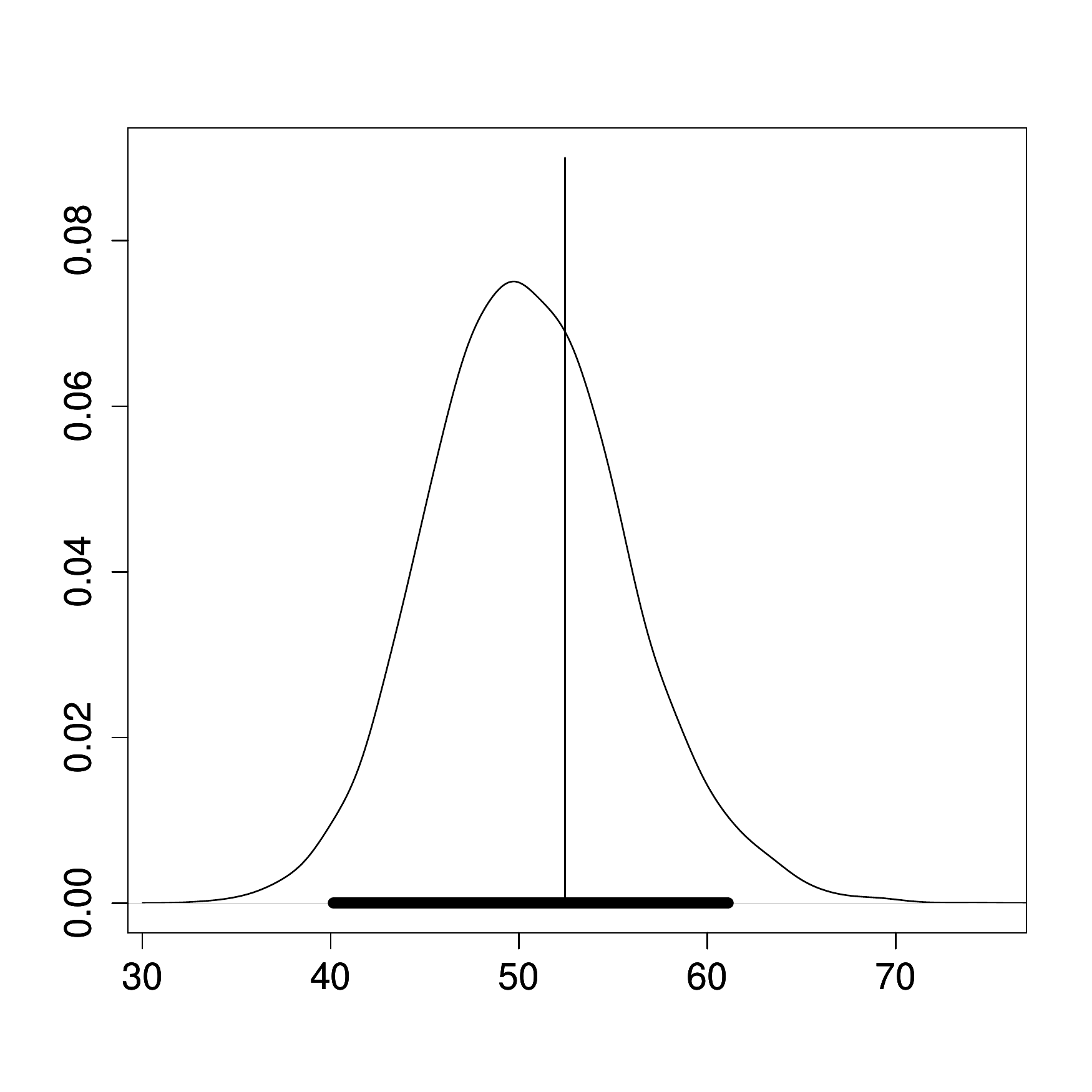}}
\hspace{2mm}
\subfigure[$\alpha=25$ and $\sigma^2_x=5$.]{ \label{fig:adeq_our1} 
\includegraphics[width=5.5cm,height=5.5cm]{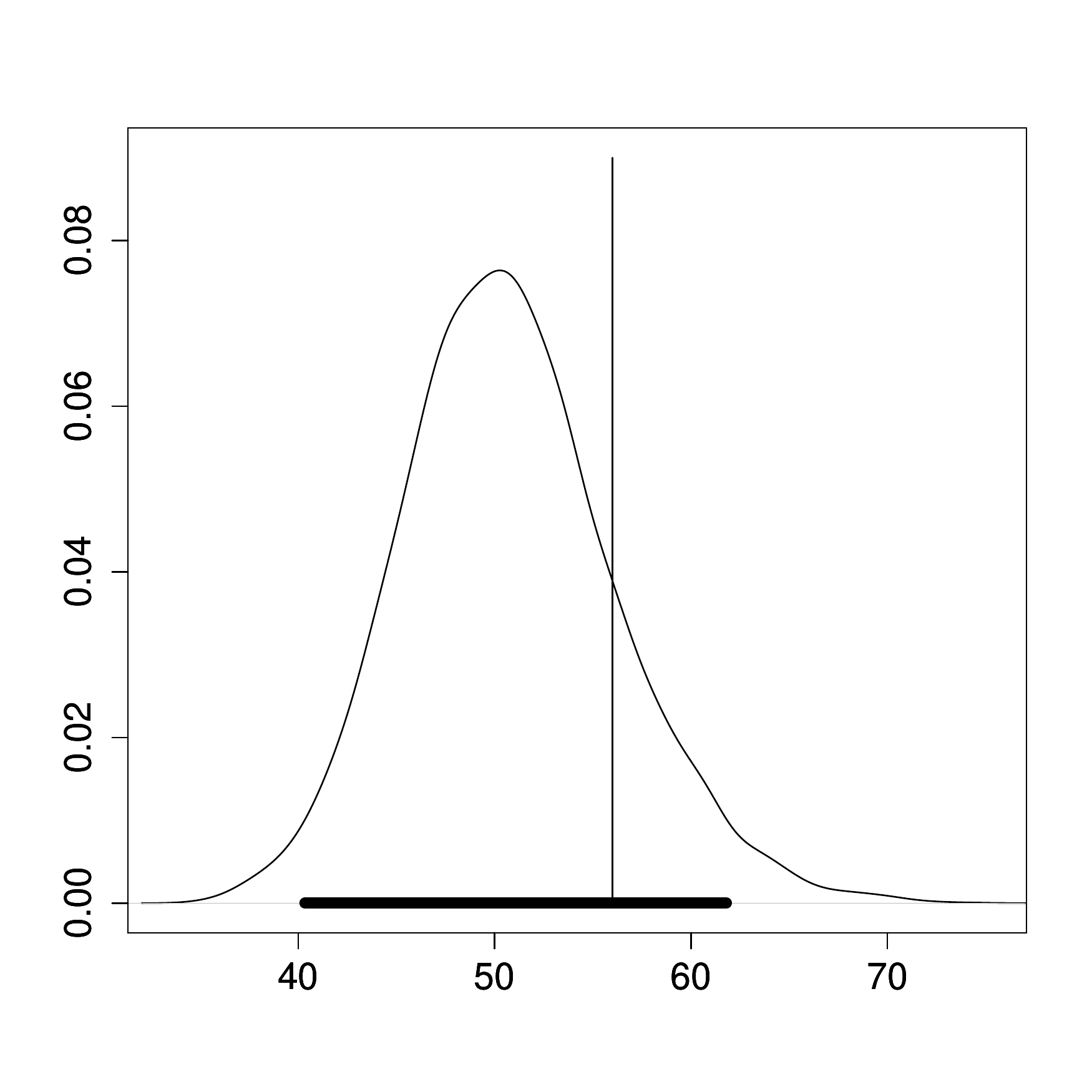}}
\vspace{2mm}
\subfigure[$\alpha=10$ and $\sigma^2_x=10$.]{ \label{fig:adeq_our3}
\includegraphics[width=5.5cm,height=5.5cm]{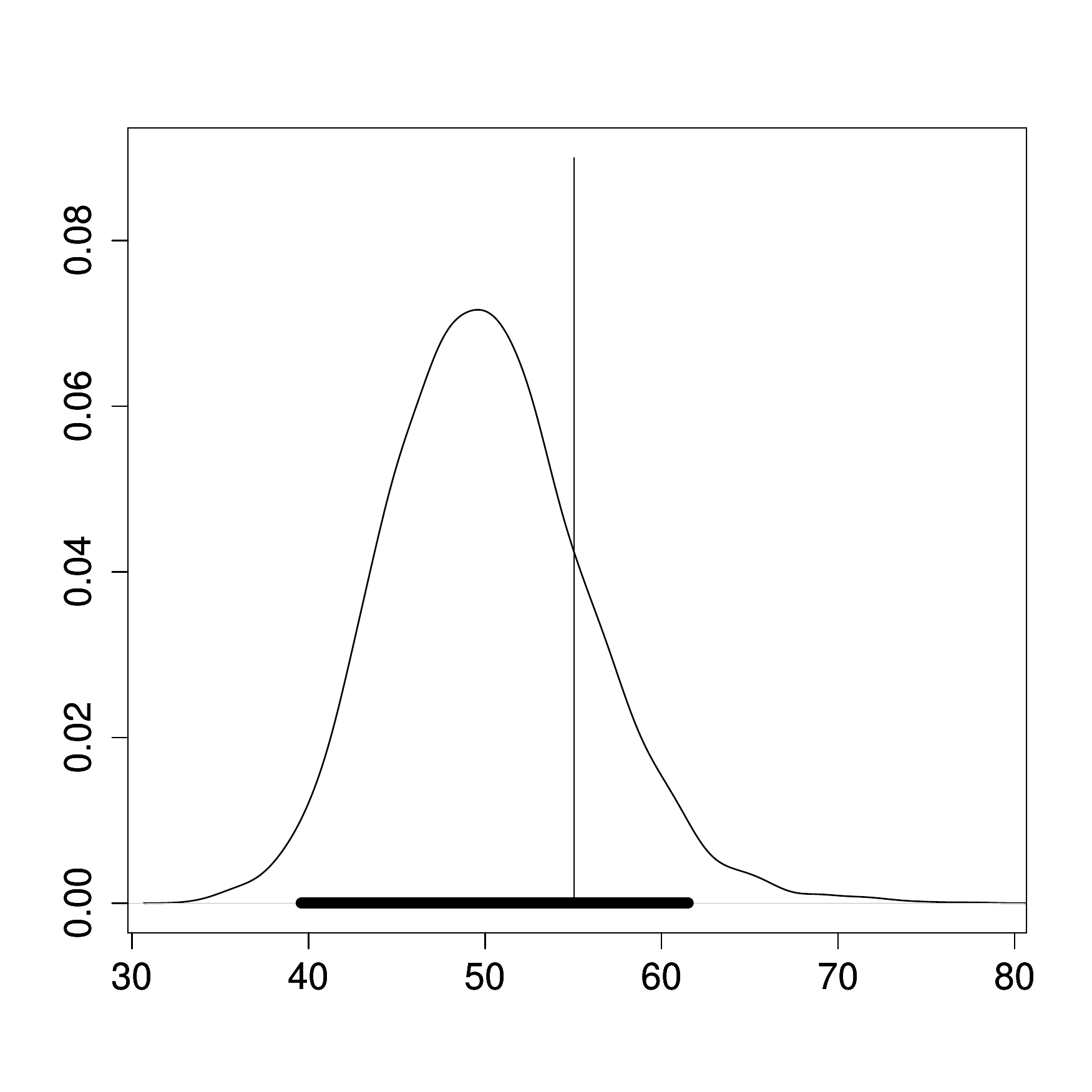}}
\vspace{2mm}
\subfigure[$\alpha=10$ and $\sigma^2_x=5$.]{ \label{fig:adeq_our4} 
\includegraphics[width=5.5cm,height=5.5cm]{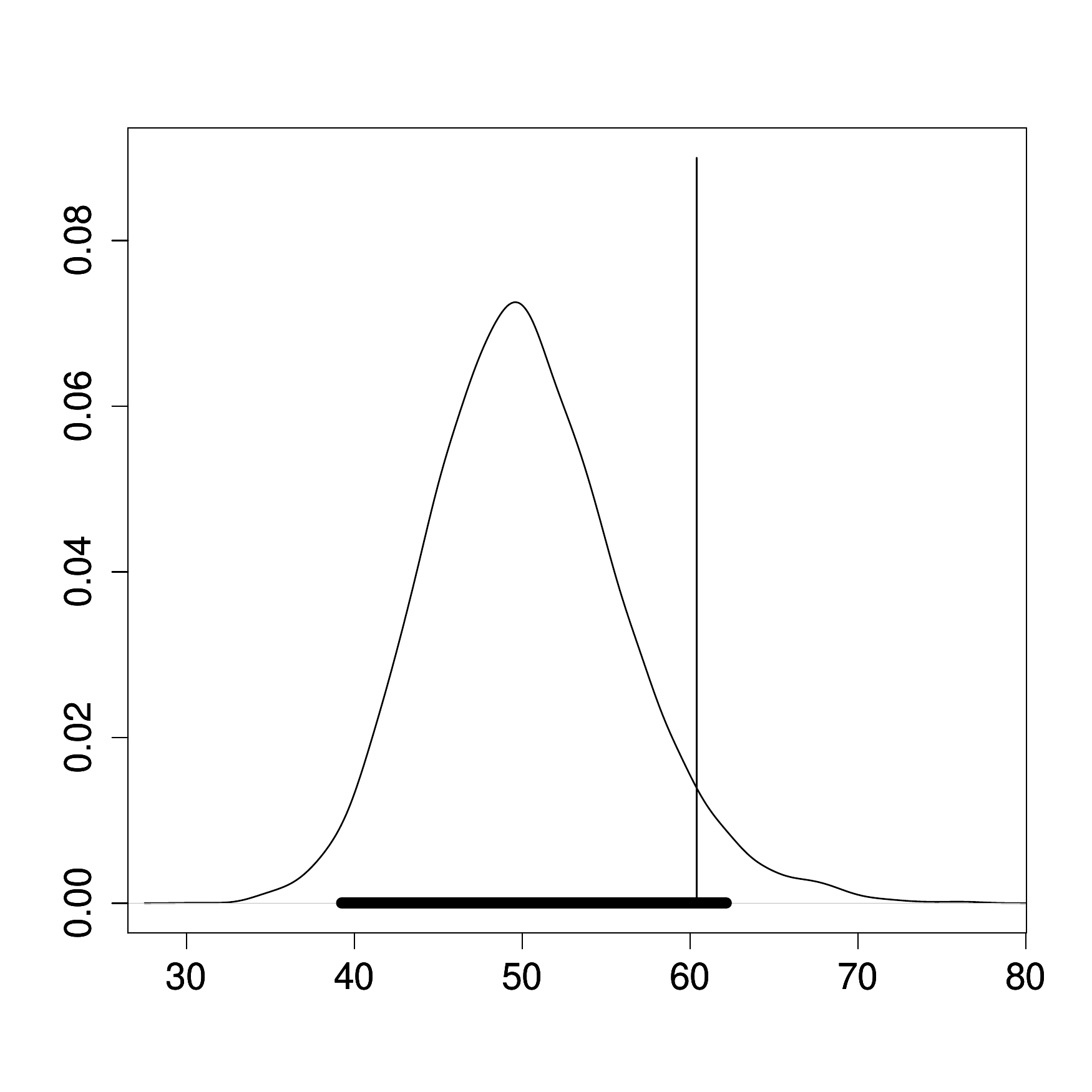}}
\caption{{\bf Model adequacy test for the chironomid data}: Shown are the posterior distributions
of $T_1(\tilde\bX)$ for various values of the prior parameters $\alpha$ and $\sigma^2_x$. 
The thick line in the bases represent the 95\% HPD intervals and the vertical lines stand for
the corresponding observed discrepancy measure $T_1(\bX)$.}
\label{fig:model_adequacy}
\end{figure}

Before applying the model adequacy test of \ctn{Bhattacharya12} we need to choose an appropriate
discrepancy measure $T(\cdot)$. Figure \ref{fig:cv1} shows that posterior distributions of some of the $x_i$ 
are skewed, while some are strongly indicative of multimodality. Considering the global mode $\tilde x^*_i$ of the posterior
distribution of $\tilde x_i$ as a convenient measure of central tendency, we use the following observed discrepancy measure:
\begin{eqnarray}
T_1(\bX) = \sum_{i=1}^{n} \frac{\vert x_{i}-\tilde x^*_i\vert}{\sqrt{Var(\tilde x_i)}}. \label{eq:model_statistic}
\end{eqnarray}
Replacing $\bX$ with $\tilde \bX$ in (\ref{eq:model_statistic}) yields the inverse reference distribution corresponding to $T_1(\bX)$.
%
Figures \ref{fig:adeq_our2}, \ref{fig:adeq_our1}, \ref{fig:adeq_our3}
and \ref{fig:adeq_our4} show the inverse reference distributions based on the IRMCMC simulations and 
the associated observed discrepancy measures
corresponding to our model with ($\alpha=25$, $\sigma^2_x=10$), ($\alpha=25$, $\sigma^2_x=5$), 
($\alpha=10$, $\sigma^2_x=10$) and ($\alpha=10$, $\sigma^2_x=5$), respectively.  
The thick, black horizontal lines represent the 95\% HPD regions of the posteriors of $T_1(\tilde \bX)$. 
The vertical lines represent the observed discrepancy measures $T_1(\bX)$.  
In all the cases $T_1(\bX)$ fall comfortably within the 95\% HPD regions of the corresponding 
inverse reference distributions, clearly leading to acceptance of our model. 
We also considered several variants of the discrepancy measure (\ref{eq:model_statistic}) 
by replacing the mode $\tilde x^*_i$ with the median, taking sum of squares instead of sum 
of absolute deviations, etc. However, all these variants led to acceptance of our model.

\begin{figure}
\centering
\subfigure[$\alpha=25$ and $\sigma^2_x=10$.]{ \label{fig:kl_like1}
\includegraphics[width=5.5cm,height=5.5cm]{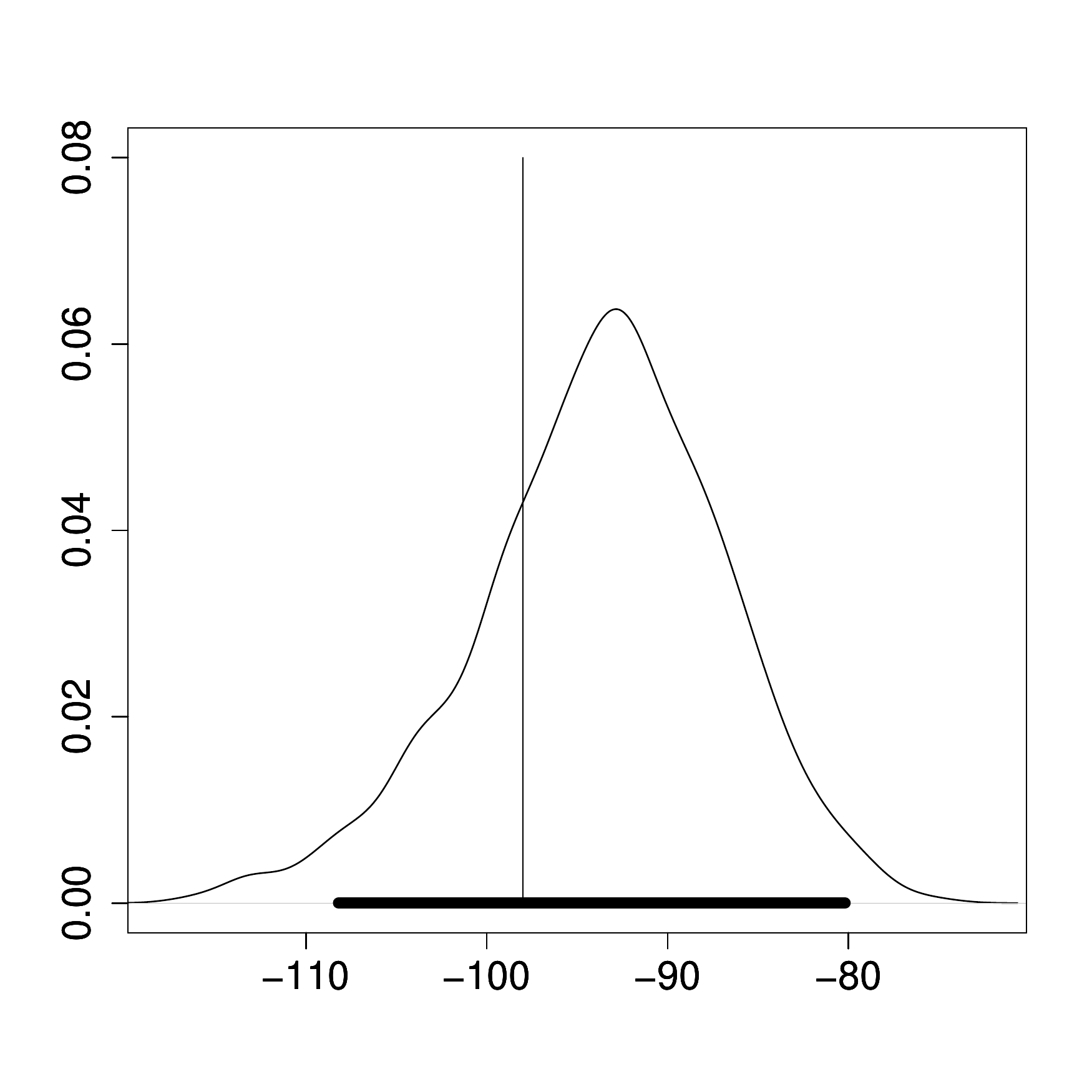}}
\hspace{2mm}
\subfigure[$\alpha=25$ and $\sigma^2_x=5$.]{ \label{fig:kl_like2} 
\includegraphics[width=5.5cm,height=5.5cm]{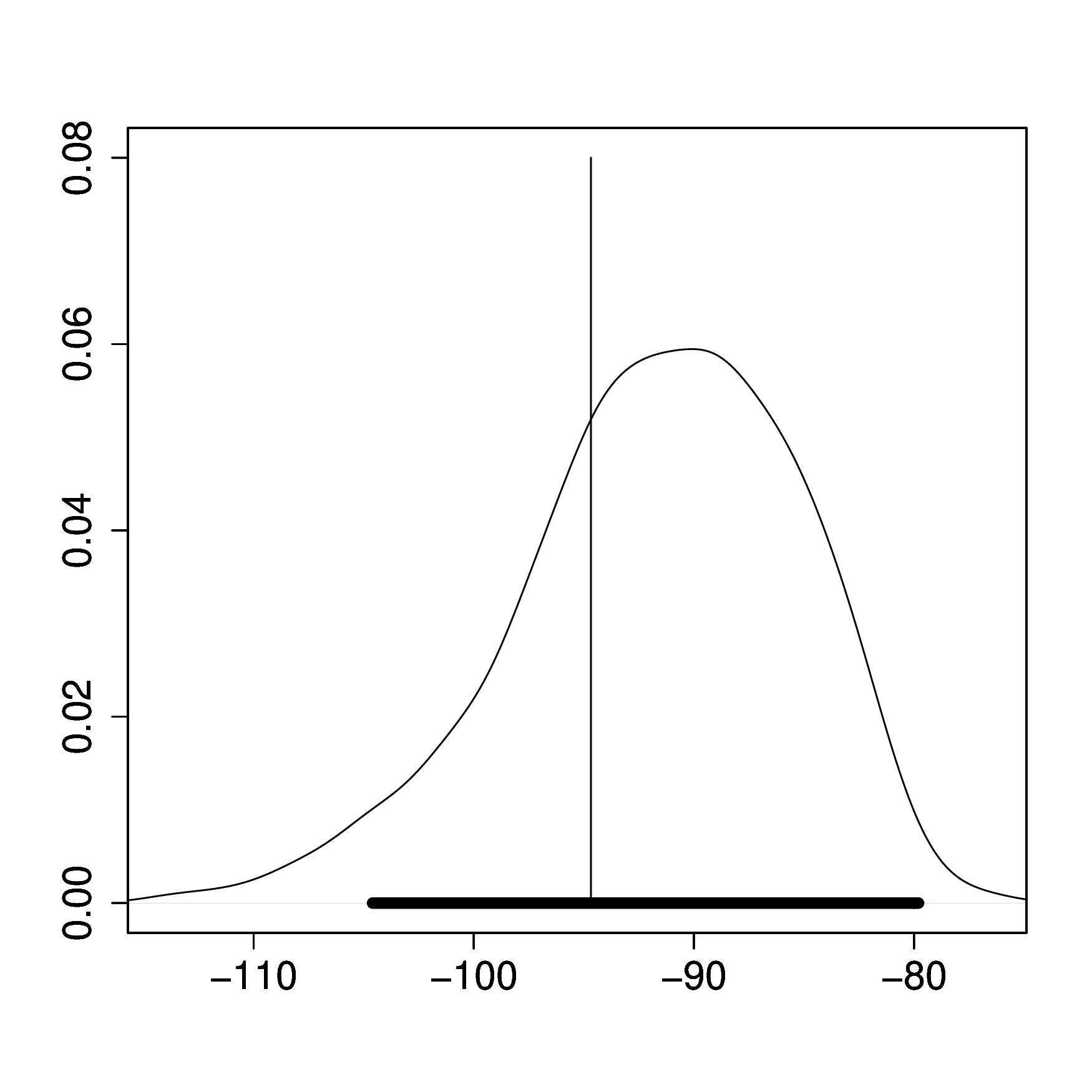}}
\vspace{2mm}
\subfigure[$\alpha=10$ and $\sigma^2_x=10$.]{ \label{fig:kl_like3}
\includegraphics[width=5.5cm,height=5.5cm]{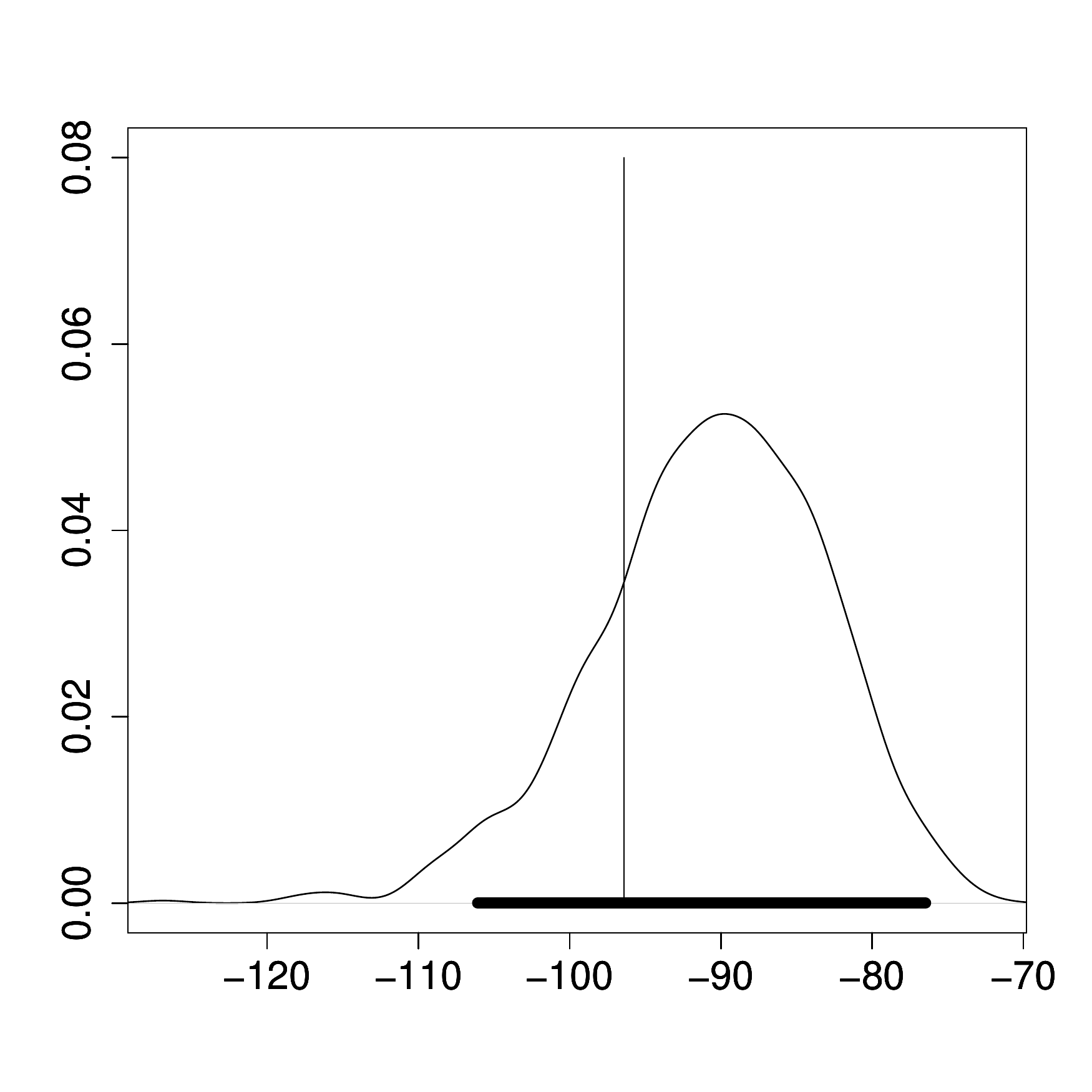}}
\vspace{2mm}
\subfigure[$\alpha=10$ and $\sigma^2_x=5$.]{ \label{fig:kl_like4} 
\includegraphics[width=5.5cm,height=5.5cm]{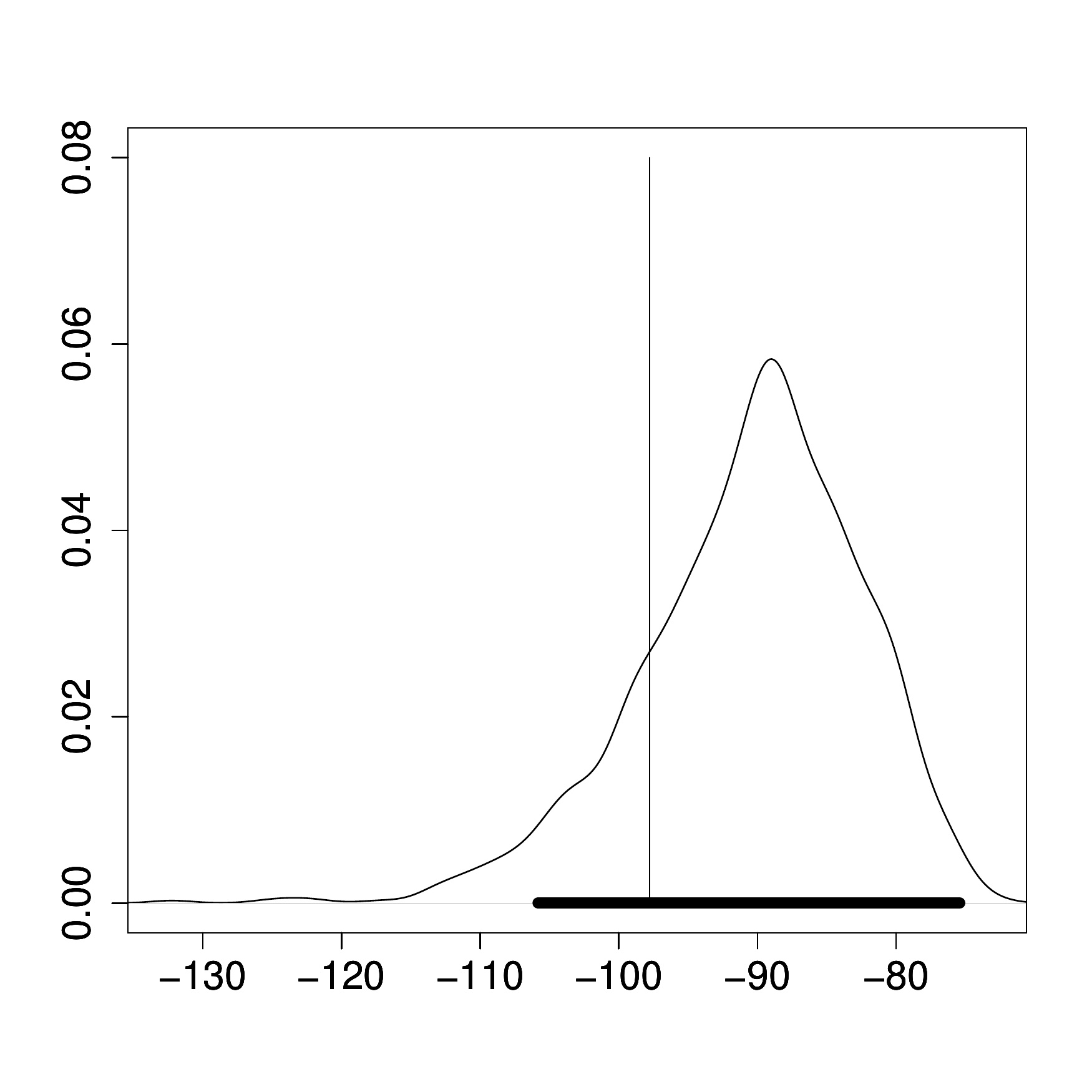}}
\caption{{\bf Model adequacy test for the chironomid data}: Shown are the posterior distributions
of $T_2(\tilde\bX)$ for various values of the prior parameters $\alpha$ and $\sigma^2_x$. 
The thick line in the bases represent the 95\% HPD intervals and the vertical lines stand for
the corresponding observed discrepancy measure $T_2(\bX)$.}
\label{fig:chironomid_kl_like}
\end{figure}

Since the cross-validation posterior distributions of $\tilde x_i$ are multimodal, it is possible 
to question our choice of the discrepancy measure that makes use of the absolute deviation. 
One plausible discrepancy measure in this case may be that associated with the logarithms of the 
cross-validation posteriors. In other words, we may choose the following discrepancy measure: 
\begin{equation} 
T_2(\tilde \bX)=\sum_{i=1}^n\log\pi(\tilde x_i\vert \bX_{-i},\bY),\quad\mbox{so that}\quad
T_2(\bX)=\sum_{i=1}^n\log\pi(x_i\vert \bX_{-i},\bY). 
\label{eq:kl_like} 
\end{equation} 
Figures \ref{fig:kl_like1}, \ref{fig:kl_like2}, \ref{fig:kl_like3}
and \ref{fig:kl_like4} display the IRMCMC-based inverse reference distributions associated with $T_2$
corresponding to our model with ($\alpha=25$, $\sigma^2_x=10$), ($\alpha=25$, $\sigma^2_x=5$), 
($\alpha=10$, $\sigma^2_x=10$) and ($\alpha=10$, $\sigma^2_x=5$), respectively.  
As with $T_1$, even with $T_2$, the observed discrepancy measure $T_2(\bX)$ falls comfortably within
the 95\% HPD region for all the four different choices of $(\alpha,\sigma^2_x)$.

In Section S-3 of the supplement we investigate the relationship of the discrepancy measure $T_1$
with other discrepancy measures that are variants of $T_2$ above.

\section{Generalization of our model and methods to the modern pollen data}
\label{sec:pollen_data}

The training data set of HWB consists of modern pollen counts on $m$ = 14 species from $n$ = 7815 
different sites of the world, which we denote as before by
$\by_i = (y_{i1}, \ldots, y_{im})$, for $i=1, \ldots, n$. 
It is important to mention that unlike in the case of the chironomid data, here most of the total counts
$y_{i\cdot}$ are missing. It is however known that the total counts in this case are typically $400$. 
Following HWB we also treat the total counts as $400$, that is, we take $y_{i\cdot}=400$, for $i=1,\ldots,7815$.

The data also
includes modern, bivariate climate variables, namely, MTCO and GDD5 
at those sites, which we
denote as $\bx_i = (x_{i1}, x_{i2})$. 
Here we standardize $x_{i1}$ and $x_{i2}$ so that their sample means and variances are 0 and 1, respectively.
As in the case of the chironomid data we model
the pollen counts $\by_i$ as zero-inflated multinomial of the same 
form as (\ref{eq:multinomial}). Also, as in (\ref{eq:response}), $\lambda_{ik}$ is assumed to follow 
$Gamma(\xi_{ik}, \frac{1}{\psi})$, where
$\xi_{ik}$ is now modelled as 
\begin{eqnarray}
\xi_{ik} = \sum_{j=1}^{M_k} N_{2}\left(\bx_i, \bbeta_{kj}, \bSigma_{k}\right),
\label{eq:haslett_alpha}
\end{eqnarray}
where $N_{2}\left(\bx_i, \bbeta_{kj}, \bSigma_{k}\right)$ represents the bivariate normal density at $\bx_i$
with mean $\bbeta_{kj}$ and covariance matrix $\bSigma_{k}$. The $(s, t)$-th element of $\bSigma_{k}$ is denoted as
$\sigma_{k,st}$, $s,t = 1,2$. 
We assume that
\begin{align}
[\bbeta_{kj}\mid \bSigma_{k}]&\stackrel{iid}{\sim}\bG; \ \ \ \ j=1,\ldots,M_k; \ \ k=1,\ldots,m\label{eq:haslett_dp1}\\
[\bG\mid \bSigma_{k}] & \sim DP(\alpha \bG_0) \label{eq:haslett_dp2}
\end{align}
Under $\bG_0$, $\bbeta_{kj}$ is assumed to follow bivariate normal with mean vector 
$\bmu_{\beta} = (\mu_{\beta 0}, \mu_{\beta 1})$ and
covariance matrix $\bSigma_{k}$, where $\bmu_{\beta}$ is a known vector. For our application
we choose $\mu_{\beta 0}=\mu_{\beta 1}=0$, matching the sample mean of the standardized
climate variables GDD5 and MTCO.
The reason that we select these prior parameters in this way
is that the species optima $\{\bbeta_{kj};j=1,\ldots,M_k\}$,
which are exchangeable, and the climate variables at which the species data are collected, 
are expected to be similar, and hence uncertainties about them are not expected to be very
different. In fact, VTK and SB also assume the same prior mean for optimum temperature and the 
temperature variable.

For the prior on $\bSigma_k$ we assume that for $i=1,2$, $\sigma_{k,ii}\sim IG(a_{0i}, b_{0i})$,
the inverse-gamma prior with mean $b_{0i}/(a_{0i}+1)$ and variance $b^2_{0i}/(a_{0i}-1)^2(a_{0i}-2)$,
for $a_{0i}>2$. Here we choose $a_{0i}=4.1$ and $b_{0i}=5.1$ for $i=1,2$ so that both the prior means are 
1, matching the sample (standardized) variances of $x_{i1}$ and $x_{i2}$, while the prior variance is
1.3. Again, the rationale
for matching the sample variances is that the species optima and the climate variables
at which the species data are obtained are expected to have similar distributions.
The prior variances of $\sigma_{k,11}$ and $\sigma_{k,22}$ are made slightly larger than the sample climate
variances since the former are unobserved unlike the latter, thus incurring relatively more uncertainty.
Denoting $\frac{\sigma_{k,12}}{\sqrt{\sigma_{k,11} \sigma_{k,22}}}$ by $\rho_{k,12}$, we 
put the $Uniform (-1,1)$ prior on $\rho_{k,12}$. 

For this pollen data example, we choose $M_k=10$ and $\alpha=1$. Unlike the chironomid example, here
setting larger values of $\alpha$ led to overfitting the pollen data by increasing the number of mixture components
in the response function (\ref{eq:haslett_alpha}). This suggests that the response surface in the pollen
data example is expected to have less number of modes than in the chironomid data case.
It is useful to remark that the choice $\alpha=1$ is so common (see, for example, \ctn{Escobar95}, \ctn{Neal00}, \ctn{Green01}, \ctn{Dahl09},
\ctn{Jensen08}, \ctn{Ishwaran01a}, \ctn{Ishwaran01c}, etc.) that it is usually considered as the default choice in the literaure on
Dirichlet process.

For the cross-validation purpose we need to select a prior for $\bx=(x_1,x_2)$, where $\bx$ corresponds
to the left out observed climate variable $\bx_i=(x_{i1},x_{i2})$. Based on the observed sample, we 
set a bivariate normal prior for $\bx$ with means $\mu_{x_1}=\mu_{x_2}=0$ and variances $\sigma^2_{x_1}=\sigma^2_{x_2}=10$ for the co-ordinates
of $\bx$. Somewhat larger variances are chosen to account for extra uncertainty in $\bx$, which is now
treated as unobserved.
Based on the observed sample, the covariance is taken as 0.8.

The joint posterior distribution and the forms of the full conditional distributions of 
the parameters can be easily calculated as in Section \ref{subsec:joint}
and Section S-1.

\section{Cross-validation of the pollen data}
\label{sec:haslett_sim}

\subsection{Implementation issues}
Application of IRMCMC in the pollen data problem is carried out by first selecting $i^{*}$ = 5353 according the criterion
presented in Section 4.2 of \ctn{Bhatta07}.
We used additive TMCMC to update $\Lambda$, $x$, and $\{\sigma_{k,11},\sigma_{k,22},\rho_{12}\}$
in blocks. In fact, we apply TMCMC to the reparameterized versions of the elements of $\bSigma_k$, that is,
using additive TMCMC we update jointly $\{\log\sigma_{k,11}, \log\sigma_{k,22},
\tan\left(\frac{\pi\rho_{12}}{2}\right)\}$.
The reparameterized versions, being supported on the entire real line, ensures free movement of our additive
TMCMC sampler, resulting in good mixing properties.
It is important to mention that updating $\bbeta_{kj}$ using the Polya urn distribution
as the proposal distribution failed to yield satisfactory mixing. We overcame the problem by
adding a TMCMC step to update the distinct components of $\bbeta_{kj};j=1,\ldots,M_k$ in a single block, after 
Metropolis-Hastings with the Polya urn proposal has been applied sequentially to $\bbeta_{kj};j=1,\ldots,M_k$.
A further step of TMCMC consisting of only two move-types with equal probabilities, either adding a single
$\epsilon\sim N(0,0.5)I_{\{\epsilon>0\}}$ to all the variables or subtracting it from all of them with equal probabilities,
using the TMCMC-based acceptance ratio to decide on the final acceptance, very significantly improved 
the mixing properties of our algorithm. 
%

With the above proposal mechanisms we generated $30,000$ MCMC samples from the posterior corresponding to
$i^*=5353$. We discarded the initial $10,000$ samples as burn-in and stored the rest of the samples 
for importance re-sampling. We implemented IRMCMC fixing $K_1=200$ and $K_2=100$, thus obtaining 
$20,000$ IRMCMC samples for each of the 7815 cross-validation posteriors. The entire exercise took
around 9 hours.

\subsection{Results of cross-validation}
\begin{figure}
\centering
\subfigure[Site 1.]{ \label{fig:modern_sitex1_1}
\includegraphics[width=4.5cm,height=4.5cm]{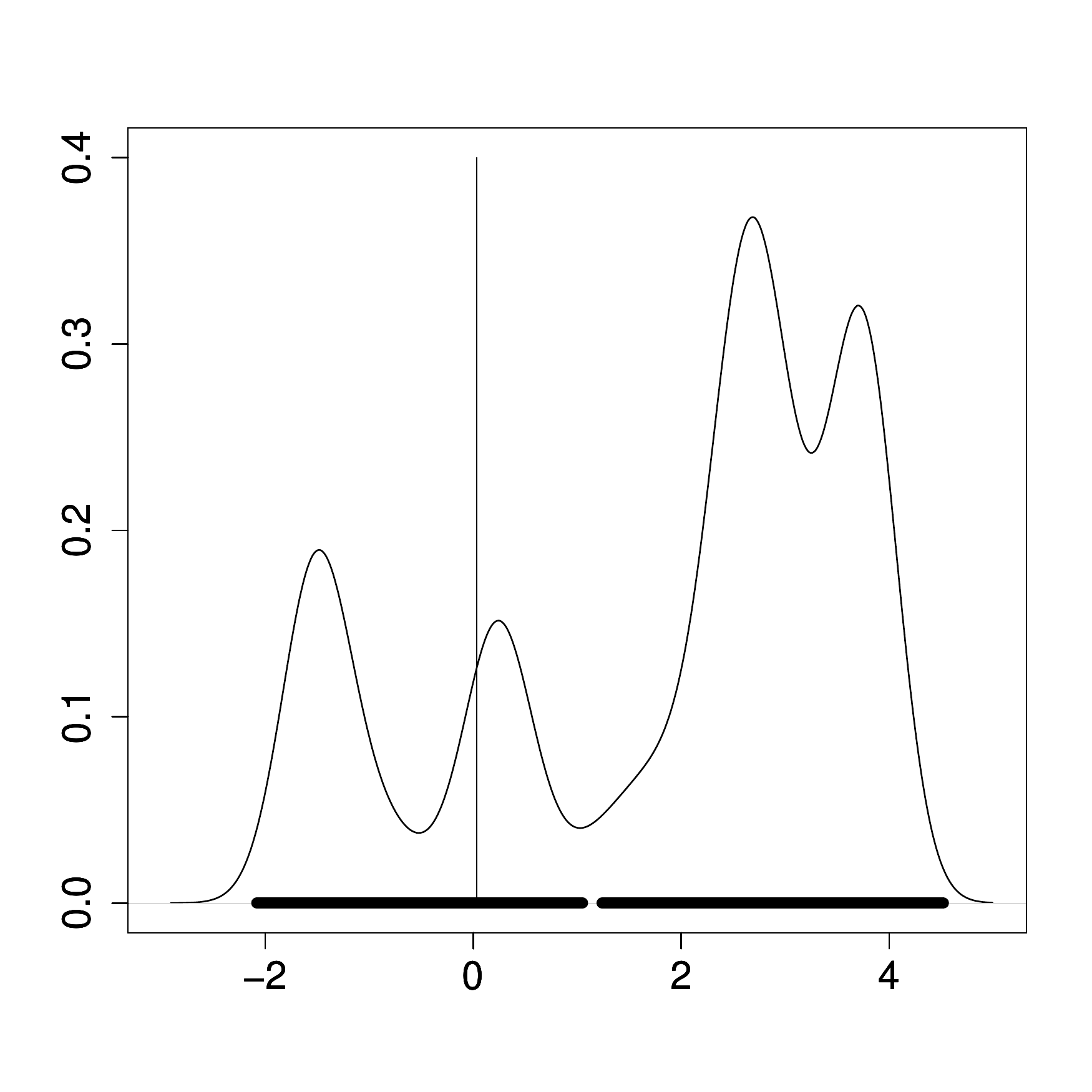}}
\hspace{2mm}
\subfigure[Site 51.]{ \label{fig:modern_sitex1_51} 
\includegraphics[width=4.5cm,height=4.5cm]{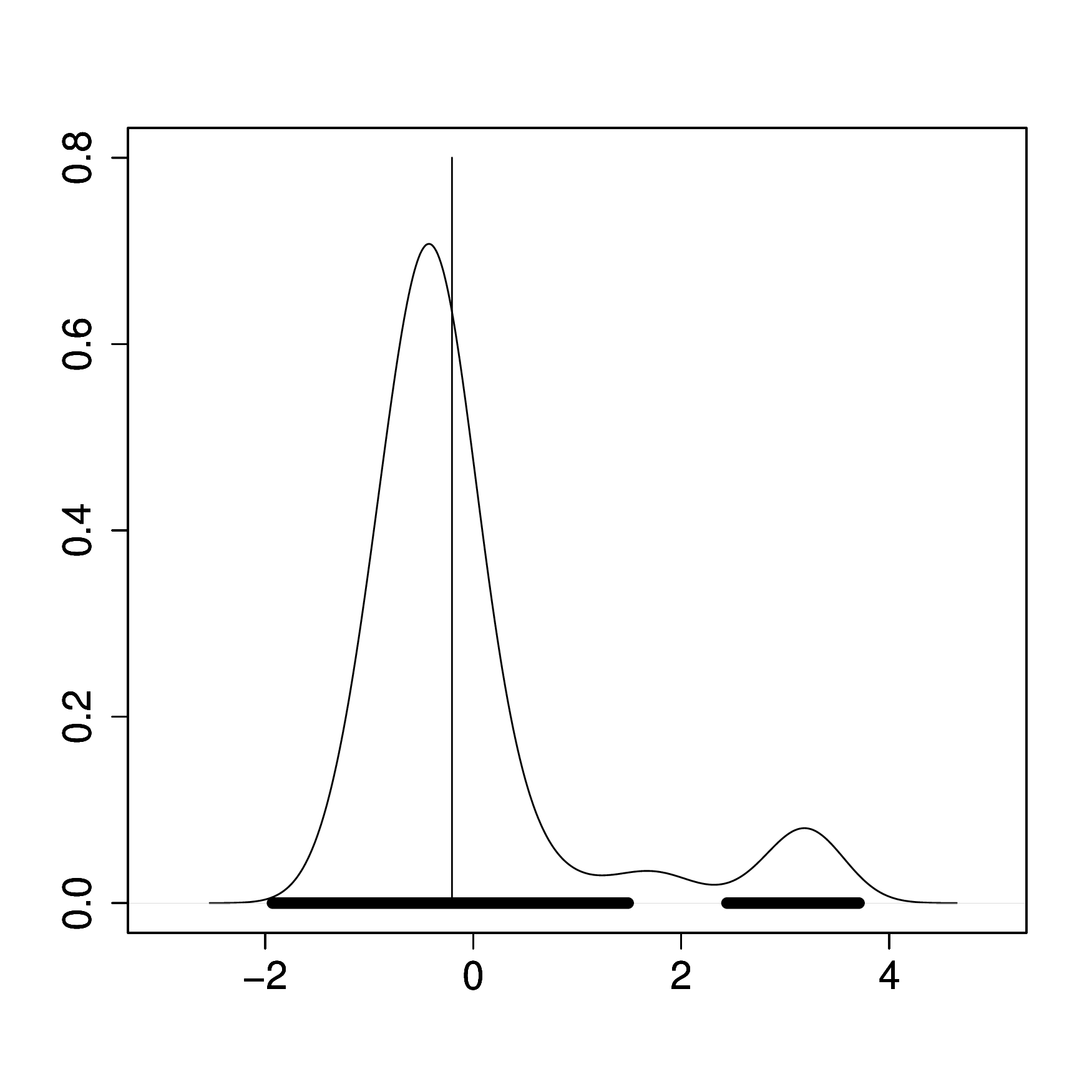}}
\hspace{2mm}
\subfigure[Site 100.]{ \label{fig:modern_sitex1_100} 
\includegraphics[width=4.5cm,height=4.5cm]{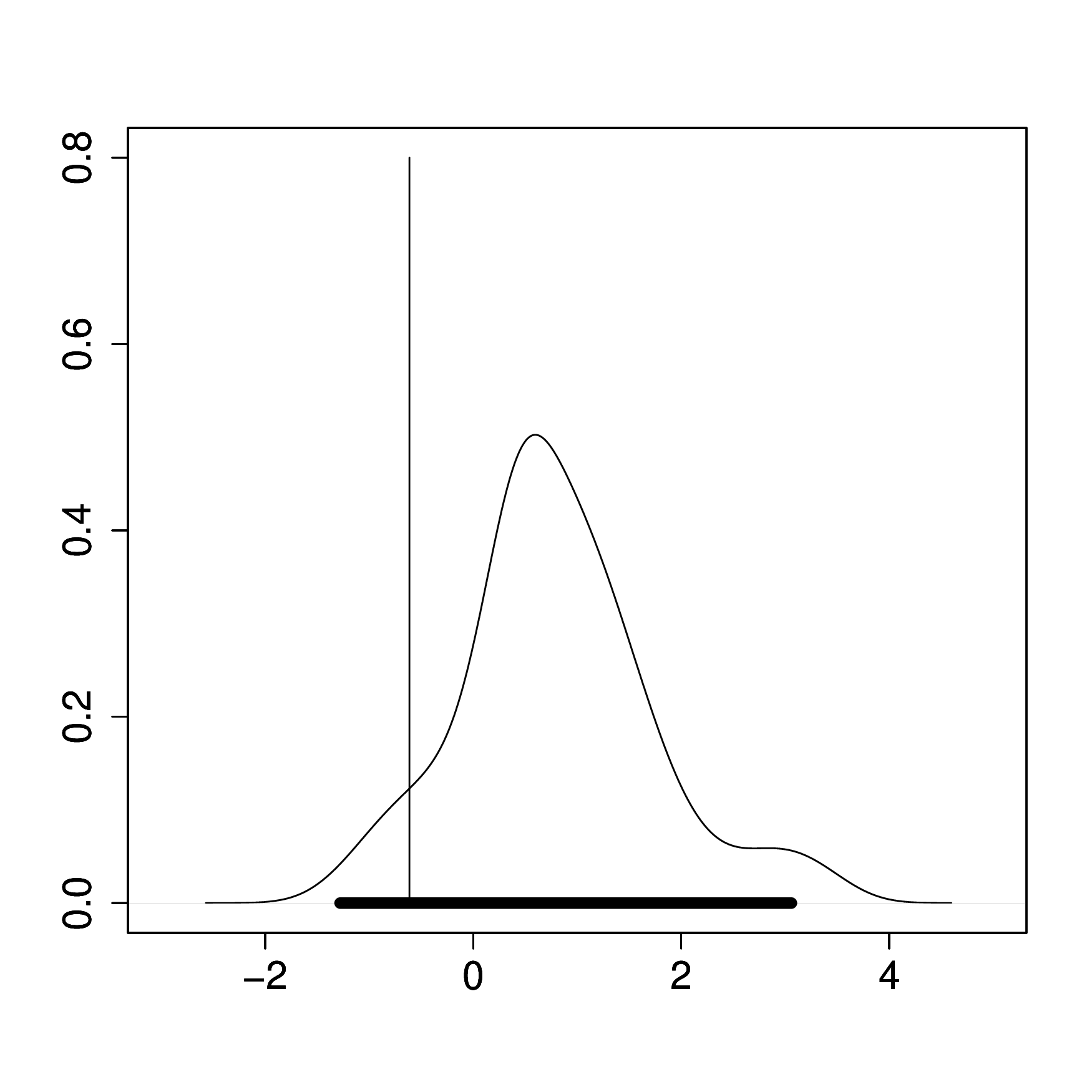}}\\
\vspace{2mm}
\subfigure[Site 200.]{ \label{fig:modern_sitex1_200} 
\includegraphics[width=4.5cm,height=4.5cm]{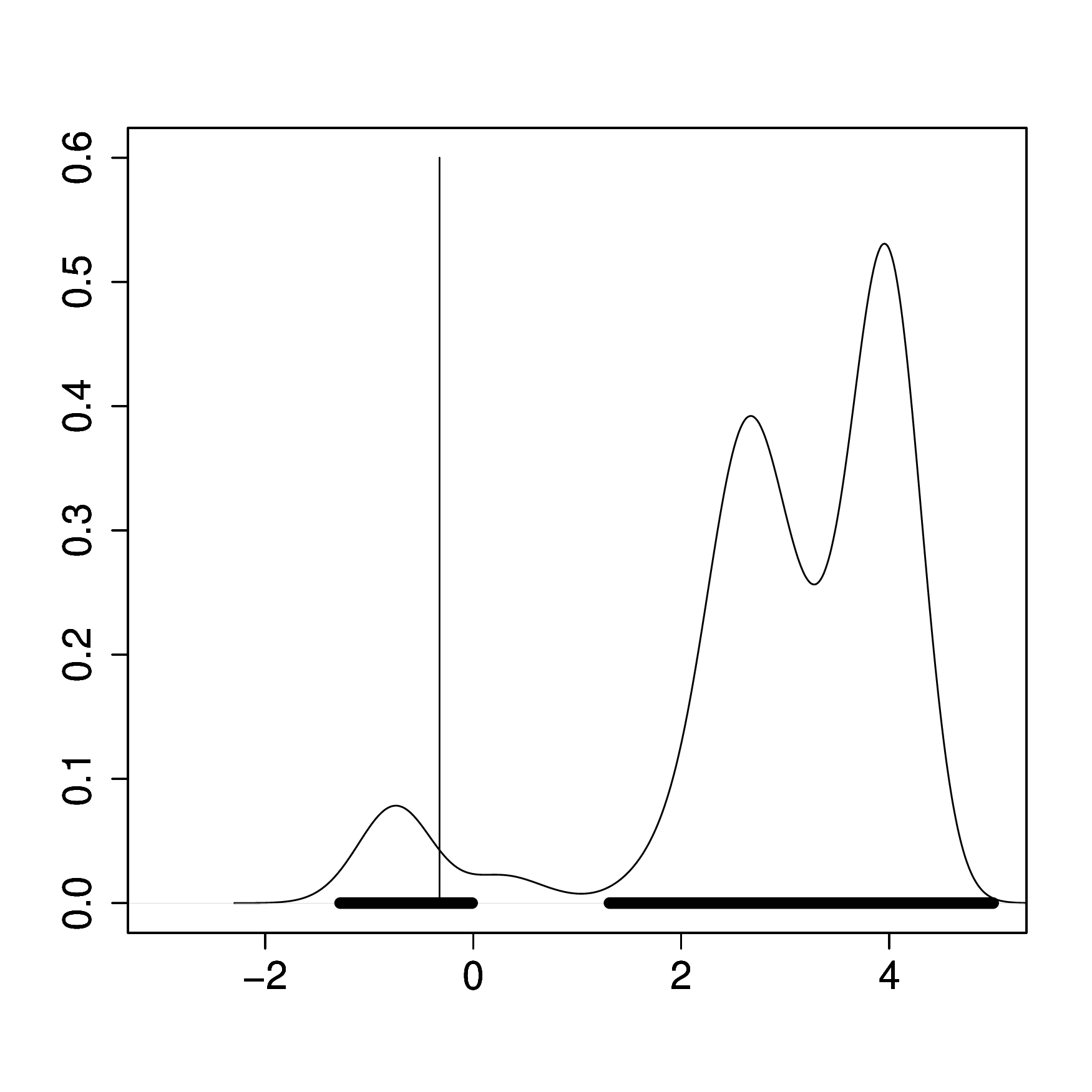}}
\hspace{2mm}
\subfigure[Site 977.]{ \label{fig:modern_sitex1_977} 
\includegraphics[width=4.5cm,height=4.5cm]{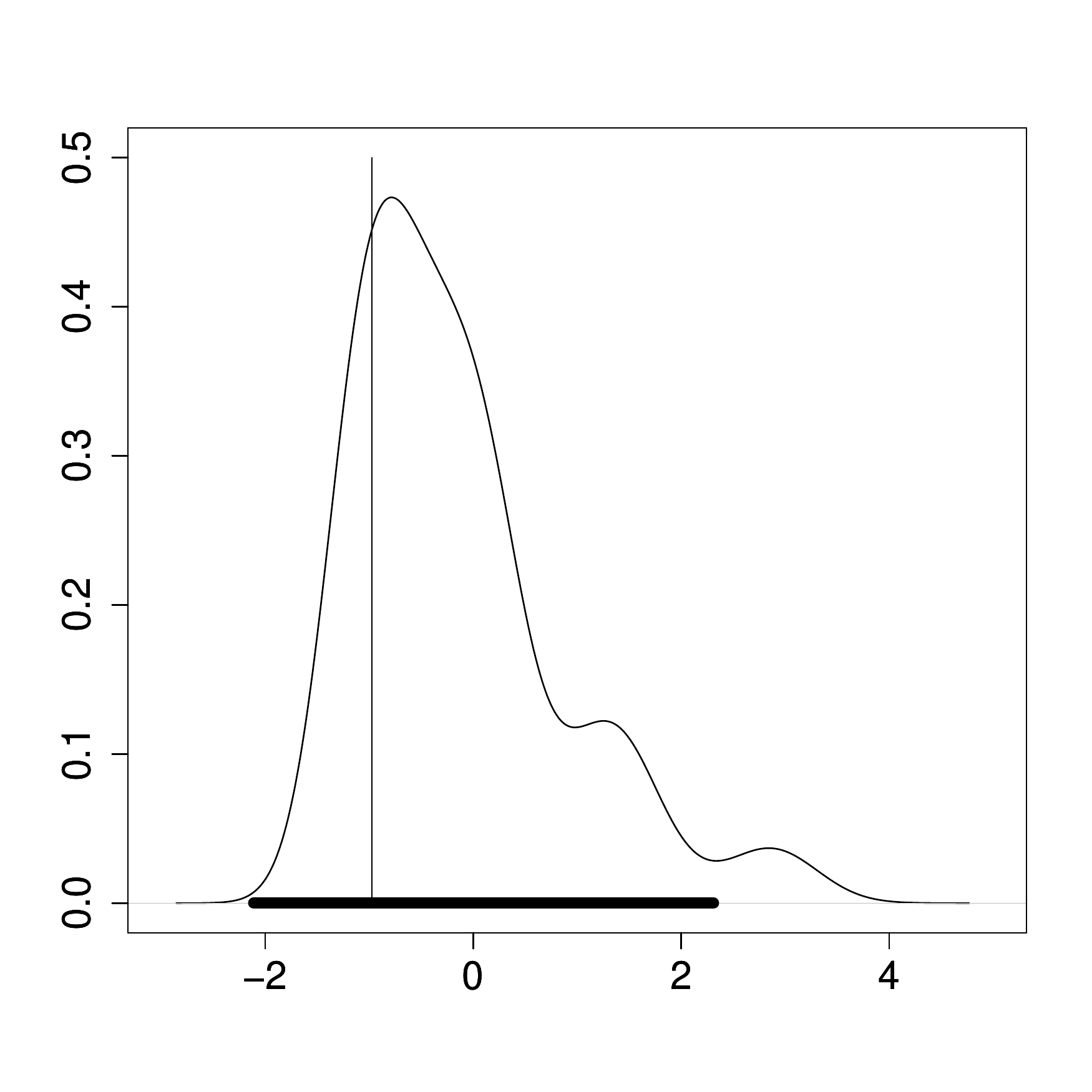}}
\hspace{2mm}
\subfigure[Site 1953.]{ \label{fig:modern_sitex1_1953}
\includegraphics[width=4.5cm,height=4.5cm]{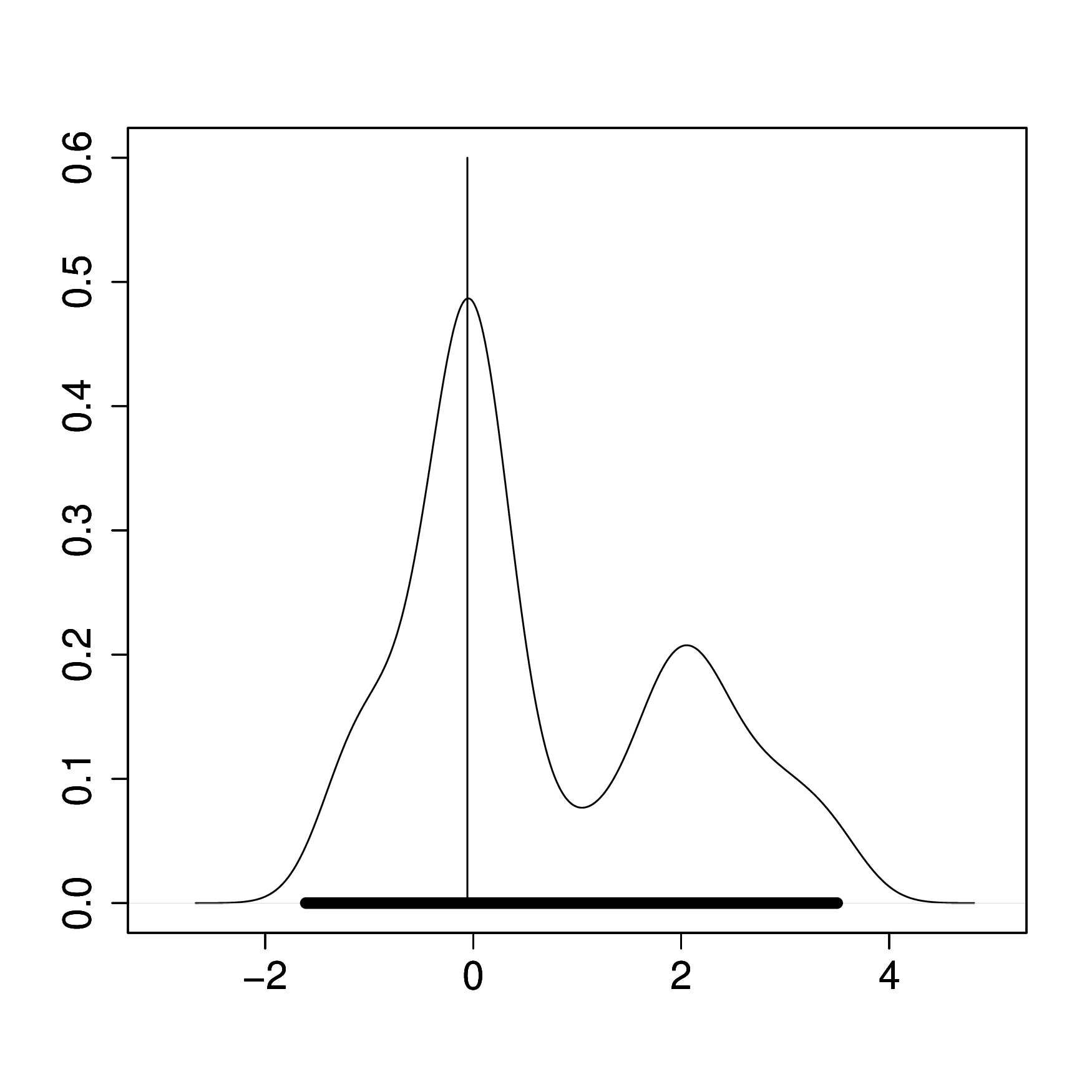}}\\
\vspace{2mm}
\subfigure[Site 5857.]{ \label{fig:modern_sitex1_5857} 
\includegraphics[width=4.5cm,height=4.5cm]{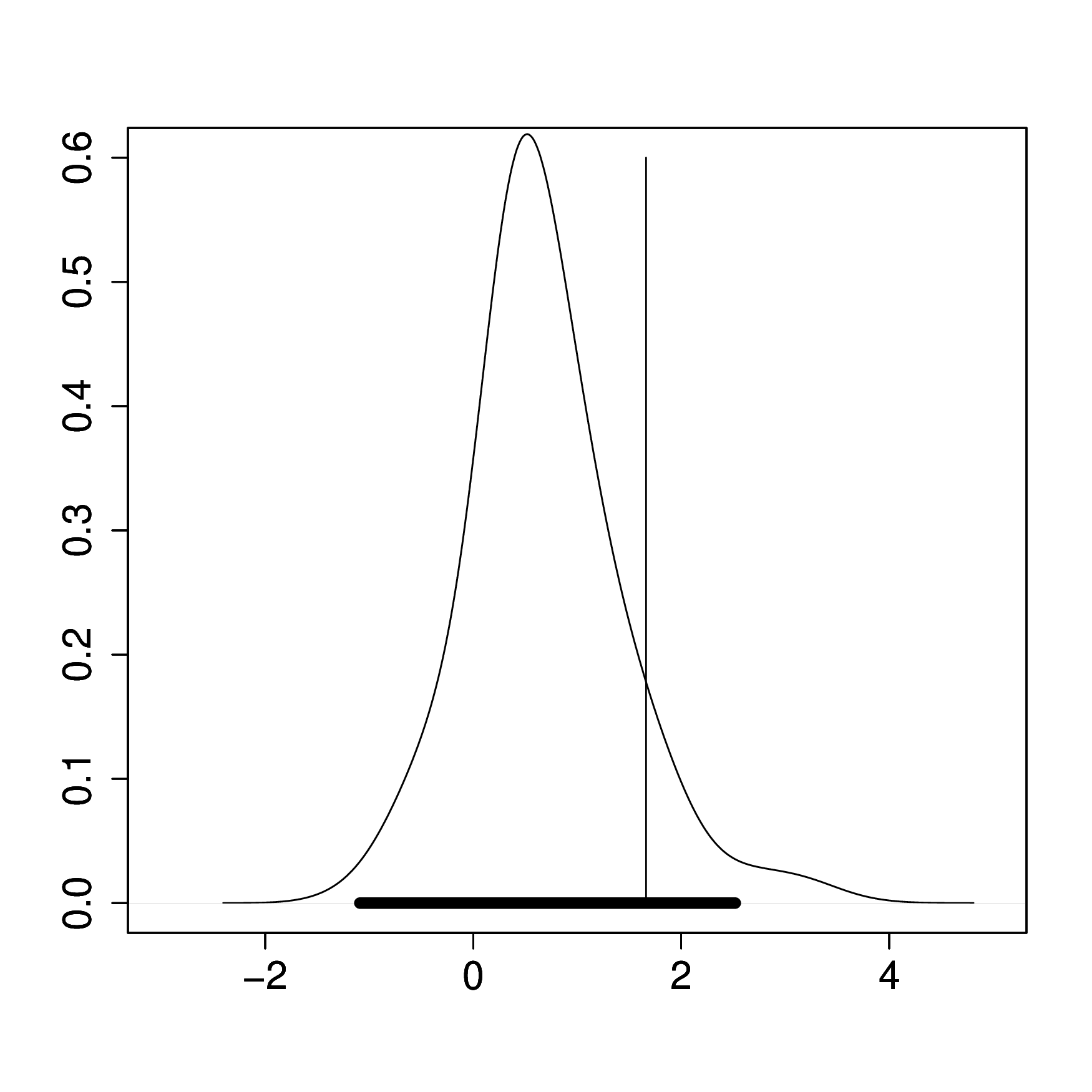}}
\hspace{2mm}
\subfigure[Site 6833.]{ \label{fig:modern_sitex1_6833} 
\includegraphics[width=4.5cm,height=4.5cm]{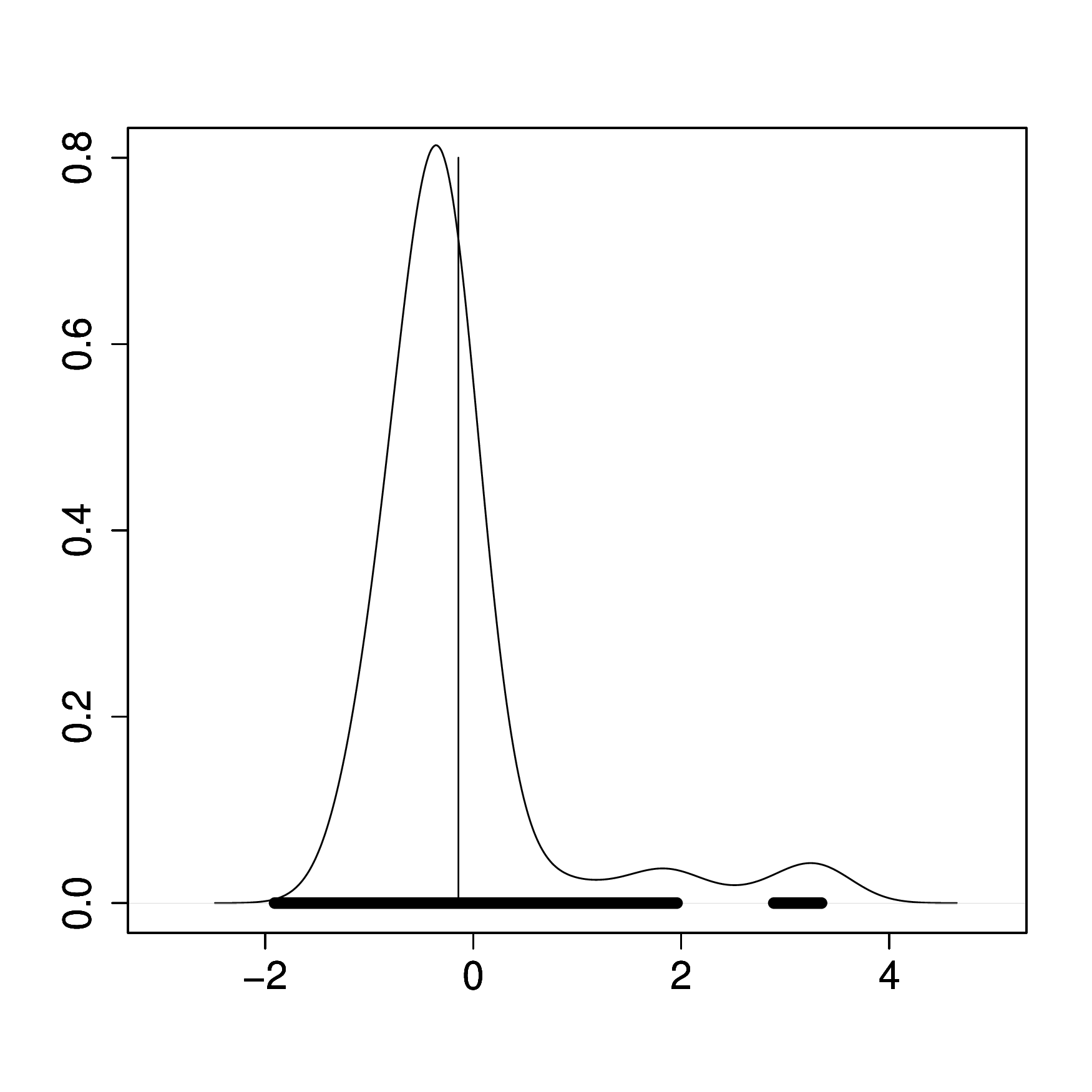}}
\vspace{2mm}
\subfigure[Site 7809.]{ \label{fig:modern_sitex1_7809} 
\includegraphics[width=4.5cm,height=4.5cm]{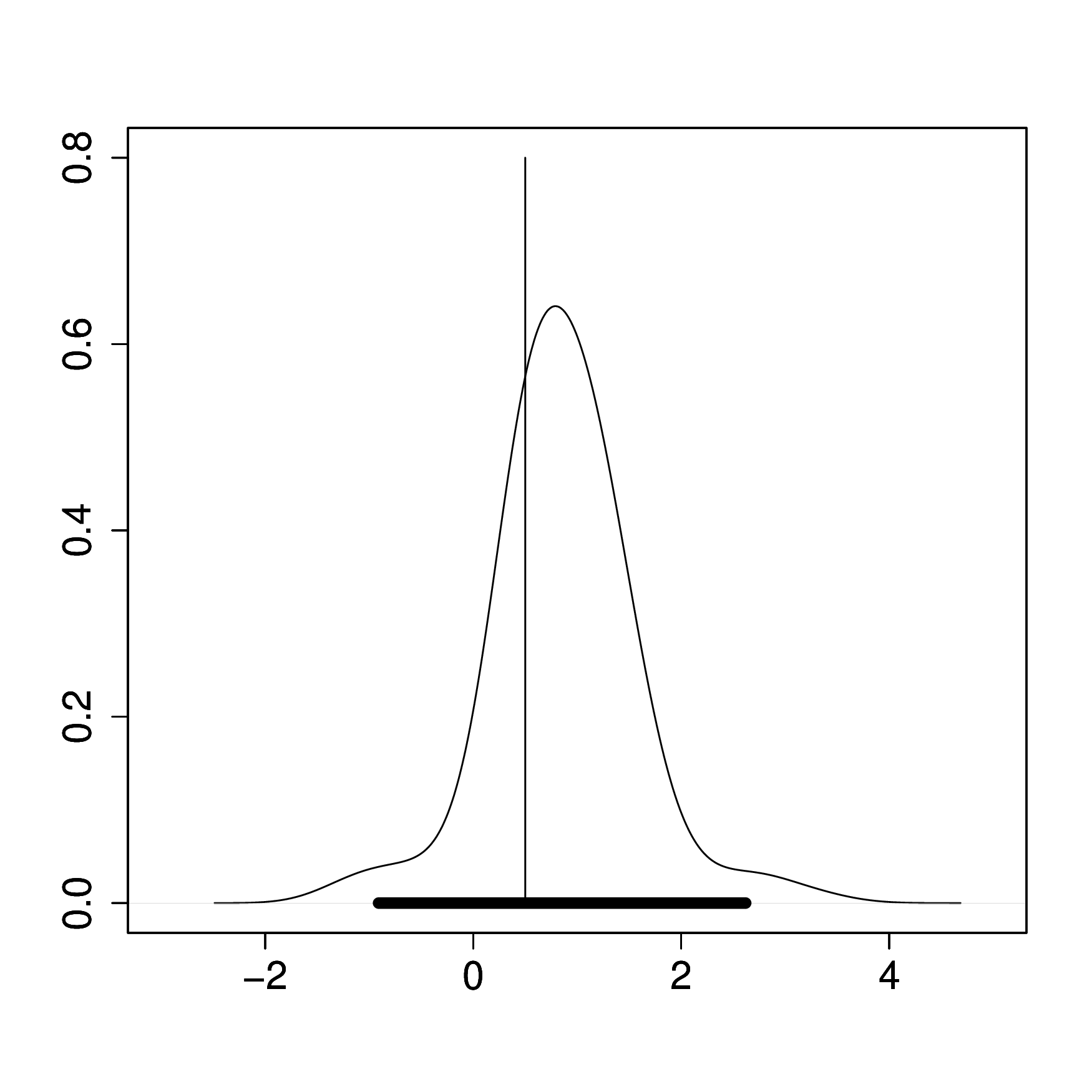}}\\
\hspace{2mm}
\caption{{\bf Pollen data:} Leave-one-out cross-validation posteriors of GDD5 for our model; 
the vertical line indicates the true (observed) value $\{x_{1i}\}$. The thick, horizontal line 
within the support of the cross-validation posterior
indicates the 95\% HPD.} 
\label{fig:modern1}
\end{figure}

\begin{figure}
\centering
\subfigure[Site 1.]{ \label{fig:modern_sitex2_1}
\includegraphics[width=4.5cm,height=4.5cm]{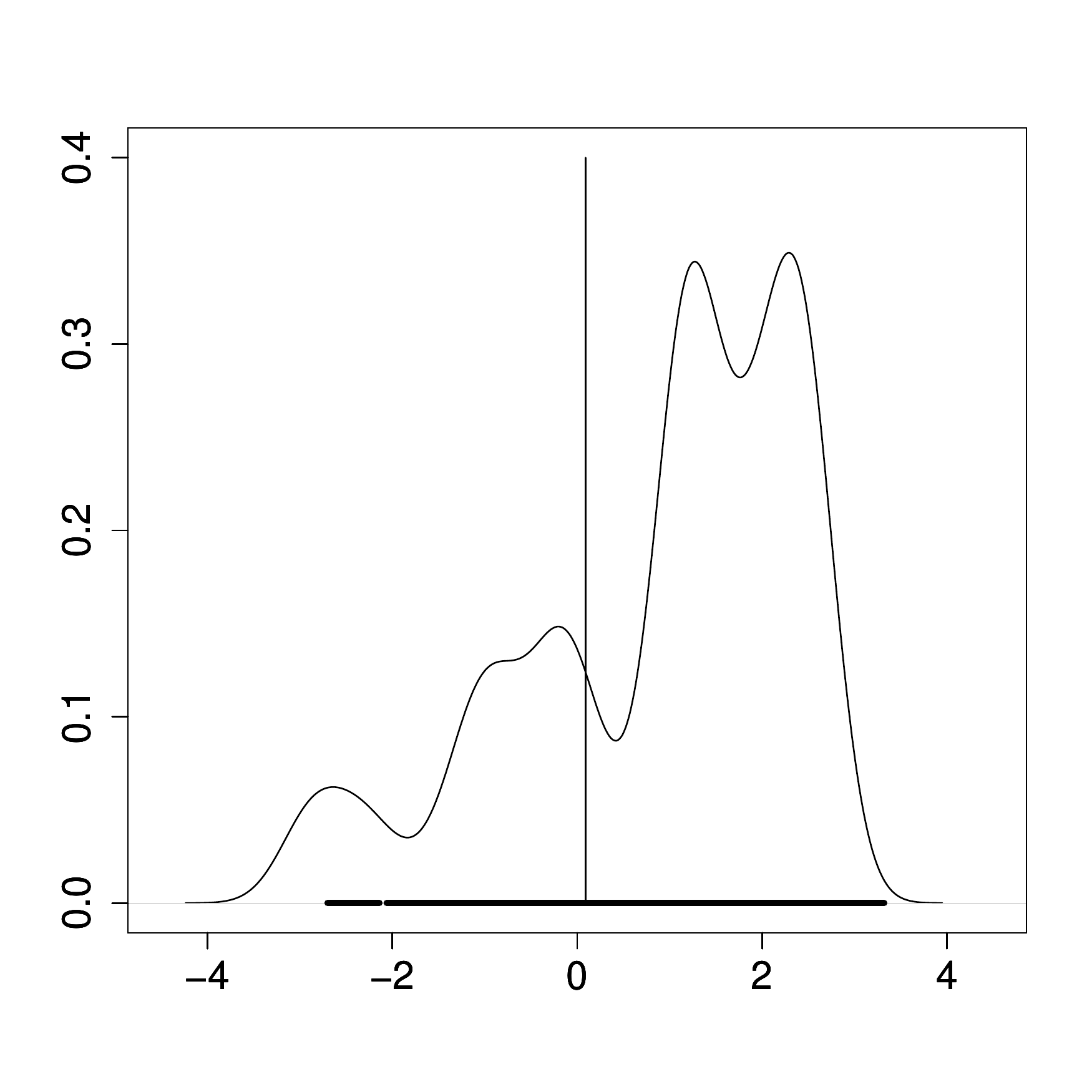}}
\hspace{2mm}
\subfigure[Site 51.]{ \label{fig:modern_sitex2_51} 
\includegraphics[width=4.5cm,height=4.5cm]{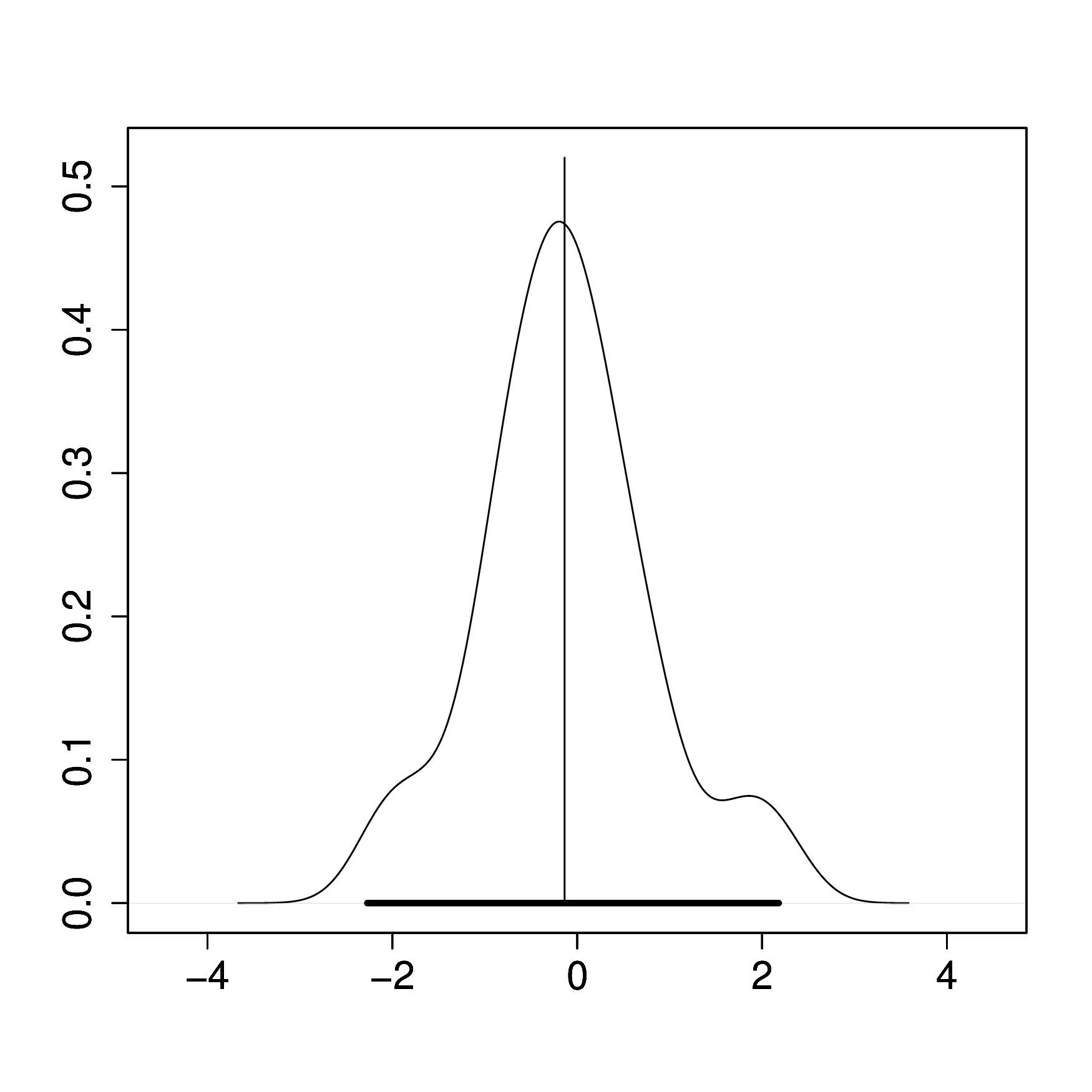}}
\hspace{2mm}
\subfigure[Site 100.]{ \label{fig:modern_sitex2_100} 
\includegraphics[width=4.5cm,height=4.5cm]{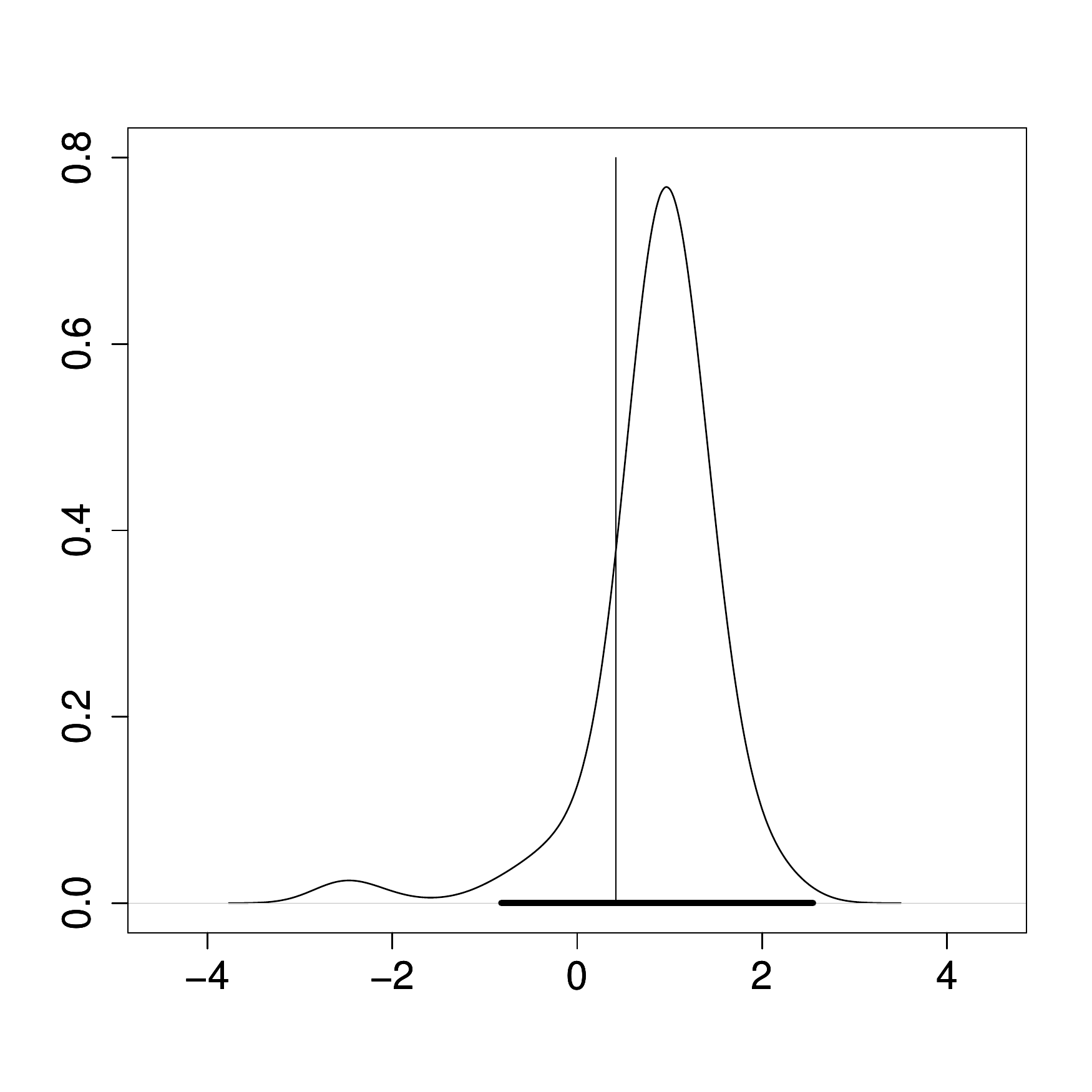}}\\
\vspace{2mm}
\subfigure[Site 200.]{ \label{fig:modern_sitex2_200} 
\includegraphics[width=4.5cm,height=4.5cm]{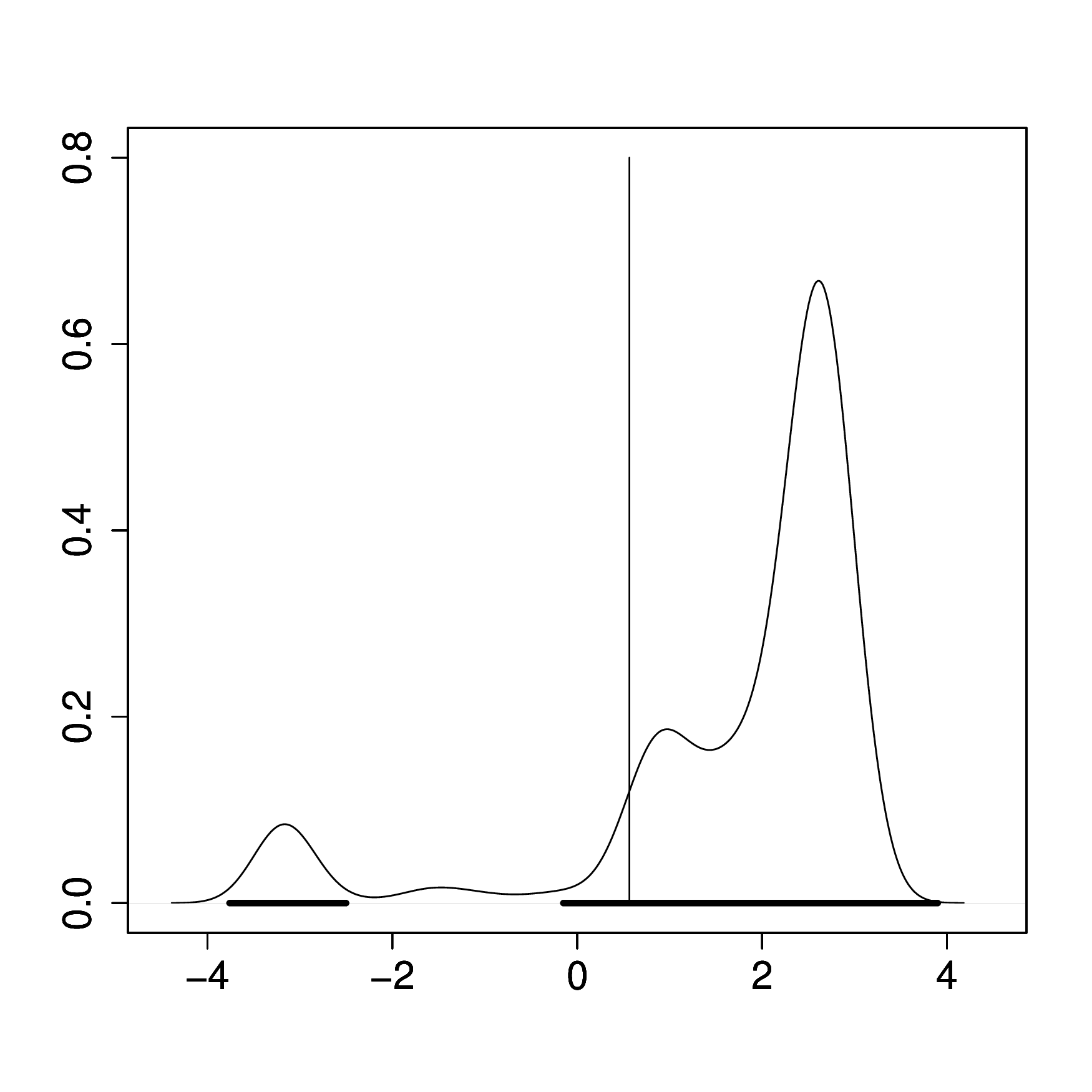}}
\hspace{2mm}
\subfigure[Site 977.]{ \label{fig:modern_sitex2_977} 
\includegraphics[width=4.5cm,height=4.5cm]{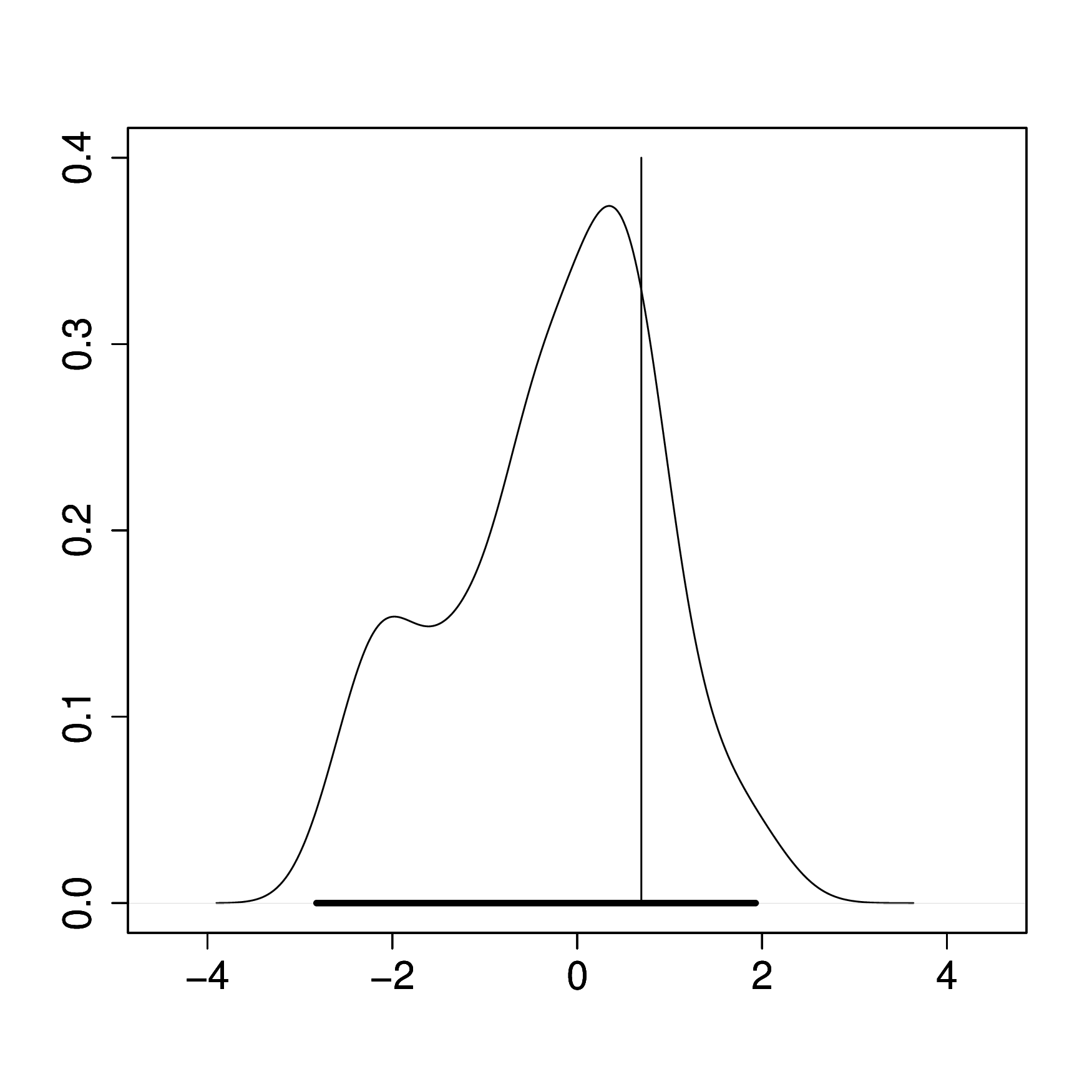}}
\hspace{2mm}
\subfigure[Site 1953.]{ \label{fig:modern_sitex2_1953}
\includegraphics[width=4.5cm,height=4.5cm]{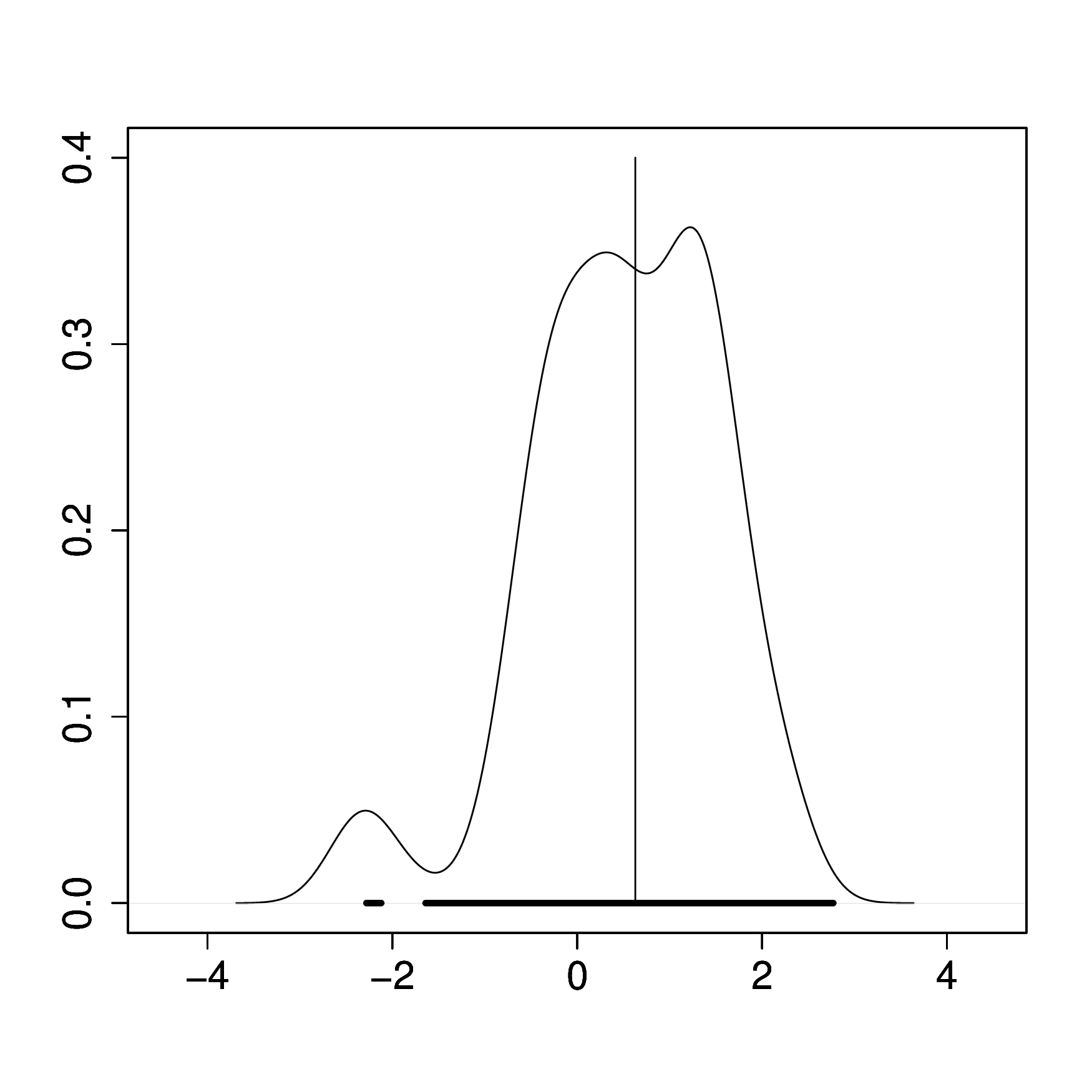}}\\
\vspace{2mm}
\subfigure[Site 5857.]{ \label{fig:modern_sitex2_5857} 
\includegraphics[width=4.5cm,height=4.5cm]{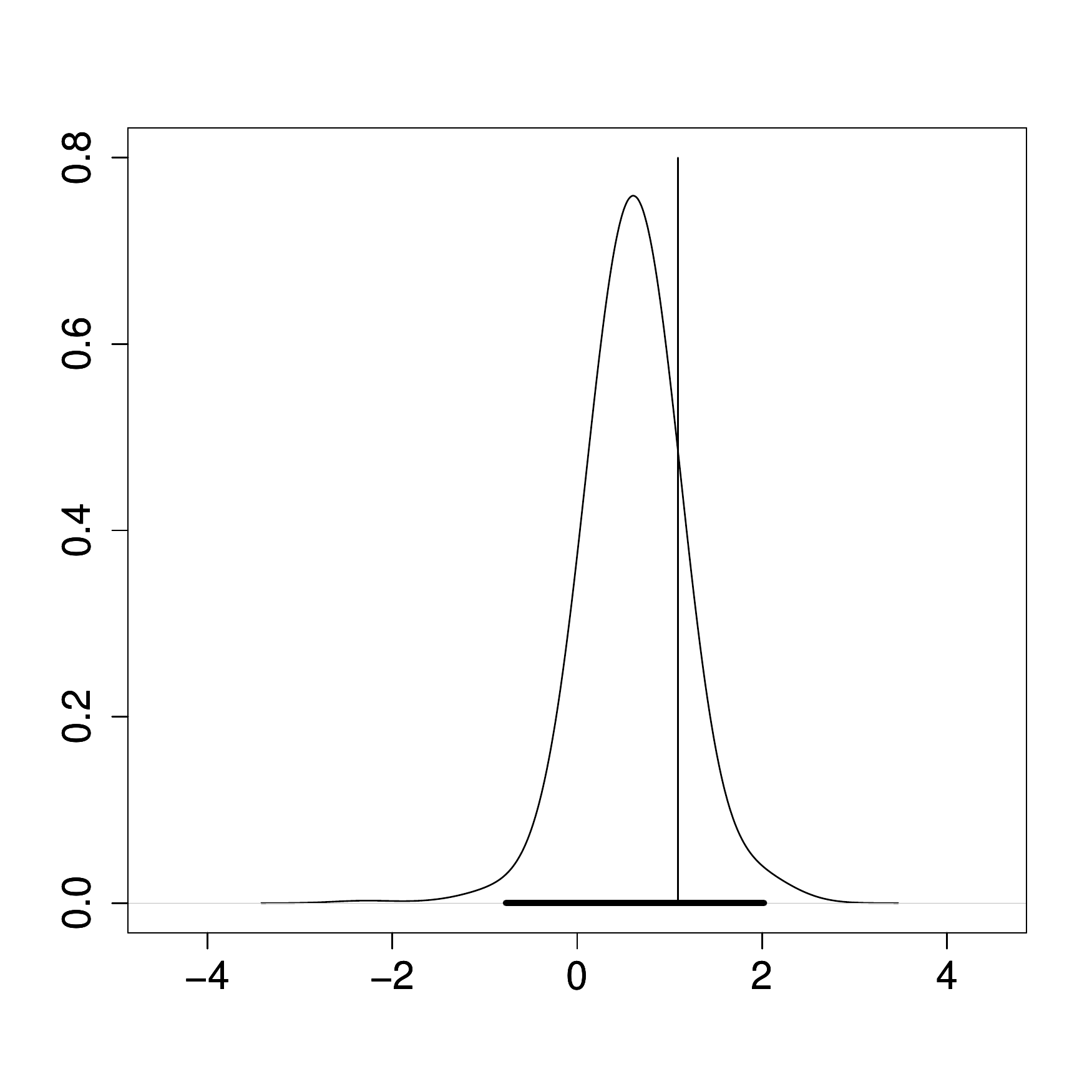}}
\hspace{2mm}
\subfigure[Site 6833.]{ \label{fig:modern_sitex2_6833} 
\includegraphics[width=4.5cm,height=4.5cm]{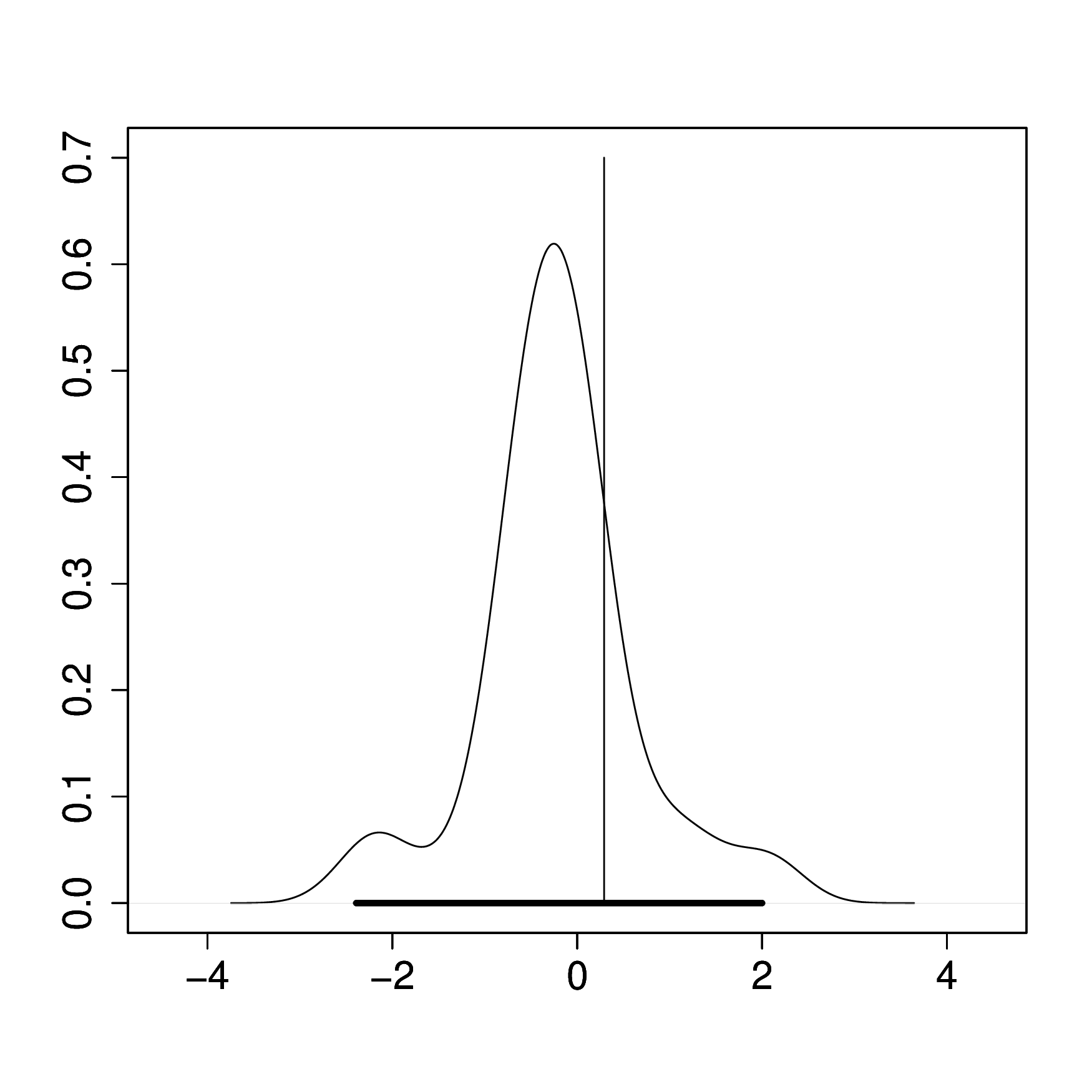}}
\vspace{2mm}
\subfigure[Site 7809.]{ \label{fig:modern_sitex2_7809} 
\includegraphics[width=4.5cm,height=4.5cm]{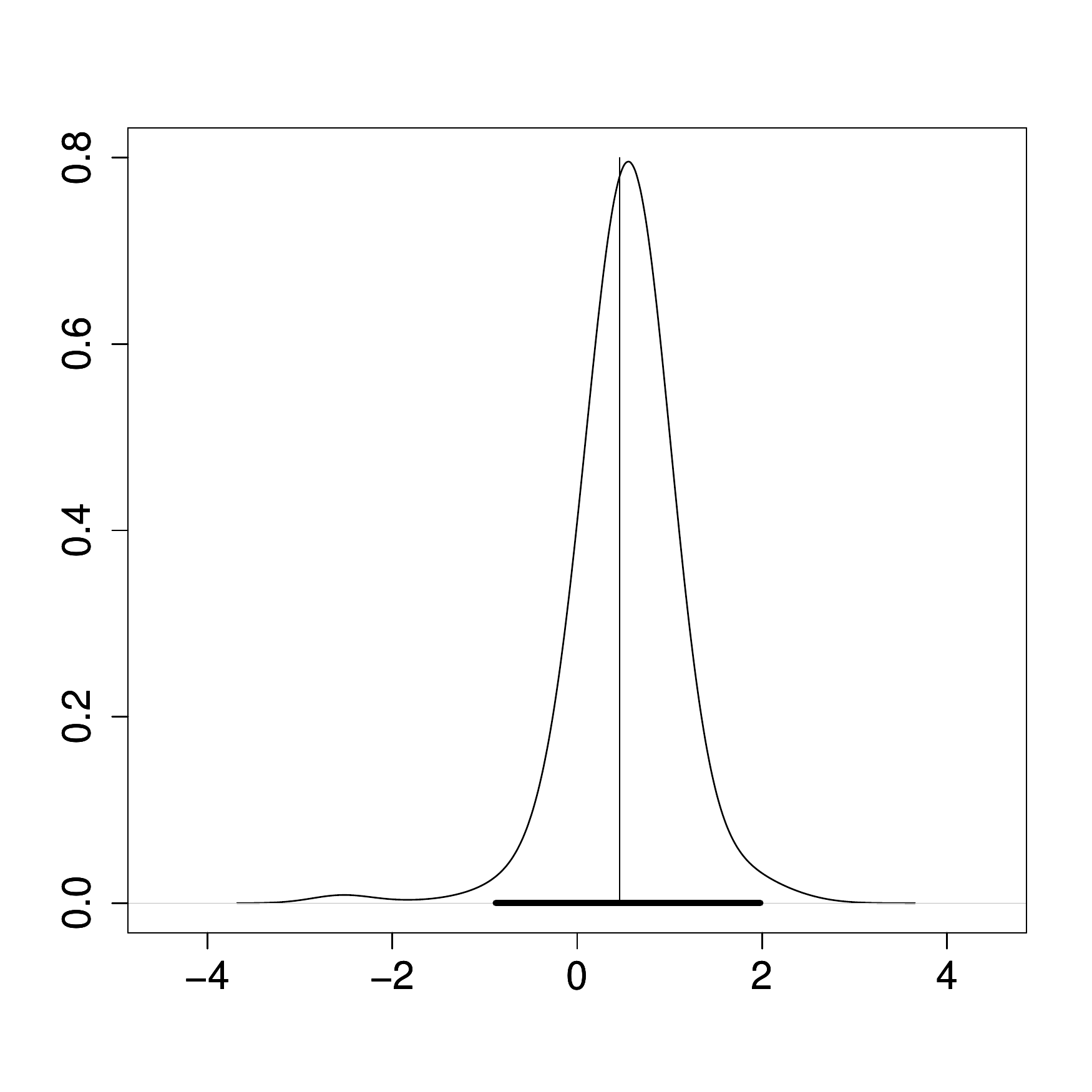}}\\
\hspace{2mm}
\caption{{\bf Pollen data:} Leave-one-out cross-validation posteriors of MTCO for our model; 
the vertical line indicates the true (observed) value $\{x_{2i}\}$. The thick, horizontal line 
within the support of the cross-validation posterior
indicates the 95\% HPD.} 
\label{fig:modern2}
\end{figure}

In about 94.60\% cases $x_1$, the co-ordinate associated with GDD5, fell within the 95\% HPD regions
of the corresponding cross-validation posteriors, and in about 94.19\% cases $x_2$, associated
with MTCO, fell within the respective 95\% HPD regions. Figures \ref{fig:modern1} and \ref{fig:modern2}
show some cross-validation posteriors associated with GDD5 and MTCO respectively, with the vertical lines
and the thick horizontal lines denoting the true (observed) climate values and the 95\% HPD intervals.
The cross-validation posteriors are highly multimodal; the degrees of multimodality seem to be 
higher in comparison to those of the chironomid example. Indeed, in this pollen case, several species are combined
to form a single category; see Appendix A of HWB for a discussion justifying amalgamation of species.
Also, some species, such as {\it Juniperus}, consist of several sub-species having contrasting climate
preferences. These issues substantially contribute to multimodality of the cross-validation posteriors.
A detailed discussion on multimodality can also be found in \ctn{Bhatta04}.

\begin{figure}
\centering
\subfigure[Site 1, Species 1.]{ \label{fig:pollen_prop1}
\includegraphics[width=4.5cm,height=4.5cm]{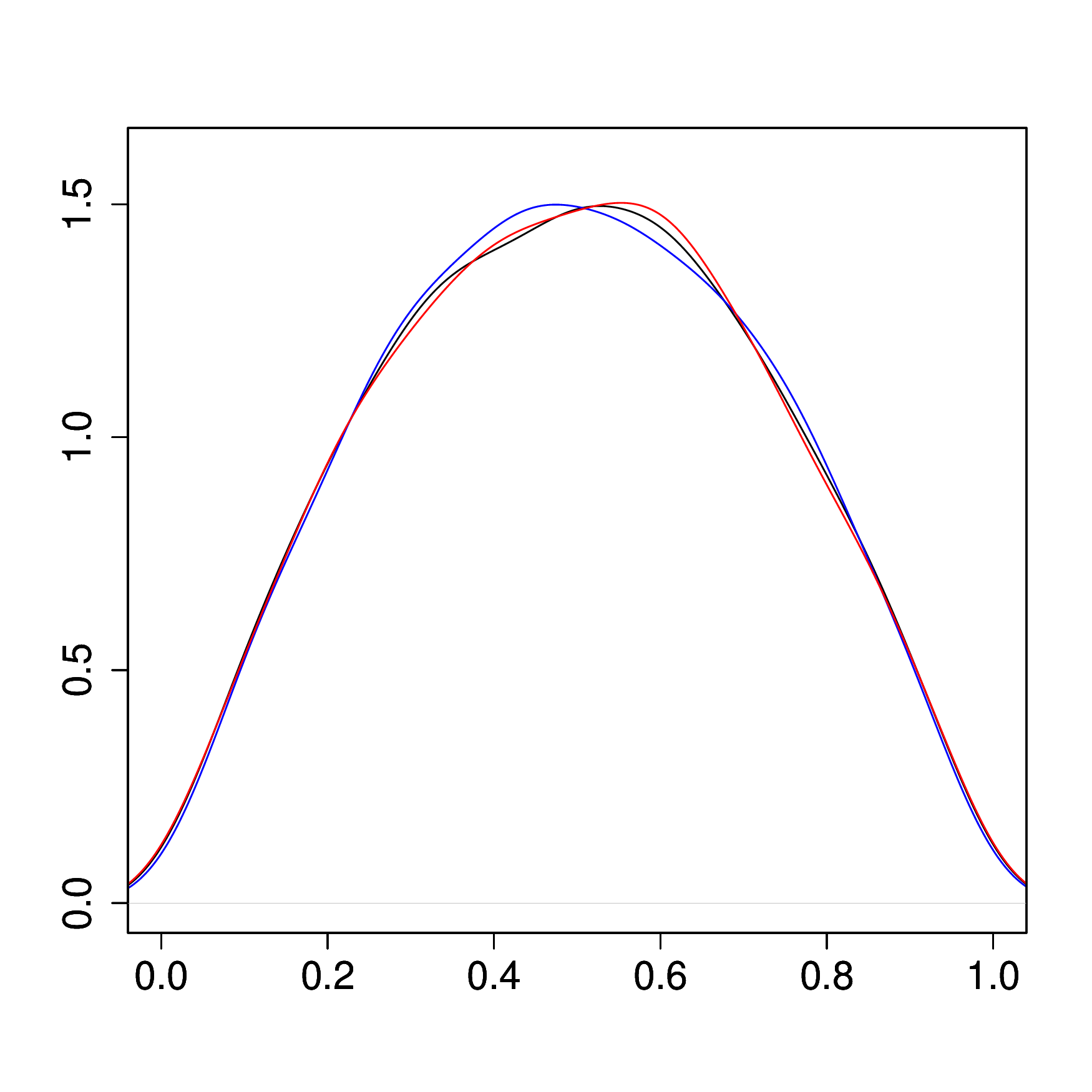}}
\hspace{2mm}
\subfigure[Site 1, Species 10.]{ \label{fig:pollen_prop2} 
\includegraphics[width=4.5cm,height=4.5cm]{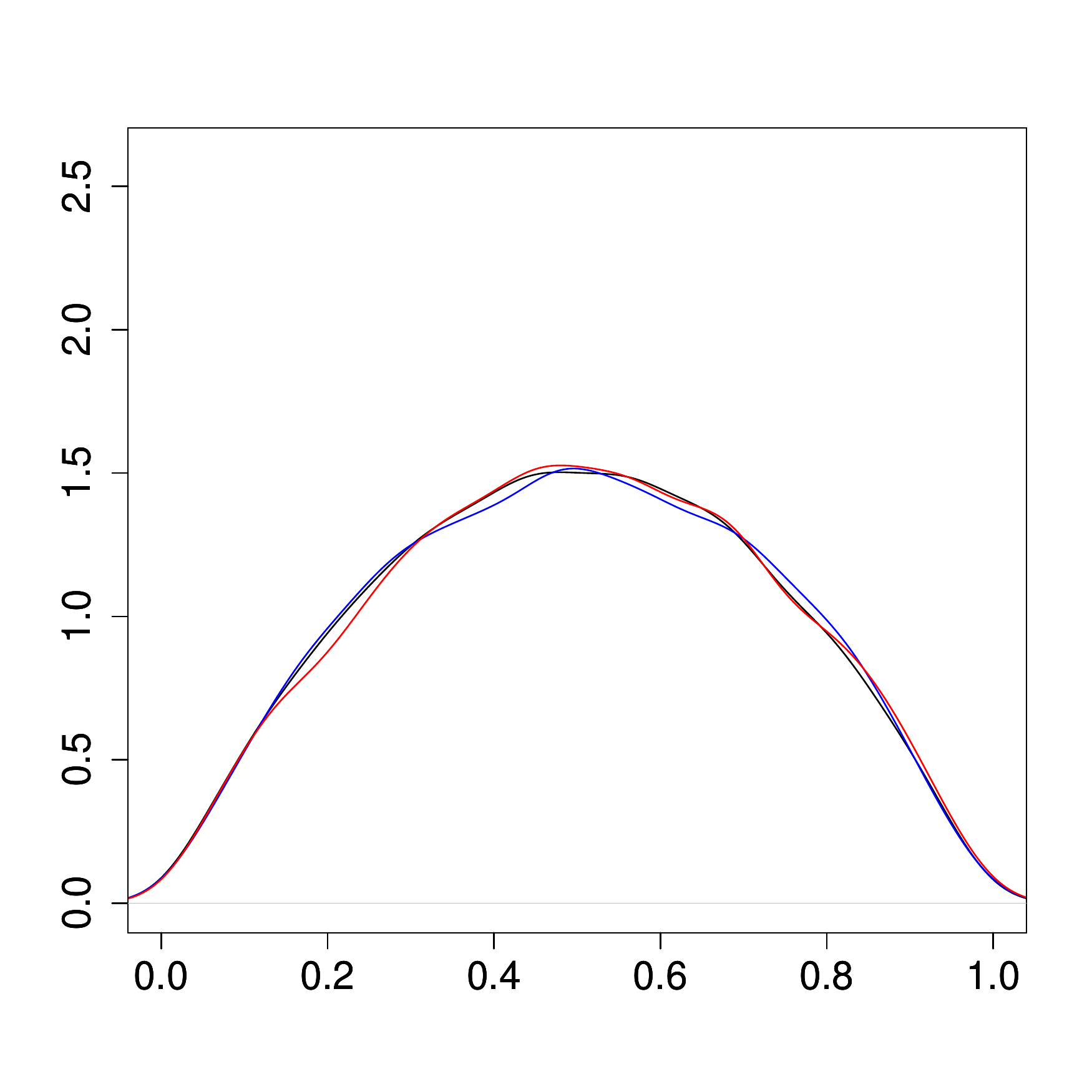}}
\hspace{2mm}
\subfigure[Site 1, Species 5.]{ \label{fig:pollen_prop3}
\includegraphics[width=4.5cm,height=4.5cm]{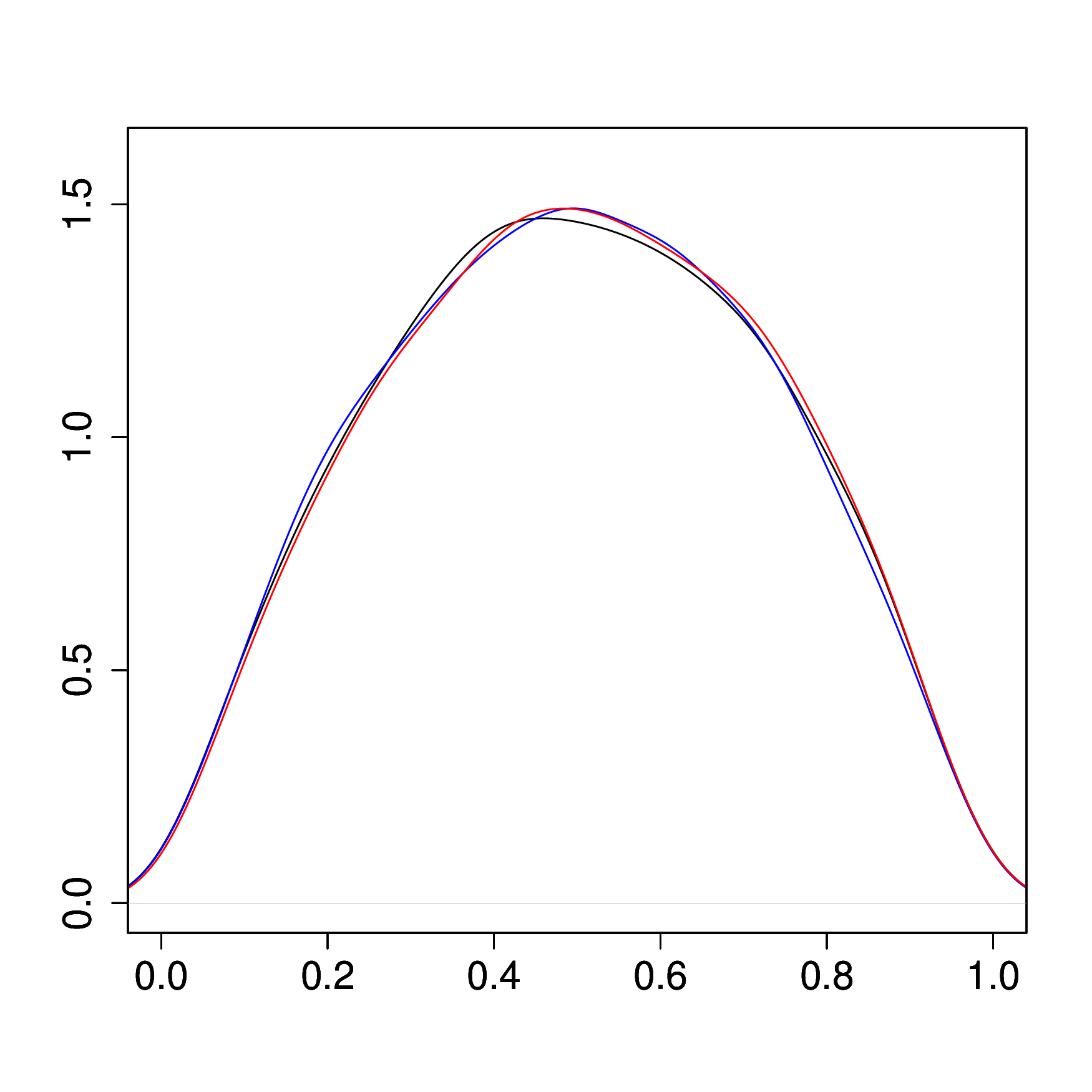}}\\
\vspace{2mm}
\subfigure[Site 100, Species 1.]{ \label{fig:pollen_prop4} 
\includegraphics[width=4.5cm,height=4.5cm]{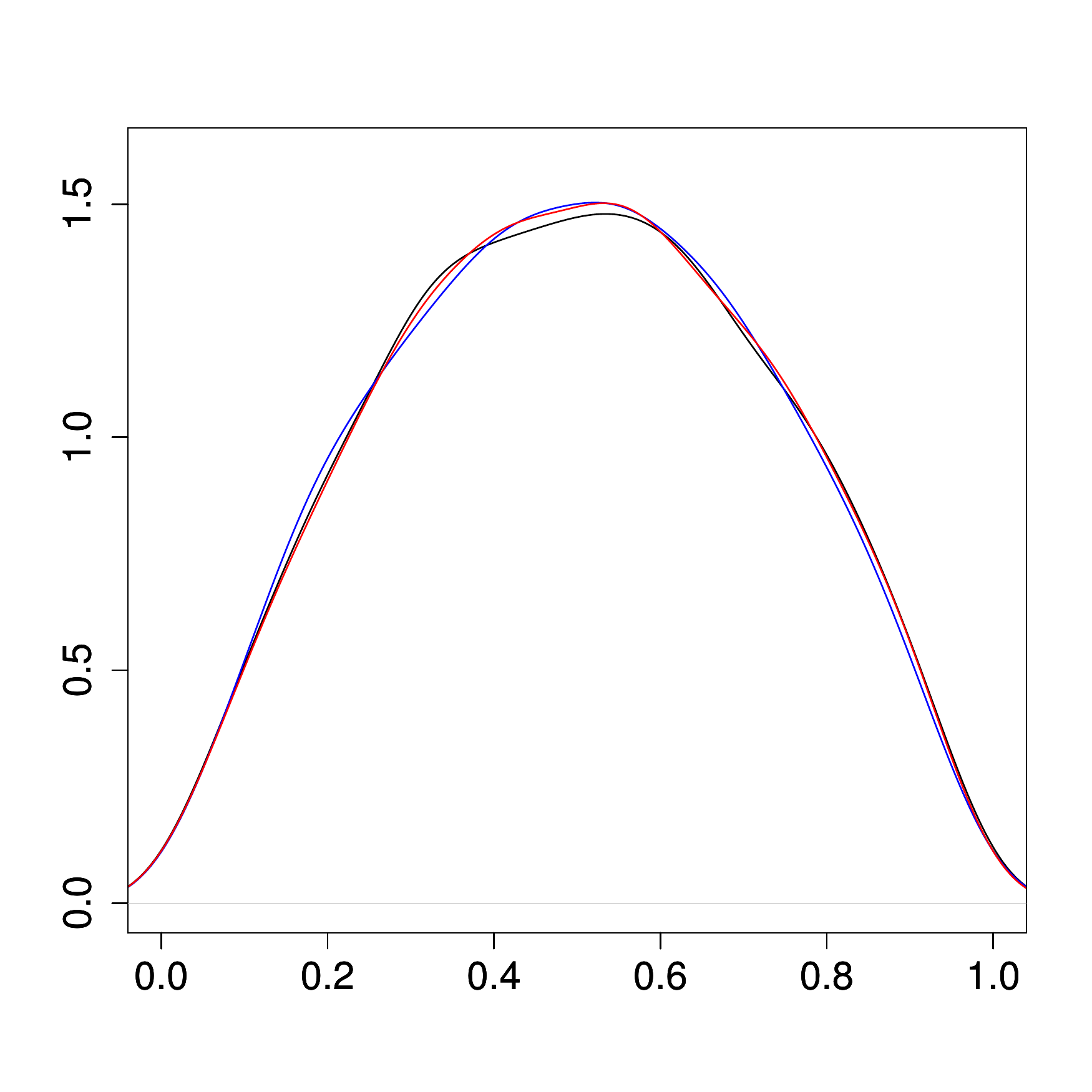}}
\hspace{2mm}
\subfigure[Site 100, Species 5.]{ \label{fig:pollen_prop5} 
\includegraphics[width=4.5cm,height=4.5cm]{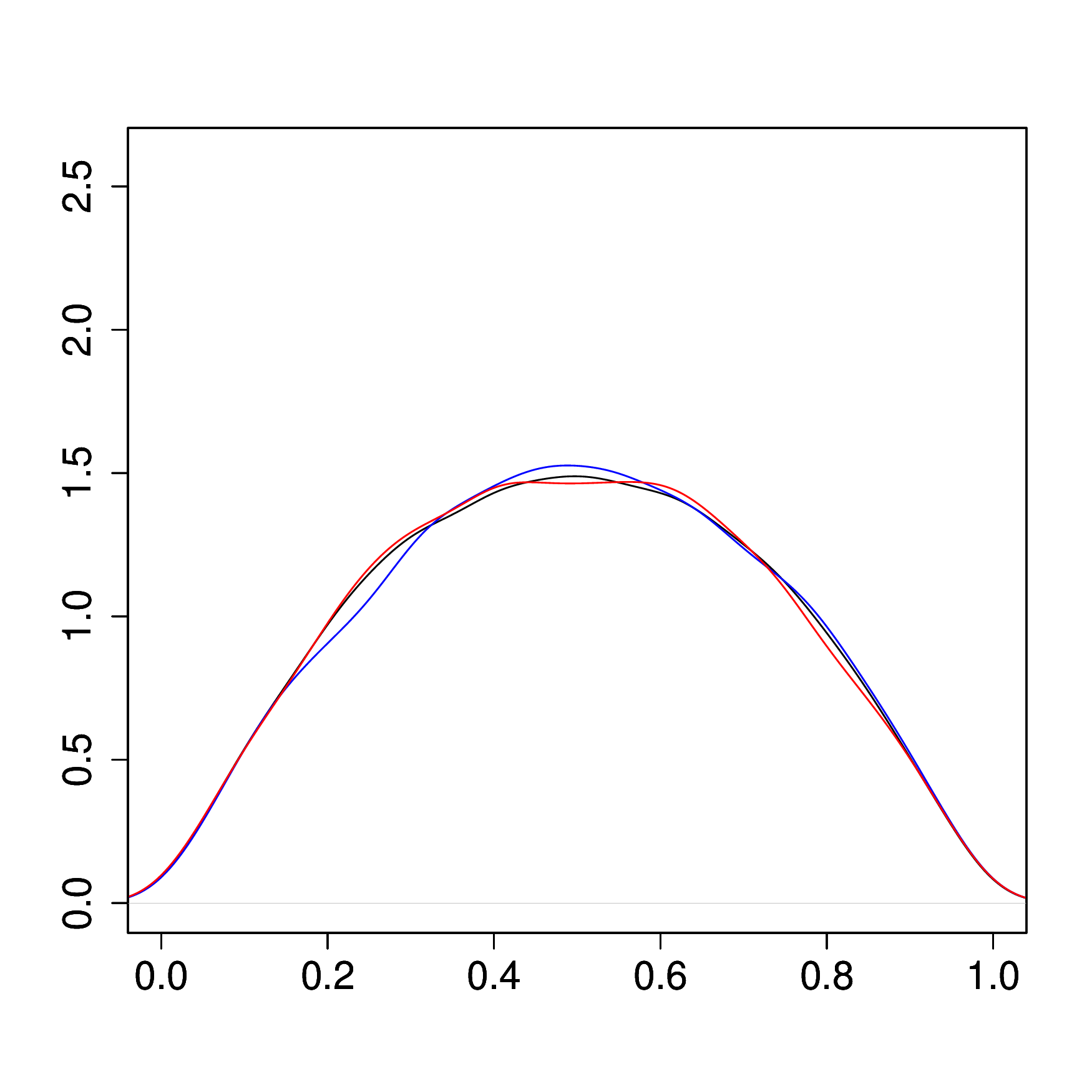}}
\hspace{2mm}
\subfigure[Site 500, Species 1.]{ \label{fig:pollen_prop6} 
\includegraphics[width=4.5cm,height=4.5cm]{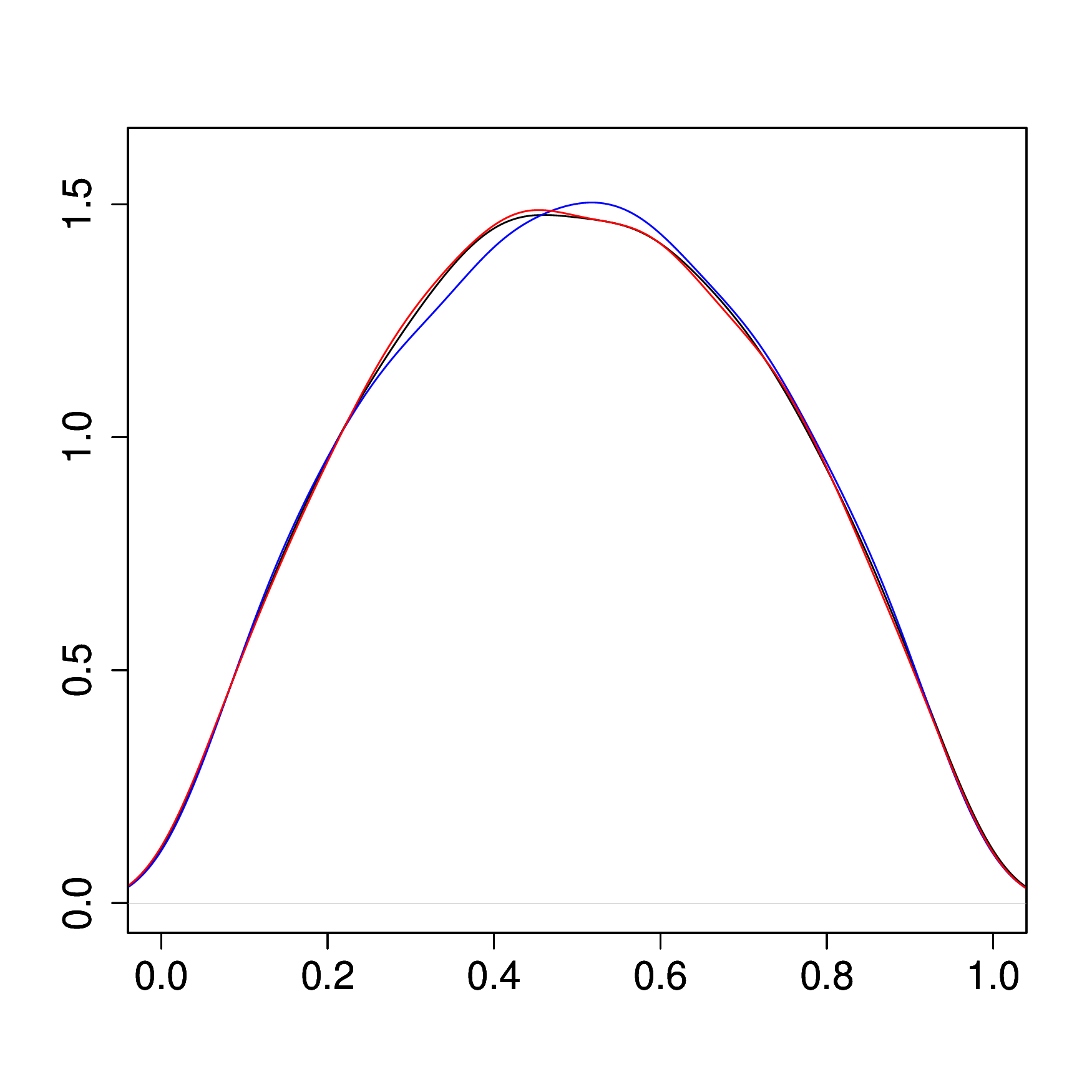}}\\
\vspace{2mm}
\subfigure[Site 500, Species 10.]{ \label{fig:pollen_prop7} 
\includegraphics[width=4.5cm,height=4.5cm]{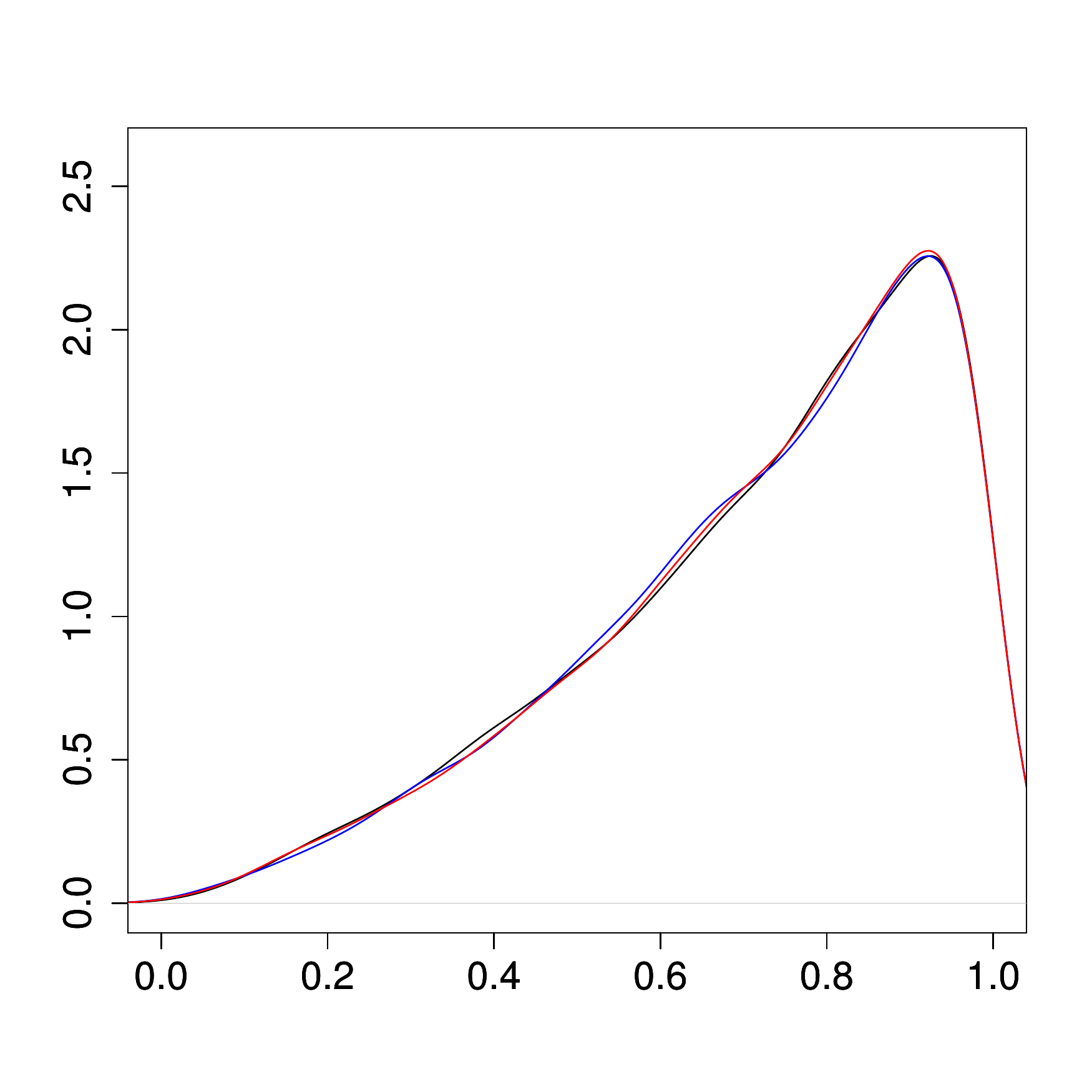}}
\hspace{2mm}
\subfigure[Site 7805, Species 10.]{ \label{fig:pollen_prop8} 
\includegraphics[width=4.5cm,height=4.5cm]{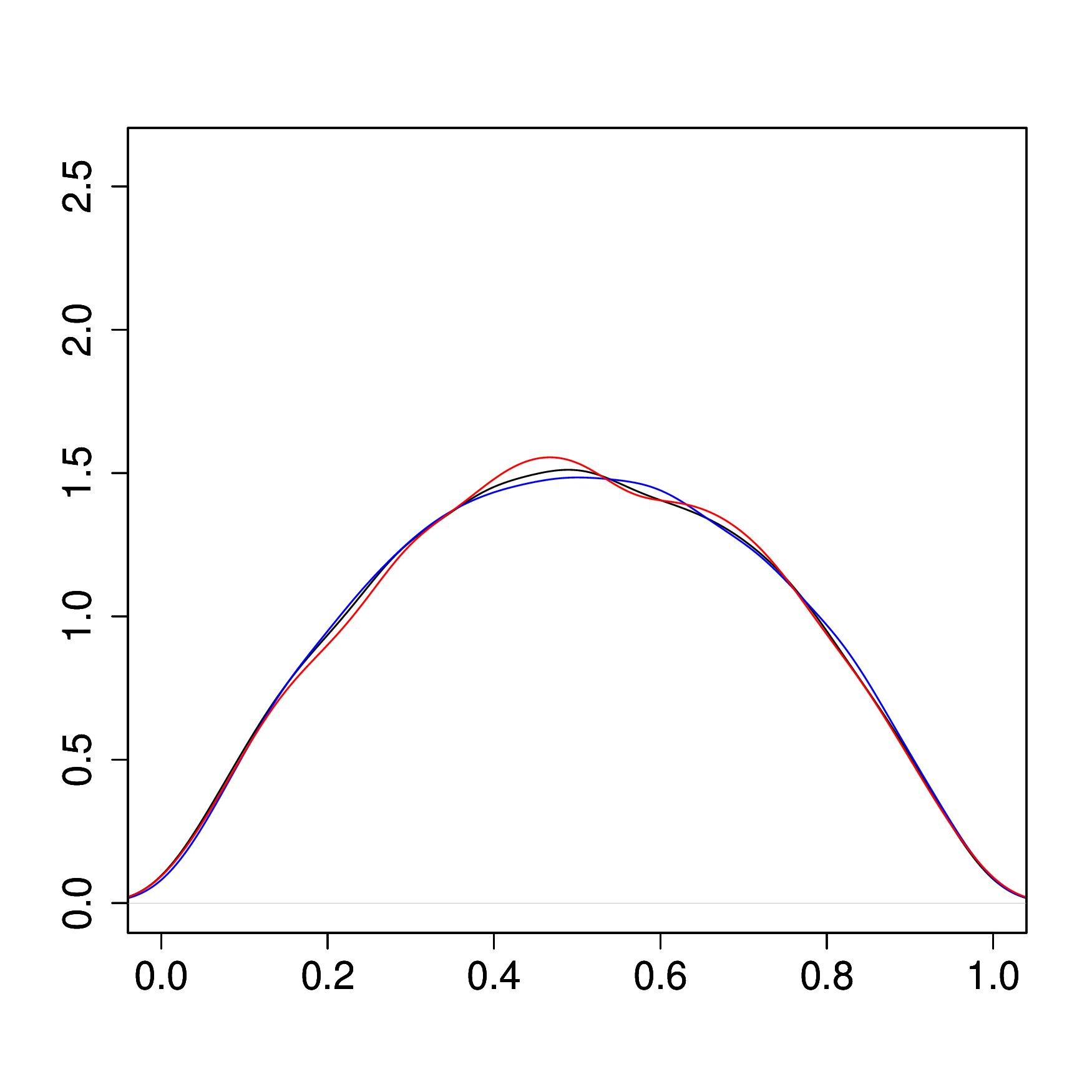}}
\hspace{2mm}
\subfigure[Site 7805, Species 5.]{ \label{fig:pollen_prop9} 
\includegraphics[width=4.5cm,height=4.5cm]{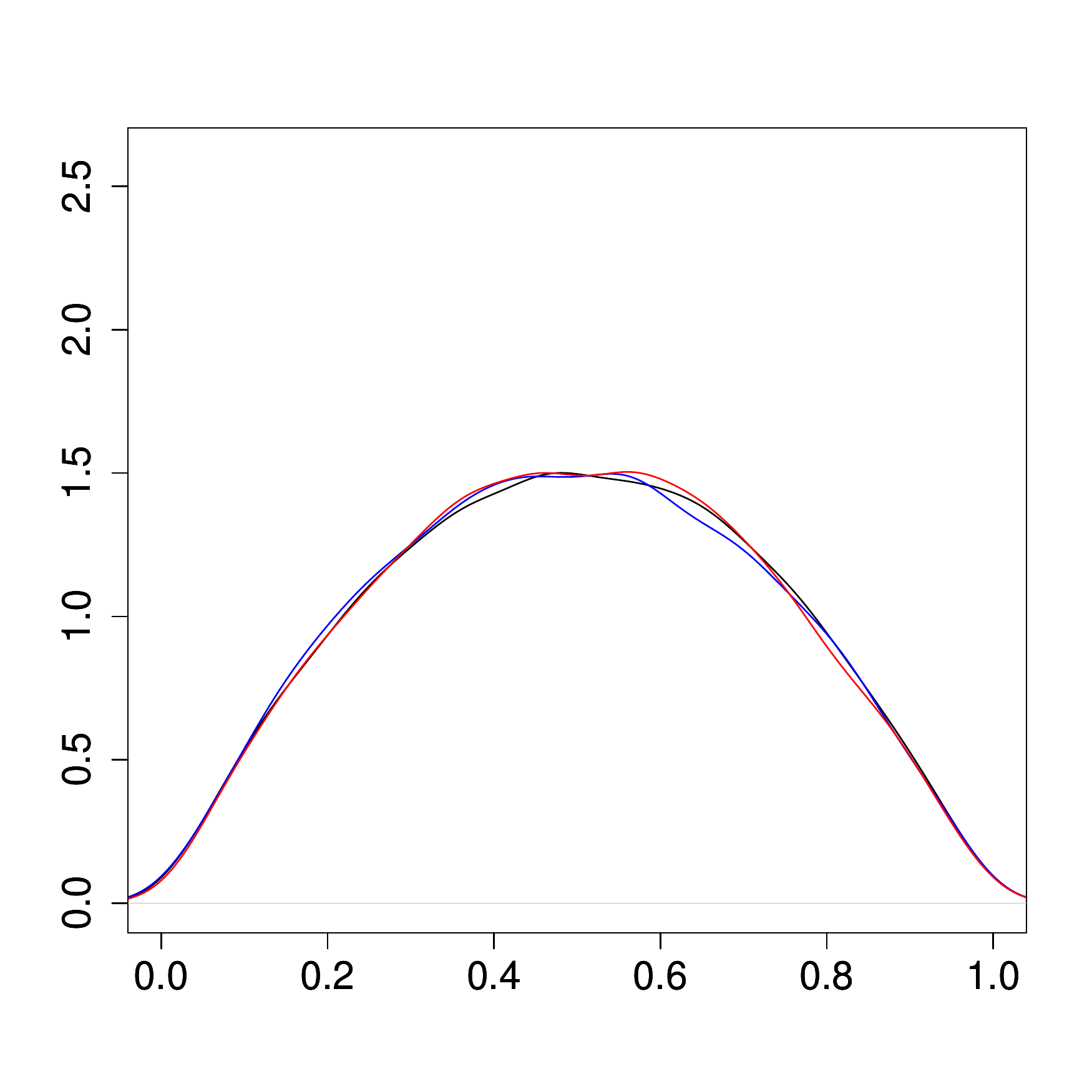}}
\caption{{\bf Pollen data:} Posterior distributions of $\pi_{ik}$ corresponding to the full MCMC run for the 
joint posterior associated with $i^*=5353$ with respect to different choices of $\alpha$ and $\sigma^2_x$.
Different colours represent posteriors with respect to different prior choices; 
black corresponds to ($\alpha=1$, $\sigma^2_{x_1}=\sigma^2_{x_2}=10$), blue to ($\alpha=5$, $\sigma^2_{x_1}=\sigma^2_{x_2}=3$),
red to ($\alpha=1$, $\sigma^2_{x_1}=\sigma^2_{x_2}=5$).} 
\label{fig:pollen_pi}
\end{figure}

Figure \ref{fig:pollen_pi} shows the posteriors of $\pi_{ik}$ associated with the pollen data, with respect
to different choices of $\alpha$ and $\sigma^2_x$. 
The posterior modes are significantly greater than zero, again vidicating the importance of
zero-inflated multinomial. 
As in the case of the chironomid data, here also the
posteriors of $\pi_{ik}$ appear to be quite robust with respect to the choices of $\alpha$, $\sigma^2_{x_1}$
and $\sigma^2_{x_2}$ (we assume $\sigma^2_{x_1}=\sigma^2_{x_2}$ for each choice).
The fact that the posteriors of $\pi_{ik}$ remain almost unchanged
even with the relatively large value of $\alpha~(=5)$ which caused our model to overfit the data, confirms that 
the overfit with $\alpha=5$ was caused solely due to increase of the number of mixture components in our Dirichlet process
based response function, and the modeling associated with $\pi_{ik}$ plays no role in it.

\subsection{Response surfaces for the pollen data}
\label{subsec:response_surface_pollen}
\begin{figure}
\begin{center}
\includegraphics[width = 15cm, height = 15cm]{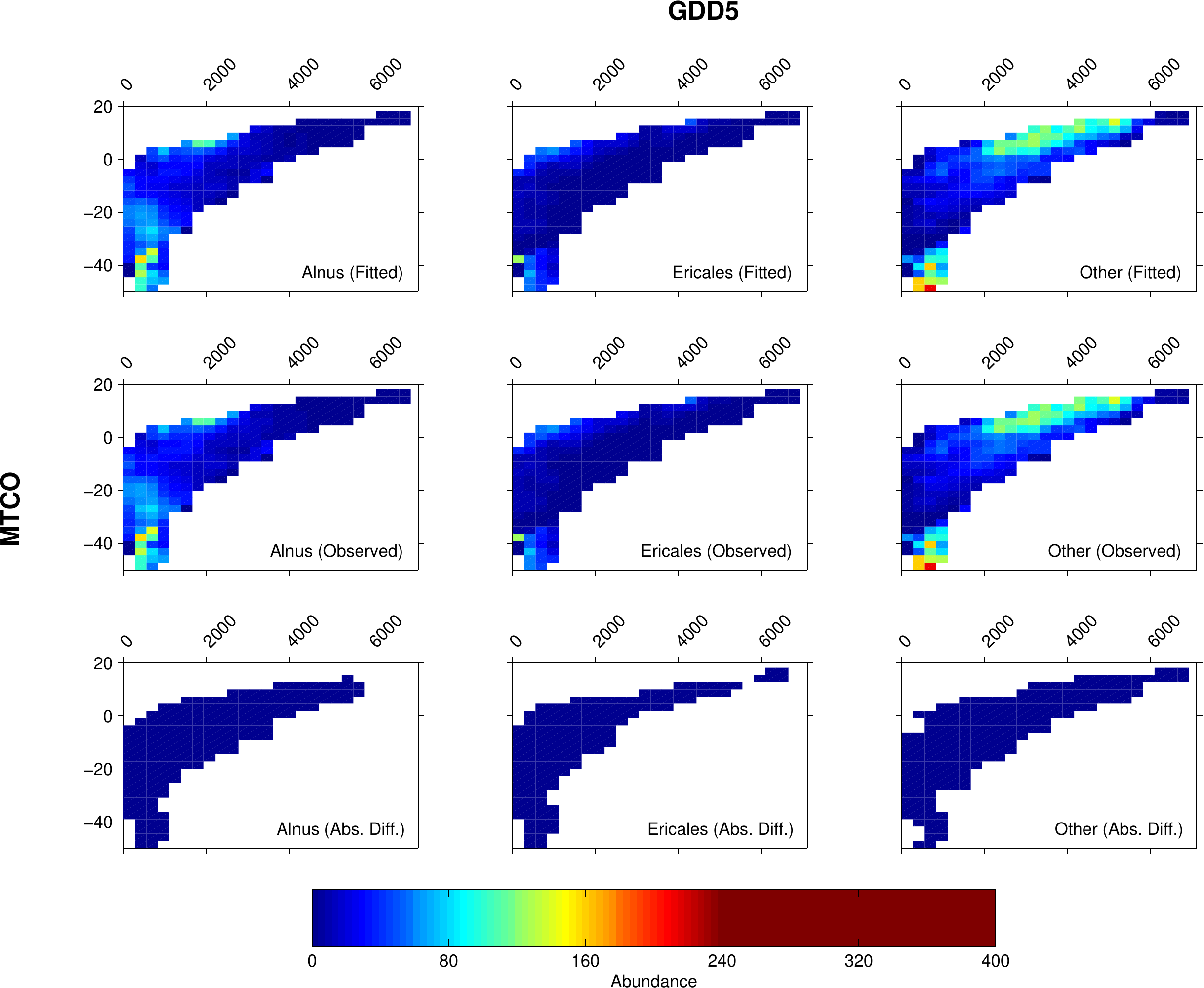}
\caption{{\bf Pollen data:} Fit of the response surfaces for the species {\it Alnus}, 
{\it Ericales} and {\it Other}.}
\label{fig:check_haslett}
\end{center}
\end{figure}

As in the chironomid case, here also we assess the fit of our model-based version of species abundances
to the observed abundances. Figure \ref{fig:check_haslett} displays three such instances, focussing attention
on the pollen species {\it Alnus}, {\it Ericales} and {\it Other}, where the last represents a combination
of the counts of many species (see Appendix A of HWB for the details). Fitting {\it Other} is expected 
to be challenging because the various species amalgamated into the single category 
may respond differently to climate changes. The first row of Figure \ref{fig:check_haslett},
which represent our fitted response surfaces for the above three species, has been constructed
as follows. As in Figure 5 of HWB we construct a support lattice which covers the entire set 
of observed two-dimensional climate points with lattice squares -- within each lattice square, 
we then take averages of the posterior medians of all $\tilde y_{ik}$ that fall within the lattice square.
The second row of Figure \ref{fig:check_haslett} represent the observed response surfaces and is
construced in the same way as the first row, but the posterior medians are replaced with the 
observed abundances. The last row shows the absolute difference in each lattice square 
between the averaged posterior medians and the averaged observed abundances. The spectra of colours
ranging from dark blue to dark red indicate progressively larger abundances ranging from $0$ to $400$.
The plots of the absolute differences in the last row are completely dominated by the dark blue hue,
indicating excellent model fit. These indicate that the response surface modeling style that we adopted here
is quite adequate.

\section{Model adequacy test for the pollen data}
\label{sec:model_glendalough}


Since in this pollen data example the climate variable is bivariate, we
consider the following discrepancy measure and its variants: 
\begin{eqnarray}
\mathcal T_1(\bX) = \sum_{i=1}^{n} (\bx_i-\tilde \bx^*_i)^{'}\bS^{-1}(\bx_i-\tilde\bx^*_i), 
\label{eq:modern_statistic}
\end{eqnarray}
where $\tilde \bx^*_{i}=(\tilde x^*_{i1},\tilde x^*_{i2})$ is the mode of the $i$-th cross-validation posterior,
and $\bS$ is the covariance matrix of $\tilde\bx$ based on the IRMCMC samples.
Obviously, the above measure can be straightforwardly extended
to functions of any number of variables. Variants of the above measure, such as square root
of the quadratic form, replacing the mode of $\tilde\bx$ with the median of $\tilde\bx$,
can be easily considered.

\begin{figure}
\begin{center}
\includegraphics[width = 6cm, height = 7cm]{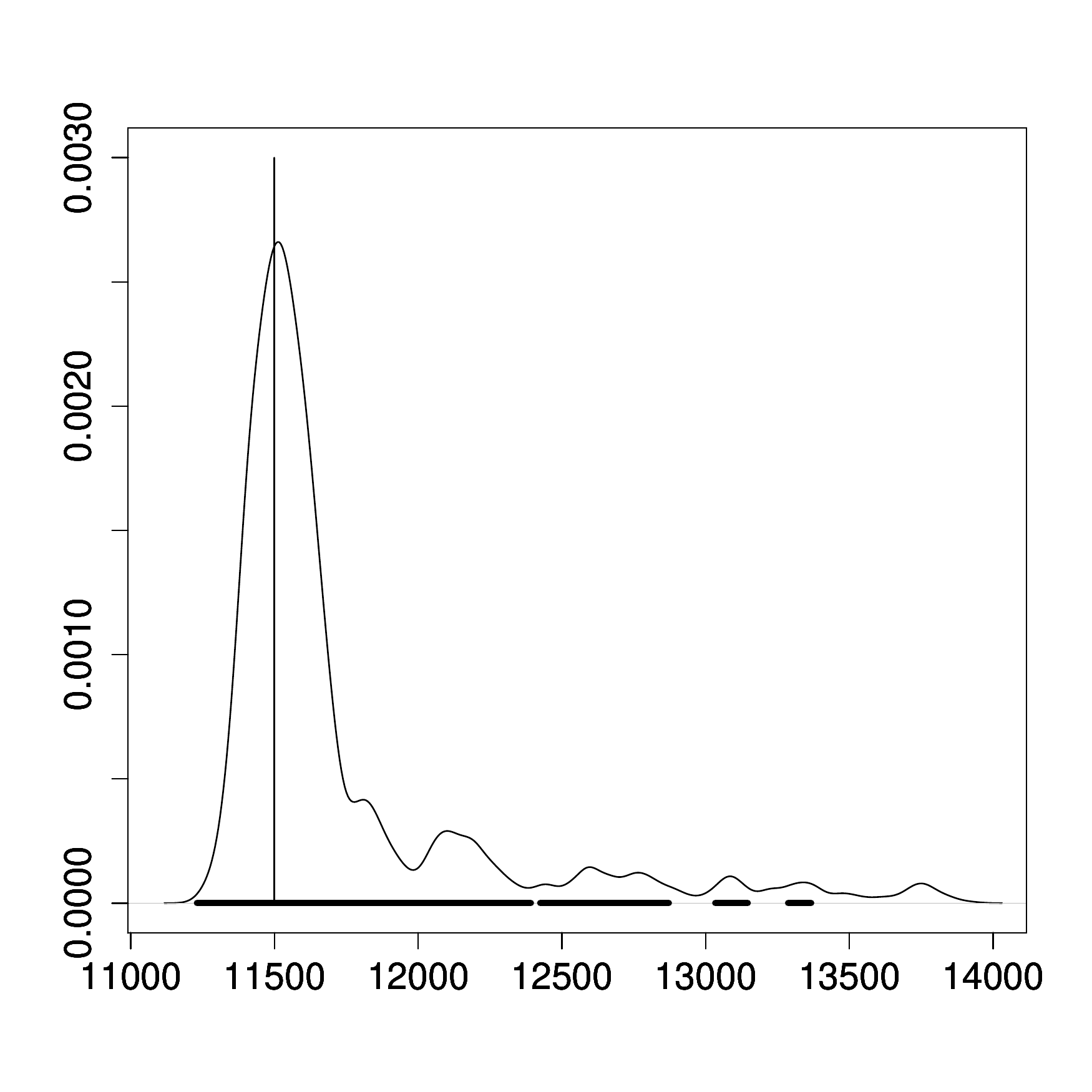}
\caption{{\bf Model adequacy test for the pollen data}: Shown is the posterior distribution
of $\mathcal T_1(\tilde\bX)$ where
the thick line in the base represents the 95\% HPD interval and the vertical line indicates
the observed discrepancy measure $\mathcal T_1(\bX)$; here $\bx^*_{i}=(\tilde x^*_{i1},\tilde x^*_{i2})$
denote the co-ordinate-wise modes of the $i$-th cross-validation posterior.}
\label{fig:ade_modern1}
\end{center}
\end{figure}

\begin{figure}
\begin{center}
\includegraphics[width = 6cm, height = 7cm]{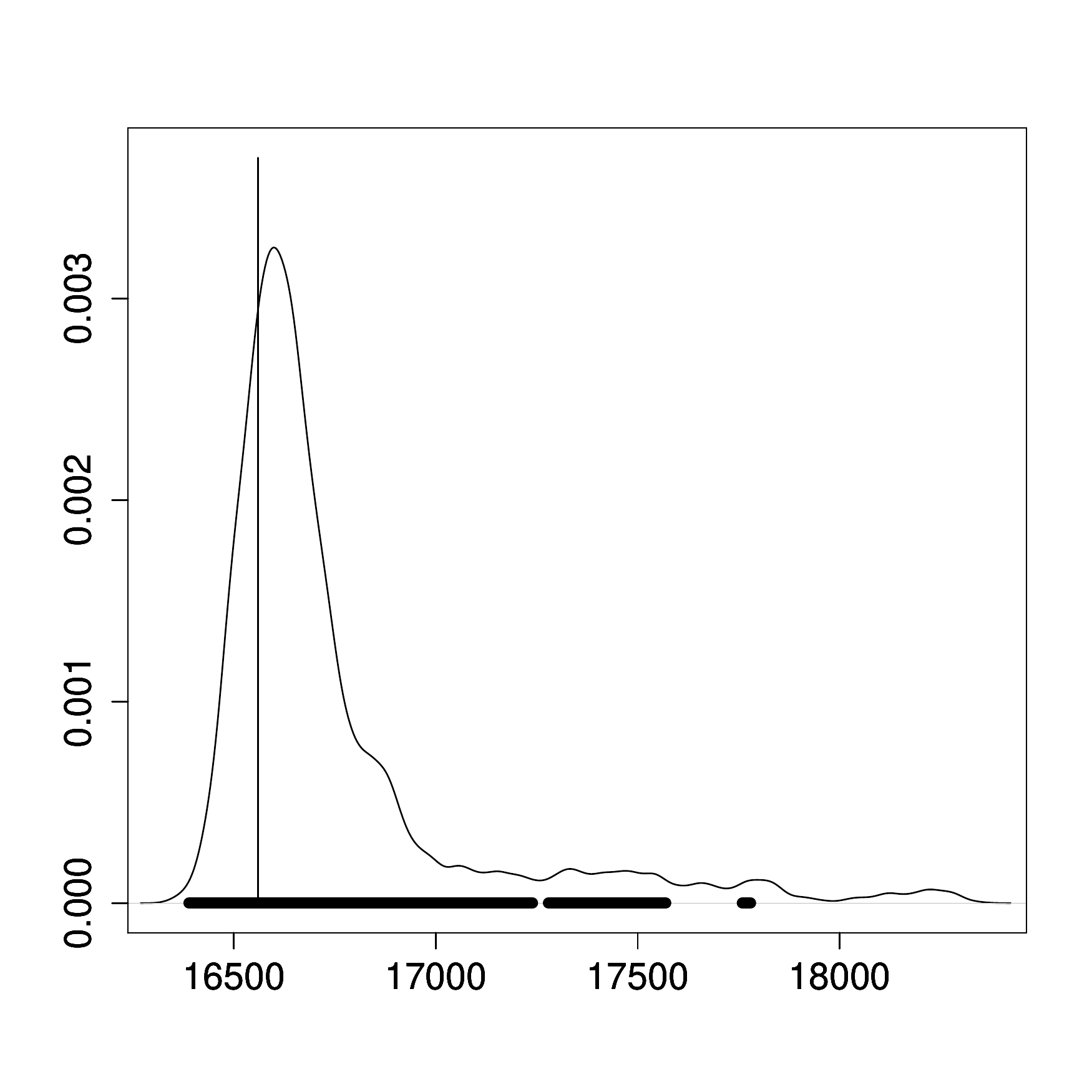}
\caption{{\bf Model adequacy test for the pollen data}: Shown is the posterior distribution
of $\mathcal T_1(\tilde\bX)$ where
the thick line in the base represents the 95\% HPD interval and the vertical line indicates
the observed discrepancy measure $\mathcal T_1(\bX)$; here $\bx^*_{i}=(\tilde x^*_{i1},\tilde x^*_{i2})$
denote the co-ordinate-wise medians of the $i$-th cross-validation posterior.}
\label{fig:ade_modern2}
\end{center}
\end{figure}

Shown in Figures \ref{fig:ade_modern1} and \ref{fig:ade_modern2} are the posterior distributions
of $\mathcal T_1(\tilde \bX)$ along with the corresponding observed discrepancy measure $\mathcal T_1(\bX)$, when 
$\tilde \bx^*_{i}=(\tilde x^*_{i1},\tilde x^*_{i2})$ are the co-ordinate-wise modes and medians,
respectively, of the $i$-th cross-validation posterior. Both the figures clearly indicate that
our model very satisfactorily passes the model adequacy test of \ctn{Bhattacharya12}.



As in the case of chironomid, here also we consider the discrepancy measure based on the sum 
of the logarithms of the cross-validation posterior distributions:
\begin{equation} 
\mathcal T_2(\tilde \bX)=\sum_{i=1}^n\log\pi(\tilde \bx_i\vert \bX_{-i},\bY),\quad\mbox{so that}\quad
\mathcal T_2(\bX)=\sum_{i=1}^n\log\pi(\bx_i\vert \bX_{-i},\bY). 
\label{eq:pollen_kl_like} 
\end{equation}

\begin{figure}
\begin{center}
\includegraphics[width = 6cm, height = 7cm]{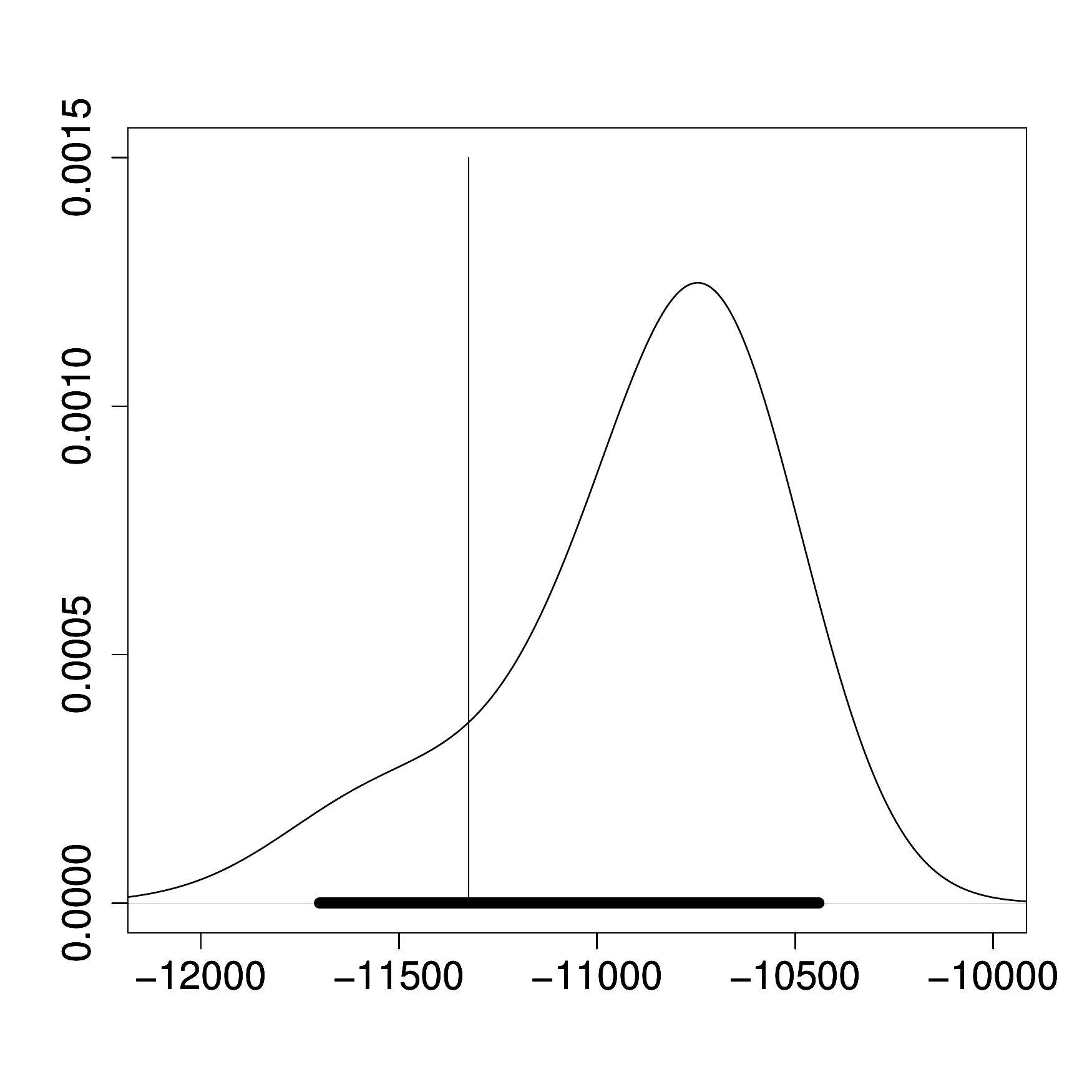}
\caption{{\bf Model adequacy test for the pollen data}: Shown is the posterior distribution
of $\mathcal T_2(\tilde\bX)$ where
the thick line in the base represents the 95\% HPD interval and the vertical line indicates
the observed discrepancy measure $\mathcal T_2(\bX)$.}
\label{fig:pollen_kl_like}
\end{center}
\end{figure}

Figure \ref{fig:pollen_kl_like} shows that the observed discrepancy measure $\mathcal T_2(\bX)$ falls
comfortably within the 95\% HPD region of the inverse reference distribution associated with $\mathcal T_2(\tilde\bX)$,
indicating that our model passes the model adequacy test even with respect to $\mathcal T_2$. 

\section{Conclusions and future work}
\label{sec:conclusions}

Our work can be considered to be the necessary stepping stone to full-fledged
palaeoclimate reconstructions. 
Indeed, the fact that the same modelling idea
is able to fit both the chironomid and the pollen data vindicates the generality
of our model; it is only natural to expect that the same model and methodologies
developed in this paper will be able to reconstruct past Holocene temperature (\ctn{Korhola02})
as well as past Irish climate (HWB). In fact, we see no reason why our model and methods
will not be appropriate for predicting and analysing past climates of any other places of interest.

A very important advantage of our model is that it is relatively simple and is 
quite cheap computationally, with TMCMC playing an important role in this regard.
For massive palaeoclimate datasets meant for climate reconstruction, this will certainly
turn out to be of great value. 

In the current work on cross-validation of modern, training data sets, we have ignored the spatial aspects
of the data sets. However, since in the training data sets the climate values are recorded, the observed climate values are expected
to have much stronger bearing on inference compared to spatial effects. It seems that the spatial (in fact, spatio-temporal)
effects will play important roles while reconstructing past climates at multiple locations, since in such cases
the past climates are unknown (see also Section 6 of HWB). Our model can be further generalized by incorporating desirable spatio-temporal effects; 
we will report this work elsewhere.

\section*{Acknowledgment}
We are sincerely grateful to the reviewers for providing detailed, constructive, comments on our paper which 
greatly improved the quality of our paper.

\newpage

\normalsize

\title{\vspace{-0.8in}
Supplement to ``An Improved Bayesian Semiparametric Model for Palaeoclimate Reconstruction:
Cross-validation Based Model Assessment"}
\author{Sabyasachi Mukhopadhyay and Sourabh Bhattacharya\thanks{
Sabyasachi Mukhopadhyay is a postdoctoral researcher in Southampton Statistical Sciences Research Institute,
University of Southampton, U. K. and Sourabh Bhattacharya 
is an Assistant Professor in
Bayesian and Interdisciplinary Research Unit, Indian Statistical
Institute, 203, B. T. Road, Kolkata 700108.
Corresponding e-mail: sourabh@isical.ac.in.}}
\date{\vspace{-0.5in}}
\maketitle

\end{comment}




\begin{center}
{\bf\Large SUPPLEMENT}
\end{center}

\setcounter{section}{0}
\setcounter{equation}{0}

\renewcommand\thefigure{S-\arabic{figure}}
\renewcommand\thetable{S-\arabic{table}}
\renewcommand\thesection{S-\arabic{section}}

\section{Updating procedure using a combination of Gibbs, Metropolis-Hastings and additive TMCMC steps}
\label{subsection:updating_procedure}

\subsection{Full conditionals of $z_{ik}$}
\label{subsubsec:fullcond_z}
If $y_{ik}\neq 0$, the full conditional distribution of $z_{ik}$ gives full mass to 0, that is,
\begin{equation}
[z_{ik}=0\mid\cdots]=1 \ \ \mbox{if} \ \ y_{ik}\neq 0.\label{eq:fullcond_z}
\end{equation}
On the other hand, if $y_{ik}=0$,
\begin{align}
[z_{ik}=1\mid\cdots]&=C \pi_{ik}
\prod_{r\neq k:z_{ir}=0}\left(\frac{\lambda_{ir}}{\sum_{\ell\neq k:z_{i\ell}=0}\lambda_{i\ell}}\right)^{y_{ir}};
\label{eq:fullcond_z_1}\\
[z_{ik}=0\mid\cdots]&= C(1-\pi_{ik})
\prod_{r\neq k:z_{ir}=0}\left(\frac{\lambda_{ir}}{\lambda_{ik}+
\sum_{\ell\neq k:z_{i\ell}=0}\lambda_{i\ell}}\right)^{y_{ir}};
\label{eq:fullcond_z_2}
\end{align}
where $C$ is such that (\ref{eq:fullcond_z_1}) + (\ref{eq:fullcond_z_2}) $=1$.

\subsection{Full conditionals of $\pi_{ik}$}
\label{subsubsec:fullcond_pi}
The full conditional of $\pi_{ik}$ is given by
\begin{equation}
[\pi_{ik}\mid\cdots]\propto\pi^{z_{ik}}_{ik}(1-\pi_{ik})^{1-z_{ik}}.
\label{eq:fullcond_pi}
\end{equation}
In other words, $\pi_{ik}\sim Beta(z_{ik}+1,2-z_{ik})$.

\subsection{Full conditionals of $\lambda_{ik}$}
\label{subsubsec:fullcond_lambda}
The full conditional distribution of $\lambda_{ik}$ is given by
\begin{equation}
[\lambda_{ik}\mid\cdots]\propto\prod_{r:z_{ir}=0}\left(\frac{\lambda_{ir}}{\sum_{\ell:z_{i\ell}=0}\lambda_{i\ell}}\right)^{y_{ir}}
\times\exp\left\{-\lambda_{ik}/\psi\right\}\lambda^{\xi_{ik}-1}_{ik}.
\label{eq:fullcond_lambda}
\end{equation}
Note that if $z_{ik}=1$, implying $y_{ik}=0$, then the above full conditional boils down to just the prior of $\lambda_{ik}$
given by the second factor of (\ref{eq:fullcond_lambda}).
So, even though (\ref{eq:fullcond_lambda}) is not amenable to straightforward sampling when $z_{ik}=0$, for $z_{ik}=1$,
one would simply sample from the $Gamma(\xi_{ik},1/\psi)$ prior of $\lambda_{ik}$.
We shall use the additive TMCMC methodology with approximately optimized scaling constants to update the entire set 
of $\lambda_{ik}$ corresponding to $z_{ik}=0$ in a single block.

\subsection{Full conditionals of $\btheta_{kj}$}
\label{subsubsec:fullcond_theta}

The full conditional distribution of $\btheta_{kj}$ is given by the following:
\begin{equation}
[\btheta_{kj}\mid\cdots]\propto\prod_{i=1}^n
\frac{\lambda^{\xi_{ik}-1}_{ik}}{\psi^{\xi_{ik}}\Gamma(\xi_{ik})}
\times[\btheta_{kj}\mid\bTheta_{-kj}],
\label{eq:fullcond_theta}
\end{equation}
where $\bTheta_{-kj}=\bTheta_k\backslash\{\btheta_{kj}\}$, and, $[\btheta_{kj}\mid\bTheta_{-kj}]$, which follows
from the Polya urn scheme, 
is given by
\begin{equation}
[\btheta_{kj}\mid\bTheta_{-kj}]
\sim \frac{\alpha G_0(\btheta_{kj})}{\alpha+M_k-1}
+\sum_{\ell=1;\ell\neq j}^{M_k}\frac{\delta_{\btheta_{k\ell}}(\btheta_{kj})}{\alpha+M_k-1}.\label{eq:polya3}
\end{equation}
It is clear that it is not straightforward to simulate from (\ref{eq:fullcond_theta}). 
Also notice that continuous distributions, for example,
normal random walk will not be appropriate in this case since $\btheta_{kj}$ has a discrete, not a continuous
distribution. Because of similar reasons TMCMC is not valid either. As a result, following \ctn{Bhatta06} 
we shall employ (\ref{eq:polya3}) as a proposal distribution for updating $\btheta_{kj}$ using a Metropolis-Hastings step.
A key advantage of using this proposal is that the factor $[\btheta_{kj}\mid\bTheta_{-kj}]$ does not appear
in the Metropolis-Hastings ratio, thus simplifying proceedings to a large extent. 

\section{IRMCMC}
\label{subsec:irmcmc}
Our proposed procedure can be stated in the following manner.
\begin{itemize}
\item[1.] Choose an initial case $i^*$. Use $[x,\bTheta,\Pi,\bLambda,Z\mid \bX_{-i^*},\bY]$ as the importance sampling density,
where $\bX_{-i^*}=\{x_1,\ldots,x_{i^*-1},x_{i^*+1},\ldots,x_n\}$.
\ctn{Bhatta07} demonstrate
that an appropriate $i^*$ may be obtained by minimizing a certain distance function. However, as shown 
in \ctn{Bhatta07}, in cases where 
the importance weights does not depend upon the count data $\bY$, this distance functions leads to that $i^*$ for which $x_{i^*}$
is the median of $\bX$. As shown below, in our case also the importance weights are independent of $\bY$, implying that
$i^*=\{i:x_i=median(\bX)\}$.
 \item[2.] From this density, sample, using MCMC,\\ 
 $(x^{(\ell)},\bTheta^{(\ell)},\Pi^{(\ell)},\bLambda^{(\ell)},Z^{(\ell)});\ell=1,\ldots,L$, for large $L$.
 \item[3.] For $i \in \left\{1,\ldots,i^{*}-1,i^{*}+1,\ldots,n\right\}$ do,
 \begin{enumerate}
  \item [a.] For each sample value $(x^{(\ell)},\bTheta^{(\ell)},\Pi^{(\ell)},\bLambda^{(\ell)},Z^{(\ell)})$, 
  compute importance weights $w_{i^{*},i}^{(\ell)}$=$w_{i^{*},i}(x^{(\ell)},\bTheta^{(\ell)},\Pi^{(\ell)},\bLambda^{(\ell)},Z^{(\ell)})$, 
  where the importance weight function is given by
  \begin{equation}
   w_{i^{*},i}(x,\bTheta,\Pi,\bLambda,Z)
   =\frac{L(i,x,\bTheta,\Pi,\bLambda,Z,x_{i^*})}{L(i^*,x,\bTheta,\Pi,\bLambda,Z,x_{i})},
   \end{equation}
   where 
\begin{equation}
L(i,x,\bTheta,\Pi,\bLambda,Z,x_{i^*})=
\prod_{k=1}^m\frac{\lambda^{\xi_{ik}(x)-1}_{ik}}{\psi^{\xi_{ik}(x)}\Gamma(\xi_{ik}(x))}
\times\prod_{k=1}^m\frac{\lambda^{\alpha_{i^*k}(x_{i^*})-1}_{i^*k}}{\psi^{\alpha_{i^*k}(x_{i^*})}\Gamma(\alpha_{i^*k}(x_{i^*}))},
\label{eq:weights1}
\end{equation}
and
\begin{equation}
L(i^*,x,\bTheta,\Pi,\bLambda,Z,x_{i})=
\prod_{k=1}^m\frac{\lambda^{\xi_{ik}(x_i)-1}_{ik}}{\psi^{\xi_{ik}(x_i)}\Gamma(\xi_{ik}(x_i))}
\times\prod_{k=1}^m\frac{\lambda^{\alpha_{i^*k}(x)-1}_{i^*k}}{\psi^{\alpha_{i^*k}(x)}\Gamma(\alpha_{i^*k}(x))},
\label{eq:weights2}
\end{equation}
The arguments corresponding to $\xi_{ik}$ and $\alpha_{i^*k}$ in (\ref{eq:weights1}) and (\ref{eq:weights2})
show the appropriate climate values (random or observed) corresponding to the response functions. Note that 
$w_{i^*,i}$ does not depend upon the count data $\bY$. As a result, following \ctn{Bhatta07}, we recommend selecting
$i^*=\{i:x_i=median(\bX)\}$.

   \item[b.]For $r \in \left\{1,\ldots,K_1\right\} $
   \begin{enumerate}
    \item [(i)]Sample $(\tilde x^{(r)},\tilde\bTheta^{(r)},\tilde\Pi^{(r)},\tilde\bLambda^{(r)},\tilde Z^{(r)})$ from 
    $(x^{(\ell)},\bTheta^{(\ell)},\Pi^{(\ell)},\bLambda^{(\ell)},Z^{(\ell)});\ell=1,\ldots,L$
    {\it without replacement}, where the probability of
     sampling $(x^{(r)},\bTheta^{(r)},\Pi^{(r)},\bLambda^{(r)},Z^{(r)})$ is proportional to $w_{i^{*},i}^{(r)}$.
     \item[(ii)]For fixed $(x,\bTheta,\Pi,\bLambda,Z)=
     (\tilde x^{(r)},\tilde \bTheta^{(r)},\tilde \Pi^{(r)},\tilde\bLambda^{(r)},\tilde Z^{(r)})$, 
     draw, using MCMC, $K_2$ times from $[x,\pi_{i1},\ldots,\pi_{im},\lambda_{i1},\ldots,\lambda_{im},z_{i1},\ldots,z_{im}\mid\cdots]$, 
     following the relevant details provided in Sections \ref{subsubsec:fullcond_z}, \ref{subsubsec:fullcond_pi},
     \ref{subsubsec:fullcond_lambda}, and Section 4.1 of our main manuscript \ctn{Sabya13b}. 
      \end{enumerate}
      \item[c.] Store the $K_1\times K_2$ draws of $x$ as the posterior for $x_{i}$ 
      as $\hat{x}_{i}^{(1)},\ldots,\hat{x}_{i}^{(K_1K_2)}$.
      \end{enumerate}
      \end{itemize}

\section{Relationship of our discrepancy measure with other discrepancy measure 
using logarithms of the cross-validation posteriors}
\label{subsec:other_discrepancies}

Consider the following variant of the discrepancy measure $D_1$ proposed in equation (16) of Section 6 of our main
manuscript \ctn{Sabya13b}: 
\begin{eqnarray} 
D_1(\tilde \bX)&=&\sum_{i=1}^n\left\vert\left\{\log\pi(\tilde x_i\vert \bX_{-i},\bY)-\log\pi(\tilde x^*_i\vert \bX_{-i},\bY)\right\}\right\vert\nonumber\\ 
&=&\sum_{i=1}^n\left\vert\log\frac{\pi(\tilde x_i\vert \bX_{-i},\bY)}{\pi(\tilde x^*_i\vert \bX_{-i},\bY)} \right\vert, \label{eq:kl_1} 
\end{eqnarray} 
so that 
\begin{eqnarray} 
D_1(\bX) &=&\sum_{i=1}^n\left\vert\log\frac{\pi(x_i\vert \bX_{-i},\bY)}
{\pi(\tilde x^*_i\vert \bX_{-i},\bY)}\right\vert.  \label{eq:kl_2} 
\end{eqnarray} 
In the above, $\tilde x^*_i$ can be either the median or the mode of the $i$-th cross-validation posterior.
We consider two cases -- in the first case we investigate the relationship between the discrepancy measure 
$D_1$, given by (\ref{eq:kl_1}) (and its variant) and $T_1$, given by (15) of our main manuscript, 
letting $\tilde x^*_i$ be the
median. In the second case, we investigate such relationships denoting the posterior mode by $\tilde x^*_i$. 

\subsection*{Case 1: $\tilde x^*_i$ is the median of the cross-validation posterior}

Following \ctn{Bhattacharya12}, under the ``0-1" loss function, we accept the model 
if the posterior probability 
$P\left(\frac{\left\vert D_1(\tilde\bX)-D_1(\bX)\right\vert}{\sqrt{Var\left\{D_1(\tilde\bX)\vert\bY\right\}}}
\leq\epsilon\right)$ exceeds $1/2$; as a rule of thumb, we may choose $\epsilon$ as the desired percentile of 
$\frac{\left\vert  D_1(\tilde \bX)\right\vert}{\sqrt{Var\{D_1(\tilde X)\vert\bY\}}}$; see \ctn{Bhattacharya12}. 

Now note that 
\begin{align} 
&P\left(\frac{\left\vert D_1(\tilde\bX)-D_1(\bX)\right\vert}{\sqrt{Var\left\{D_1(\tilde\bX)\vert\bY\right\}}}\leq\epsilon\right)\notag\\ 
&= P\left(\frac{\left\vert\sum_{i=1}^n\left\vert g(\tilde x_i)-g(\tilde x^*_i)\right\vert 
-\sum_{i=1}^n\left\vert g(x_i)-g(\tilde x^*_i)\right\vert\right\vert}
{\sqrt{Var\left\{D_1(\tilde\bX)\vert\bY\right\}}}\leq\epsilon\right), 
\label{eq:discrepancy1} 
\end{align} 
where, for any $x$, $g(x)=\log\pi(x\vert\bX_{-i},\bY)$. 

Taylor's series expansion up to the first order about $\tilde x^*_i$ yields 
\begin{align}
\left\vert g(\tilde x_i)-g(\tilde x^*_i)\right\vert &= \left\vert\tilde x_i-\tilde x^*_i\right\vert
\left\vert g'(u_i)\right\vert;\notag\\
\left\vert g(x_i)-g(\tilde x^*_i)\right\vert &= \left\vert x_i-\tilde x^*_i\right\vert \left\vert g'(v_i)\right\vert,\notag
\end{align}
where $u_i$ lies between $\tilde x_i$ and $\tilde x^*_i$ and $v_i$ lies between $x_i$ and $\tilde x^*_i$.
We now assume that $g'(\cdot)$ is continuous and that for $i=1,\ldots,n$, $u_i$ and $v_i$ are contained in a
small interval so that $g'(\cdot)$ is approximately constant in that interval thanks to continuity.
Such an assumption can be expected to hold in practice if the observed climate data $x_i$, after suitable
scaling if required, have small empirical variance, so that they lie close together. 
The posterior medians then are also expected to be close to each other, that is, they are expected
to lie in a small interval. The assumption that $g'(\cdot)$ is continuous on small intervals is 
expected to hold very generally.

It then holds that
for $i=1,\ldots,n$, 
$\left\vert g'(u_i)\right\vert\approx\left\vert g'(v_i)\right\vert\approx c~(>0)$.
Also, $Var\left\{D_1(\tilde\bX)\vert\bY\right\}\approx c^2Var\left\{T(\tilde\bX)\vert\bY\right\}$.
Hence, (\ref{eq:discrepancy1}) becomes
\begin{align}
&P\left(\frac{\left\vert D_1(\tilde\bX)-D_1(\bX)\right\vert}{\sqrt{Var\left\{D_1(\tilde\bX)\vert\bY\right\}}}\leq\epsilon\right)
\approx P\left(\frac{\left\vert T_2(\tilde\bX)-T_2(\bX)\right\vert}{\sqrt{Var\left\{T_2(\tilde\bX)\vert\bY\right\}}}\leq\epsilon\right),
\label{eq:discrepancy2}
\end{align}
where 
\begin{equation}
T_2(\tilde\bX)=\sum_{i=1}^n\left\vert \tilde x_i-\tilde x^*_i\right\vert\hspace{4mm}\mbox{and}\hspace{4mm}
T_2(\bX)=\sum_{i=1}^n\left\vert x_i-\tilde x^*_i\right\vert.
\label{eq:discrepancy3}
\end{equation}
The difference between $T_2$ above and $T_1$ given by (15) of our main manuscript 
is that the latter involves
scaling of each term of the summation by the posterior standard deviation of $\tilde x_i$.
If we scale each term of the summation in $D_1$ by $\sqrt{Var\{g(\tilde x_i)\}}$ and denote the
modified discrepancy measure by $D^*_1$, then again by
invoking the Taylor's series expansion $g(\tilde x_i)=g(\tilde x^*_i)+(\tilde x_i-\tilde x^*_i)g'(u_i)$,
we obtain $\sqrt{Var\{g(\tilde x_i)\}}\approx c\sqrt{Var(\tilde x_i)}$, so that (after cancelling $c$ in the ratios)
\begin{align}
&P\left(\frac{\left\vert D^*_1(\tilde\bX)-D^*_1(\bX)\right\vert}{\sqrt{Var\left\{D^*_1(\tilde\bX)\vert\bY\right\}}}\leq\epsilon\right)
\approx P\left(\frac{\left\vert T_1(\tilde\bX)-T_1(\bX)\right\vert}{\sqrt{Var\left\{T_1(\tilde\bX)\vert\bY\right\}}}\leq\epsilon\right),
\label{eq:discrepancy4}
\end{align}
showing that the discrepancy measures 
\begin{align}
D^*_1(\tilde \bX)&=
\sum_{i=1}^n\frac{\left\vert\left\{\log\pi(\tilde x_i\vert \bX_{-i},\bY)
-\log\pi(\tilde x^*_i\vert \bX_{-i},\bY)\right\}\right\vert}
{\sqrt{Var\left\{\log\pi(\tilde x_i\vert \bX_{-i},\bY)\right\}}}
\hspace{4mm}\mbox{and}\notag\\
T_1(\tilde\bX) &= \sum_{i=1}^{n} \frac{\vert \tilde x_{i}-\tilde x^*_i\vert}{\sqrt{Var(\tilde x_i)}},
\label{eq:discrepancy5}
\end{align}
are approximately equivalent for the purpose of goodness-of-fit test of \ctn{Bhattacharya12}.

\subsection*{Case 2: $\tilde x^*_i$ is the mode of the cross-validation posterior}

When $\tilde x^*_i$ is the mode of the $i$-th cross-validation posterior, we can consider the
following discrepancy measure
\begin{align}
D_2(\tilde \bX)&=
\sum_{i=1}^n\frac{\left\vert\left\{\log\pi(\tilde x_i\vert \bX_{-i},\bY)
-\log\pi(\tilde x^*_i\vert \bX_{-i},\bY)\right\}\right\vert^{1/2}}
{\{Var\left\{\log\pi(\tilde x_i\vert \bX_{-i},\bY)\right\}\}^{1/2}}.
\label{eq:discrepancy6}
\end{align}

Taylor's series expansion around the mode yields
\begin{align}
g(\tilde x_i)&=g(\tilde x^*_i)+\frac{(\tilde x_i-\tilde x^*_i)^2}{2}g''(u^*_i)\hspace{4mm}\mbox{and}\notag\\
g(x_i)&=g(\tilde x^*_i)+\frac{(x_i-\tilde x^*_i)^2}{2}g''(v^*_i),\notag
\end{align}
where $u^*_i$ lies between $\tilde x_i$ and $\tilde x^*_i$, and $v^*_i$ lies between $x_i$ and $\tilde x^*_i$. 
Now, assuming that $g''(\cdot)$ is continuous in a small interval containing $u^*_i$ and $v^*_i$ for 
$i=1,\ldots,n$, implies $\left\vert g''(u^*_i)\right\vert\approx \left\vert g''(v^*_i)\right\vert\approx 
c^*~(>0)$, 
for $i=1,\ldots,n$.
As in the previous case, here also we use the approximation $Var\{g(\tilde x_i)\}\approx c^2Var(\tilde x_i)$, 
using a first order Taylor's series expansion around the posterior median,
instead of the posterior mode. 
This yields
\begin{align}
D_2(\tilde \bX) &\approx \frac{\sqrt{c^*}}{c}T_1(\tilde\bX)\hspace{4mm}\mbox{and}\hspace{4mm}
\sqrt{Var\left\{D_2(\tilde \bX)\vert\bY\right\}} \approx \frac{\sqrt{c^*}}{c}\sqrt{Var\left\{T_1(\tilde\bX)\vert\bY\right\}},\notag
\end{align}
showing that approximate probability equality of the form (\ref{eq:discrepancy4}) holds with $D^*_1$ 
replaced with $D_2$. Hence, when $\tilde x^*_i$ are posterior modes, the discrepancy measures 
$D_2$ and $T_1$ are approximately equivalent
for the goodness-of-fit test of \ctn{Bhattacharya12}. 

It is also clear that the discrepancy measure 
$$D_3(\tilde \bX)=
\sum_{i=1}^n\frac{\left\vert\left\{\log\pi(\tilde x_i\vert \bX_{-i},\bY)
-\log\pi(\tilde x^*_i\vert \bX_{-i},\bY)\right\}\right\vert}
{Var\left\{\log\pi(\tilde x_i\vert \bX_{-i},\bY)\right\}}$$
is approximately equivalent to 
$$T_3(\tilde\bX) = \sum_{i=1}^{n} \frac{\left(\tilde x_{i}-\tilde x^*_i\right)^2}{Var(\tilde x_i)},$$
when $\tilde x^*_i$ is the mode.

\begin{comment}

\renewcommand\baselinestretch{1.3}
\normalsize
\bibliographystyle{ECA_jasa}
\bibliography{irmcmc}

\renewcommand\baselinestretch{1.3}
\normalsize
\bibliographystyle{ECA_jasa}
\bibliography{irmcmc}

\end{document}